\documentclass[reprint,amsmath,amssymbGaps,nofootinbib,aps,prd]{revtex4-1}

\usepackage{amsmath} 
\usepackage{amssymb}
\usepackage{amsfonts}
\usepackage{amstext}
\usepackage{graphicx}
\usepackage{bm}
\usepackage{color} 
\usepackage{transparent}
\usepackage{mathtools}
\usepackage{slashed}
\usepackage{hyperref}

\newcommand{\dal}{\Box \phi}
\newcommand{\dpp}{\left(\nabla \phi \right)^2}

\begin{document}

\title{Observational Constraints on the \\ Regularized 4D Einstein-Gauss-Bonnet Theory of Gravity}

\author{Timothy Clifton}
 \email{t.clifton@qmul.ac.uk}
\author{Pedro Carrilho}
\author{Pedro G. S. Fernandes}
\author{David J. Mulryne}
\affiliation{School of Physics and Astronomy, Queen Mary University of London, Mile End Road, London, E1 4NS, UK}


\begin{abstract}

In this paper we study the observational constraints that can be imposed on the coupling parameter, $\hat \alpha$, of the regularized version of the 4-dimensional Einstein-Gauss-Bonnet theory of gravity. We use the scalar-tensor field equations of this theory to perform a thorough investigation of its slow-motion and weak-field limit, and apply our results to observations of a wide array of physical systems that admit such a description. We find that the LAGEOS satellites are the most constraining, requiring $\vert \hat \alpha \vert \lesssim 10^{10} \,{\rm m}^2$. This constraint suggests that the possibility of large deviations from general relativity is small in all systems except the very early universe ($t<10^{-3}\, {\rm s}$), or the immediate vicinity of stellar-mass black holes ($M\lesssim100\, M_{\odot}$). We then consider constraints that can be imposed on this theory from cosmology, black hole systems, and table-top experiments. It is found that early universe inflation prohibits all but the smallest negative values of $\hat \alpha$, while observations of binary black hole systems are likely to offer the tightest constraints on positive values, leading to overall bounds $0 \lesssim \hat \alpha \lesssim 10^8 \, {\rm m}^2$. 

\end{abstract}

\maketitle

\section{Introduction}\label{intro}

Gauss-Bonnet terms routinely occur in low-energy effective actions for gravity, and in particular show up as the leading-order correction in heterotic string theory \cite{zwiebach1985curvature, nepomechie1985low}, from the large-N expansion of dual field theories in the AdS-CFT correspondence \cite{maldacena1999large, gubser1998gauge, witten1998anti}, and as a result of considering quantum fields in curved space-times \cite{birrell1984quantum}. They are also known to be the unique quadratic combination of the Riemann tensor that leads to ghost-free, second-order field equations \cite{Lovelock_original}. As such, they are of broad theoretical interest.

In $D=4$ dimensions, the consequences of making such a correction to the gravitational action have often been studied by coupling to a scalar field dilaton with canonical kinetic term to a Gauss-Bonnet term. Black holes in theories with linear \cite{Sotiriou:2013qea, Sotiriou:2014pfa, Saravani:2019xwx, Delgado:2020rev}, quadratic \cite{Doneva:2017bvd, Silva:2017uqg, Antoniou:2017acq, Cunha:2019dwb, Collodel:2019kkx}, and exponential couplings \cite{Kanti:1995vq, Kleihaus:2011tg, Kleihaus:2015aje, Cunha:2016wzk, Blazquez-Salcedo:2017txk} have all been well studied in this context, as have their cosmologies \cite{
Nojiri:2005vv, Jiang:2013gza, Kanti:2015pda, Chakraborty:2018scm, Odintsov:2018zhw, Odintsov:2019clh, Odintsov:2020zkl}. In this paper we study a variant of these theories that results from the dimensional regularization of gravity with a Gauss-Bonnet term \cite{previous, otherone}, using a method originally introduced to study Einstein's theory in lower dimensions \cite{2Dpaper}, and with the motivation of constraining quantum corrections to the gravitational action.

Interest in the dimensional regularization of gravity with a Gauss-Bonnet term was recently spurred by the development of {\it 4D Einstein-Gauss-Bonnet} (4DEGB) gravity \cite{original, EGB2DIM, EGB1, EGB2, Wei:2020ght, Jin:2020emq, EGB3, EGB4, EGB5, EGB6, EGB7, EGB8, EGB9, EGB10, EGB11, EGB12, EGB13, EGB14, EGB15, EGB16, EGB17, EGB18, EGB19, EGB20, EGB21, EGB22, EGB23, EGB24, EGB25, EGB26, EGB27, EGB28, EGB29, EGB30, EGB31, EGB32, Guo:2020zmf, HosseiniMansoori:2020yfj, Shu:2020cjw, Mahapatra:2020rds, Bonifacio:2020vbk, Jusufi:2020yus, Ge:2020tid, Churilova:2020mif, Arrechea:2020evj, Kumar:2020sag, Ghosh:2020cob, Yang:2020jno, Lu:2020mjp, Ma:2020ufk, Jusufi:2020qyw, Konoplya:2020der, Qiao:2020hkx, Haghani:2020ynl, Liu:2020lwc, Samart:2020sxj, Aoki:2020lig, Dadhich:2020ukj, Aoki:2020iwm, Easson:2020mpq, Hennigar:2020zif, Singh:2020mty, Lin:2020kqe, yang2020weak, feng2020theoretical, hennigar2020lower}. This theory was originally thought to contain only a single spin-2 degree of freedom \cite{original}, a claim that was later shown to be untrue \cite{EGB31}. Nevertheless, the proposed mechanism remains of interest, and can be used to study the consequences of a Gauss-Bonnet term on 4-dimensional gravitational physics. We consider the observational consequences of the regularized version of 4DEGB theory \cite{previous, otherone}; a scalar-tensor theory that belongs to the Horndeski class \cite{Horndeski:1974wa, HorndeskiReview}, and which shares solutions with the original version of 4DEGB \cite{EGBST1, EGBST2, Lin:2020kqe}.

In order to introduce this theory let us begin with the Einstein-Gauss-Bonnet theory, which is generated from the following Lagrangian:
\begin{equation} \label{love}
\mathcal{L} =  R + \alpha \, \mathcal{G} \, ,
\end{equation}
where $\alpha$ is a constant, and where
\begin{equation} \label{gb}
\mathcal{G}=R_{\alpha \beta \mu \nu} R^{\alpha \beta \mu \nu} - 4 R_{\mu \nu} R^{\mu \nu} +R^2
\end{equation}
is the Gauss-Bonnet invariant. Unfortunately, the integral of the Gauss-Bonnet term is a topological invariant in four dimensions, and so the extra term in Eq. (\ref{love}) does not contribute to the field equations in $D=4$. 

The novelty introduced in Ref. \cite{original}, which was aimed at avoiding this limitation, was to re-scale the coupling constant by
\begin{equation} \label{alpha}
\alpha = \frac{\hat{\alpha}}{D-4} \, ,
\end{equation}
where $\hat{\alpha}$ is a new constant. Taking the limit $D \rightarrow 4$ then means that $\alpha$ diverges as the new terms that occur in the field equations would otherwise vanish. A well defined set of scalar-tensor field equations for the theory that results from such a procedure were eventually written down in Refs. \cite{previous, otherone}, where counterterms were introduced into the action in order to remove divergences.

The regularized theory that results from this process is an interesting one, for a number of reasons. First, it coincides precisely with the theory obtained from performing a suitable Kaluza-Klein reduction of Gauss-Bonnet theory in higher dimensions \cite{EGBST1, EGBST2}. Second, it also coincides precisely with the effective action that one obtains for the trace anomaly that results from the broken conformal symmetry of massless fields in quantum theory \cite{Riegert:1984kt,Komargodski:2011vj} (up to arbitrary conformally invariant functional as an integration constant \cite{shapiro1994gauge}). This means that it can be viewed as both a dimensionally-reduced theory {\it and} as a gravitational theory that displays known quantum corrections.

From a phenomenological perspective, this theory is also extremely interesting. It is known to be free from Ostrogradski instabilities, and to admit simple black hole and cosmological solutions that reduce to those of General Relativity (GR) in a suitable limit. The black hole solutions admit a repulsive force near their centre, which has consequences for the singularity problem, while presenting logarithmic corrections to the Hawking-Bekenstein entropy area formula (as predicted by some quantum theories of gravity such as Loop Quantum Gravity and String Theory \cite{Rovelli:1996dv, Kaul:2000kf}). The cosmologies also allow for accelerating solutions, which could be of potential interest for the physics of the early universe. Along with all of this, the theory contains only a single additional free parameter $\hat{\alpha}$, which makes it particularly well defined.

In this paper we consider the observational constraints that can be imposed on the regularized 4DEGB theory of gravity. In Section \ref{sec:fe} we re-introduce the field equations from Refs. \cite{previous, otherone, EGBST1, EGBST2}. We then use them to derive the form of gravitational interactions in the slow-motion, weak-field limit of the theory in Section \ref{sec:weak}. The dynamical problem of two gravitationally interacting massive bodies is considered in Section \ref{sec:2bd}, where constraints are imposed on $\hat{\alpha}$ from observations made in relevant physical systems. In Section \ref{sec:rad} we proceed to study the constraints that can be imposed from observations of the propagation of electromagnetic and gravitational radiation. Finally, in Section \ref{sec:other} we consider order-of-magnitude constraints that can be imposed on regularized 4DEGB from cosmology, black hole physics, and table top experiments of gravity.  Overall, we find that the LAGEOS satellites provide the tightest bounds on the coupling parameter $\hat{\alpha}$, but that observations based on early universe physics or gravitational waves from black hole mergers are ultimately likely to give even tighter constraints. 

We use units such that $G=c=1$, and use Greek letters to denote space-time indices.

\section{Gravitational Field Equations and the Post-Newtonian Expansion}
\label{sec:fe}

In this section we will state the field equations of the theory derived in Refs. \cite{previous, otherone, EGBST1, EGBST2, Komargodski:2011vj}, and explain the slow-motion, weak-field expansion we will apply to them.

\subsection{Gravitational Field Equations}
\label{sec:field}

The action that describes our theory is given by \cite{previous, otherone, EGBST1, EGBST2, Komargodski:2011vj}
\begin{align} \nonumber
S=\int_{\mathcal{M}} d^{4} x \sqrt{-g}\big[&R-\hat{\alpha}\big(\phi \mathcal{G}-4 G^{\mu \nu} \nabla_{\mu} \phi \nabla_{\nu} \phi\\&   -4 \square \phi(\nabla \phi)^{2}-2(\nabla \phi)^{4}\big)\big]+S_{m},
\label{eq:action}
\end{align}
where $\phi$ is an extra scalar gravitational degree-of-freedom and $S_m$ describes the matter content. The field equations of the action \eqref{eq:action} are
\begin{equation} \label{feqs0}
    G_{\mu \nu} =  \hat{\alpha} \hat{\mathcal{H}}_{\mu \nu} +8\pi T_{\mu \nu} \, ,
\end{equation}
where $\hat{\alpha}$ is the re-scaled coupling constant from (\ref{alpha}), and where $\hat{\mathcal{H}}_{\mu \nu}$ is given by
\begin{widetext}
\begin{equation} \label{feqs}
\begin{aligned}
-\frac{1}{4} \hat{\mathcal{H}}_{\mu\nu} =& \frac{1}{2} G_{\mu \nu} \left(\nabla \phi\right)^2 - P_{\mu \alpha \nu \beta} \left( \nabla^{\alpha} \phi \nabla^{\beta}\phi - \nabla^{\beta} \nabla^{\alpha} \phi \right) + \left( \nabla_{\alpha} \phi \nabla_{\mu} \phi - \nabla_{\alpha} \nabla_{\mu} \phi \right) \left( \nabla^{\alpha} \phi \nabla_{\nu} \phi - \nabla^{\alpha} \nabla_{\nu} \phi \right) \\
&+ \left(\nabla_{\mu} \phi \nabla_{\nu} \phi - \nabla_{\nu} \nabla_{\mu} \phi \right) \square \phi +\frac{1}{4} g_{\mu \nu} \left( 2 \left( \square \phi \right)^2 - \left( \nabla \phi\right)^4 + 2 \nabla_{\beta} \nabla_{\alpha} \phi \left( 2 \nabla^{\alpha} \phi \nabla^{\beta} \phi - \nabla^{\beta} \nabla^{\alpha} \phi \right) \right) \ , 
\end{aligned}
\end{equation}
where $P_{\alpha \beta \mu \nu} \equiv *R*_{\alpha \beta \mu \nu} = R_{\alpha \beta \mu \nu}+g_{\alpha \nu} R_{\beta \mu}-g_{\alpha \mu} R_{\beta \nu}+g_{\beta \mu} R_{\alpha \nu}-g_{\beta \nu} R_{\alpha \mu}+\frac{1}{2}\left(g_{\alpha \mu} g_{\beta \nu}-g_{\alpha \nu} g_{\beta \mu}\right) R$, is the double dual of the Riemann tensor. The propagation equation for $\phi$ is
\begin{equation} \label{scalareq}
\begin{aligned}
&R^{\mu \nu} \nabla_{\mu} \phi \nabla_{\nu} \phi - G^{\mu \nu}\nabla_\mu \nabla_\nu \phi - \dal \dpp +(\nabla_\mu \nabla_\nu \phi)^2 
- (\dal)^2 - 2\nabla_\mu \phi \nabla_\nu \phi \nabla^\mu \nabla^\nu \phi = \frac{1}{8}\mathcal{G} \, .
\end{aligned}
\end{equation}
It can be seen from these equations that this theory belongs to the Horndeski class \cite{Horndeski:1974wa, HorndeskiReview}, with $G_2=8 \hat{\alpha} X^2$, $G_3=8 \hat{\alpha} X$, $G_4=1+4 \hat{\alpha} X$ and $G_5 = 4 \hat{\alpha} \ln X$ (where $X=-\frac{1}{2} \nabla_{\mu} \phi \nabla^{\mu} \phi$). It can also be seen that the trace of Eq. \ref{feqs} taks the particularly simple form $R+\frac{1}{2} \hat{\alpha} \mathcal{G} = -T$.
\end{widetext}

\subsection{Post-Newtonian Expansion}
\label{sec:pn}

We want to investigate the behaviour of the theory given in Section \ref{sec:field} under a post-Newtonian expansion. This is an expansion of the metric about Minkowski space, where the gravitational field is assumed to be weak and the motion of matter is assumed to be slow compared to the speed of light.

We therefore consider a metric that can be written as
\begin{equation}
g_{\mu \nu} = \eta_{\mu \nu} + h_{\mu \nu} \, ,
\end{equation}
where $h_{\mu \nu}$ is a perturbation. We proceed by using an expansion parameter $\eta$, which is taken to be of the typical order-of-magnitude of the 3-velocity $v$ of a body in the system under consideration (in most cases of interest $\eta \sim v/c \sim 10^{-5}$ to $10^{-4}$).

The different components of $h_{\mu \nu}$ can then be expanded as
\begin{align*}
h_{00} &= h^{(2)}_{00} + h^{(4)}_{00} + O\left(\eta^5\right) \\
h_{0i} &= h^{(3)}_{0i} + O\left( \eta^4 \right) \\
h_{ij} &= h^{(2)}_{ij} +O\left( \eta^3 \right) \, ,
\end{align*}
where Latin indices run over the three spatial dimensions, and where numbers in brackets indicate the order of an object with respect to the parameter $\eta$. Different components of the perturbation are expanded to different orders in $\eta$ due to the role each of them plays in the field equations and the equations of motion of particles and bodies. Here we have included the terms that are required to reproduce the leading-order Newtonian equations, and the next-to-leading-order post-Newtonian terms, for objects with time-like or null trajectories.

Matter fields in this approach are expanded such that
\begin{equation} \nonumber 
v=v^{(1)}, \qquad \rho = \rho^{(2)}, \qquad p = p^{(4)}, \qquad \Pi = \Pi^{(2)} \, ,
\end{equation}
where $\rho$ is the density of mass, $p$ is the isotropic pressure, and $\Pi$ is the internal energy per unit mass (such that energy density is given by $\tilde{\mu} = \rho (1+\Pi)$).

We also need to use the fact that time derivatives add an extra order of $\eta$, compared to spatial derivatives of the same quantity, such that
\begin{equation}
\frac{\partial / \partial t}{ \partial / \partial x^i} \sim \eta \, .
\end{equation}
This equation encodes the ``slow motion'' aspect of the expansion.  We take this rule to apply to {\it all} fields, not just those directly associated with matter, which extends this notion from the motion of bodies themselves to the gravitational fields they generate.

The only remaining object that needs to be perturbed is the scalar field $\phi$ that appears in Eqs. (\ref{feqs}) and (\ref{scalareq}), which we now write as
\begin{equation}
\phi = \bar{\phi} + \delta \phi \,
\end{equation}
where $\bar{\phi}\sim \eta^0 \sim 1$ is the constant background value of $\phi$, and where $\delta \phi \sim \eta^2$ is a perturbation.

Everything described in this section is entirely standard in the post-Newtonian approach to weak-field gravity, and is explained in great detail in (for example) Ref. \cite{poisson2014gravity}, to which the reader can refer for further explanation and justification. We will use this formalism in the following section to construct the slow-motion, weak-field metric for 4DEGB gravity.

\section{Weak Field Gravity}
\label{sec:weak}

The first task to perform in assessing the observational viability of 4DEGB in the weak field regime is to expand the field equations (\ref{feqs0}), in the smallness parameter $\eta$. The results can then be solved order-by-order, to build up a perturbative description of the gravitational field that can be used to compute observables.

\subsection{Solving the Perturbed Field Equations}

The leading-order part of the $00$ field equation (\ref{feqs0}) occurs at order $\eta^2$, and can be written
\begin{equation}
\nabla^2 h^{(2)}_{00} = -8 \pi \, \rho \,,
\end{equation}
which under asymptotically flat boundary conditions integrates to
\begin{equation}
h^{(2)}_{00} = 2\, U  = 2 \int \frac{\rho}{\vert {\bf x} - {\bf x'} \vert} d^3 x' \,,
\end{equation}
where the last equality serves to define the Newtonian gravitational potential $U=U({\bf x})$, and where the mass density should be taken to be a function of the primed coordinate position (such that $\rho=\rho({\bf x}')$). This metric perturbation is sufficient to describe all gravitational physics at the leading-order Newtonian level of approximation, for bodies following time-like geodesics.

To determine the trajectories of rays of light to Post-Newtonian order we require $h_{ij}^{(2)}$, as well as $h^{(2)}_{00}$. This can be determined from the leading-order part of the $ij$ field equations (\ref{feqs0}), which are at order $\eta^2$ and read
\begin{equation} \label{ij2}
\nabla^2 h^{(2)}_{ij} = -8 \pi \, \rho \, \delta_{ij} \,,
\end{equation}
and which have the solution
\begin{equation}
h^{(2)}_{ij} = 2\, U \, \delta_{ij} = h^{(2)}_{00} \, \delta_{ij} \,.
\end{equation}
Equation (\ref{ij2}) is derived by choosing a gauge such that $\displaystyle 2 h^{\mu}_{\phantom{\mu} i , \mu} - h^{\mu}_{\phantom{\mu} \mu , i} = 0$, which can be retrospectively shown to be satisfied for the solutions we find.

The post-Newtonian equations of motion require knowledge of $h^{(3)}_{0i}$ and $h^{(4)}_{00}$, as well as $h_{00}^{(2)}$ and $h_{ij}^{(2)}$. The equation for $h^{(3)}_{0i}$ can be found by taking the leading-order contribution to the $0i$ field equations (\ref{feqs0}):
\begin{equation} \label{0i3}
\nabla^2 h^{(2)}_{0i} +U_{,0i} = 16 \pi \, \rho \, v_i \, ,
\end{equation}
where the gauge condition $2 h^{\mu}_{\phantom{\mu} 0 , \mu} - h^{\mu}_{\phantom{\mu} \mu , 0} = - h_{00,0}$ has been used, as well as the lower-order solutions above. Again, this gauge condition can be retrospectively verified by the solutions it generates. The asymptotically flat solution to Eq. (\ref{0i3}) is
\begin{align}
h^{(3)}_{0i} &= -\frac{7}{2} V_i -\frac{1}{2}W_i
\end{align}
where the post-Newtonian gravitational potentials $V_i$ and $W_i$ are given by 
\begin{align}
V_i &= \int \frac{\rho \, v_i}{\vert {\bf x} - {\bf x'} \vert} d^3 x' \\
W_i &= \int \frac{\rho \, [{\bf v \cdot (x - x')}] (x_i-x'_i)}{\vert {\bf x} - {\bf x'} \vert^3} d^3 x' \, ,
\end{align}
and where various manipulations have been performed. The mass density $\rho$ and the 3-velocity field ${\bf v}$ in these equations should be taken to be functions of the primed coordinate position ${\bf x}'$.

It now remains to determine $h^{(4)}_{00}$. In order to do this we use the scalar field propagation equation (\ref{scalareq}), which has its first non-trivial part at order $\eta^4$:
\begin{equation} \label{weakscalar}
\delta \phi_{,ij} \, \delta \phi_{, ij} - \left( \nabla^2 \delta \phi \right)^2 = U_{,ij} \, U_{, ij} - \left( \nabla^2 U \right)^2 \, .
\end{equation}
Imposing sensible boundary conditions, this equation admits the solutions
\begin{equation} \label{sres}
\delta \phi = \pm U \, .
\end{equation}
This is remarkably simple, and gives the interesting interpretation that the value of the gravitational scalar $\phi$ at a point in space-time is simply equal to the value of the Newtonian gravitational potential (up to a sign).


Using Eq. (\ref{sres}), we can write the $00$ field equation (\ref{feqs0}) at order $\eta^4$ as
\begin{equation} \label{fe004}
\nabla^2 \left( h_{00}^{(4)} +2 U^2 -4 \Phi_1 - 4 \Phi_2 -2 \Phi_3 - 6 \Phi_4 \right)
= \pm {\hat \alpha} \, \mathcal{G}^{(4)} \, ,
\end{equation}
where
\begin{align}
\Phi_1 & =  \int \frac{\rho \, v^2}{\vert {\bf x} - {\bf x'} \vert} d^3 x' \\
\Phi_2 &= \int \frac{\rho \, U}{\vert {\bf x} - {\bf x'} \vert} d^3 x'  \\
\Phi_3 &=   \int \frac{\rho \, \Pi}{\vert {\bf x} - {\bf x'} \vert} d^3 x' \\
\Phi_4 &=  \int \frac{p}{\vert {\bf x} - {\bf x'} \vert} d^3 x' 
\end{align}
and where the order $\eta^4$ part of the Gauss-Bonnet invariant (\ref{gb}) is written as
\begin{align*}
\mathcal{G}^{(4)} &= 8 \left( U_{,ij} \, U_{, ij} - \left( \nabla^2 U \right)^2 \right) \, .
\end{align*}
Integrating Eq. (\ref{fe004}) gives the solution
\begin{equation} \label{h004}
{
h_{00}^{(4)} = -2 U^2 +4 \Phi_1 + 4 \Phi_2 + 2 \Phi_3 + 6 \Phi_4 \mp \left( \frac{\hat{\alpha}}{4\pi} \right) \Phi_{\mathcal{G}} }\, .
\end{equation}
where we have introduced the new potential
\begin{equation} \label{phig}
\Phi_{\mathcal{G}} = \int \frac{\mathcal{G}^{(4)}}{\vert {\bf x} - {\bf x'} \vert} d^3 x'  \, .
\end{equation}
This equation represents a new type of gravitational potential that is sourced by the Gauss-Bonnet invariant itself (in the same way that the mass density $\rho$ sources the Newtonian potential $U$). The $\mp$ sign in Eq. (\ref{h004}) has its origin in the $\pm$ that occurs in Eq. (\ref{sres}).

Comparison of the results above with the PPN test metric gives the gravitational parameters of this theory as
\begin{equation}
\beta =\gamma = 1 
\end{equation}
and
\begin{equation}
\xi = \alpha_1=\alpha_2=\alpha_3=\zeta_1=\zeta_2=\zeta_3=\zeta_4=0 \, ,
\end{equation}
exactly as in GR \cite{Will:1993ns}. The only difference is the appearance of the potential $\Phi_{\mathcal{G}}$ in Eq. (\ref{h004}), which has no counterpart in the standard PPN metric. In what follows we will determine the effects that this new post-Newtonian gravitational potential has on observables, and use these results to place observational constraints on the coupling parameter $\hat{\alpha}$.

\subsection{The $N$-Body Problem}

Let us now consider a collection of point sources, as the origin of the gravitational field.
The energy density of such a group of sources can be written as
\begin{equation} \label{ed}
\tilde{\mu} = \sum_A {m_A} \frac{d\tau_A}{dt} \frac{\delta^3({\bf r}-{\bf r}_A)}{\sqrt{-g}} \, .
\end{equation}
where $m_A$ is the rest mass of particle $A$, $\tau_A$ is the proper time along its world-line, and we have taken $p =0$.

If we now recall that $\tilde{\mu} = \rho (1+\Pi)$, we find that we can write
\begin{equation} \label{unbody}
U = \sum_A \frac{m_A}{\vert {\bf x} - {\bf x}_A\vert} \, ,
\end{equation}
and
\begin{align}
\Phi_1 &= \sum_A \frac{m_A v_A^2}{\vert {\bf x} - {\bf x}_A\vert} \\
\Phi_2 &= \sum_A \frac{m_A U_A}{\vert {\bf x} - {\bf x}_A\vert} \\
\Phi_3 &= -\frac{1}{2} \Phi_1 - 3 \Phi_2 \\[10pt]
\Phi_4 &= 0 \, .
\end{align}
where $U_A$ is the value of $U$ at particle $A$, without including its own infinite contribution to this quantity. 

The remaining vector post-Newtonian potentials are given by
\begin{align}
V_i &= \sum_A \frac{m_A v_{Ai}}{\vert {\bf x} - {\bf x}_A \vert} \\
W_i &= \sum_A \frac{m_A {\bf v}_A \cdot ({\bf x} - {\bf x}_A)(x-x_A)_i}{\vert {\bf x} - {\bf x}_A\vert^3} \,.
\end{align}

It now only remains to calculate $\Phi_{\mathcal{G}}$, which is done in Appendix \ref{appa}, and gives
\begin{equation} \label{gnbody}
\Phi_{\mathcal{G}} = -32 \pi \left( \frac{1}{2} \vert \nabla U \vert^2 - \psi_1 \right) \, ,
\end{equation}
where
\begin{align} \label{gn1}
\vert \nabla U \vert^2 &= \sum_{A,B} \frac{m_A m_B ({\bf x}-{\bf x}_A) \cdot ({\bf x}-{\bf x}_B)}{ \vert {\bf x}-{\bf x}_A \vert^3 \vert {\bf x}-{\bf x}_B \vert^3} \\ \label{gn2}
\psi_1 &= \sum_{A, \, B \neq A} \frac{m_A m_B ({\bf x}-{\bf x}_A) \cdot ({\bf x}_A-{\bf x}_B)}{ \vert {\bf x}-{\bf x}_A \vert^3 \vert {\bf x}_A-{\bf x}_B \vert^3} \, ,
\end{align}
and where infinite contributions have again been removed. The potentials listed in Eqs. (\ref{unbody})-(\ref{gnbody}) can be substituted back into the expressions derived in the previous section to find the relevant expressions for the components of the metric perturbation $h_{\mu \nu}$ at each order of interest.

\section{2-Body Dynamics}
\label{sec:2bd}

We will find below that a promising route for constraining the 4DEGB theory with observations involves the bound orbits of two massive bodies. We will therefore begin this section by calculating the Lagrangian and Hamiltonian formulation of this problem.

\subsection{Relativistic Lagrangian and Hamiltonian}

To begin with, let us consider a single time-like particle, which we will label $1$. The Lagrangian that can be used for investigating the motion of this particle can be written $\displaystyle L_1 = - m_1 {d\tau_1}/{dt}$, where $m_1$ is its mass and $\tau_1$ is the proper time measured along its worldline. 
The Lagrangian $L_1$ gives the force on particle $1$ by taking the partial derivative with respect to the field point:
$
{\bf F}_1 = \left( {\partial L_1}/{\partial {\bf x} }\right)\vert_{{\bf x} = {\bf x}_1} \, .
$
If instead we want a Lagrangian that will be valid for more than one particle then we can construct an $L$ for which the force on the $i$th particle will be given by
$
{\bf F}_i = {\partial L}/{\partial {\bf x}_i } \, .
$

For a two-body system this Lagrangian is given to the required order of accuracy by
\begin{align}
\label{eq:2bpotential}
L =& -(m_1 +m_2) +\frac{1}{2} m_1 v_1^2 + \frac{1}{2} m_2 v_2^2 + \frac{m_1 m_2}{r_{12}} \\
&+ \frac{m_1m_2}{2r_{12}} \left[ 3 (v_1^2+v_2^2) - 7 {\bf v}_1 \cdot {\bf v}_2 - ({\bf v}_1 \cdot {\bf n}_{12})({\bf v}_2 \cdot {\bf n}_{12}) \right] \nonumber \\
&+ \frac{1}{8} (m_1 v_1^4 + m_2 v_2^4)-\frac{m_1m_2 (m_1+m_2)}{2 r_{12}^2} \left(1 \pm \frac{4 \hat{\alpha}}{r_{12}^2} \right) \, ,\nonumber 
\end{align}
where $r_{12}= \vert {\bf x}_1 - {\bf  x}_2 \vert$ and ${\bf n}_{12}= ({\bf x}_1-{\bf x}_2)/r_{12}$, and which can be seen to reduce to the usual expression from GR in the case $\hat{\alpha}=0$ \cite{landau1975classical}.

This Lagrangian can be conveniently re-written by choosing a frame such that $m_1 {\bf v}_1 + m_2 {\bf v}_2 = 0$, which corresponds to the centre-of-momentum frame at Newtonian-level accuracy. If we also define the total and reduced masses by $M = m_1+m_2$ and $\mu = {m_1m_2}/{M}$, and the relative velocity by ${\bf v} = {\bf v}_1 -{\bf v}_2$ then we get
\begin{equation}
{
L=L^{(2)} +L^{(4)}  }
\end{equation}
where $L^{(2)}$ (the Lagrangian at the Newtonian order of approximation) is given by
\begin{equation}
L^{(2)} =- M + \frac{1}{2} \mu v^2 + \frac{\mu M}{r} \, ,
\end{equation}
and $L^{(4)}$ (the first post-Newtonian correction) is
\begin{align}
L^{(4)} =&   \frac{\mu M}{2r} \left[ \left( 3 + \frac{\mu}{M} \right) v^2 + \frac{\mu}{M} \frac{({\bf v} \cdot{\bf r})^2}{r^2} \right] \nonumber
\\&+\frac{\mu}{8} \left( 1 - 3 \frac{\mu}{M} \right) v^4- \frac{\mu M^2}{2r^2} \left( 1 \mp \frac{4 \hat{\alpha}}{r^2} \right) \, ,
\end{align}
where we have written ${\bf r}={\bf x}_1 - {\bf  x}_2$ and $r= \vert {\bf r} \vert$.

Using the expressions above for the two-body Lagrangian, we can construct the corresponding Hamiltonian and write it as
\begin{equation}
{
H= H^{(2)}+H^{(4)} }\, ,\label{H}
\end{equation}
where the Newtonian-level contribution is given by
\begin{equation}
H^{(2)} = M + \frac{p^2}{2 \mu} - \frac{\mu M}{r}
\end{equation}
and the first post-Newtonian contribution is given by
\begin{align} \nonumber
H^{(4)} =&- \frac{1}{2r} \left[ 3 \frac{(M-2\mu)}{\mu} p^2 +7 p^2 + ({\bf p} \cdot \hat{{\bf r}})^2 \right] \nonumber
\\&-\frac{(M-3\mu)}{8 M \mu^3} p^4
+ \frac{M^2 \mu}{2 r^2} \left(1 \mp \frac{4 \hat{\alpha}}{r^2} \right) \, ,
\end{align} 
where the relative momentum is ${\bf p} = {\bf p}_1 = -{\bf p}_2$, with magnitude $p=\vert {\bf p} \vert$. We will now use this Hamiltonian to calculate the periapsis advance for two bodies in closed orbits.

\subsection{Advance of Periapsis}

We can calculate the advance of periapsis using the Hamilton-Jacobi approach, which has an action
\begin{align}\nonumber
S(r,\phi,t) &= S_r(r)+S_{\varphi}(\varphi) + S_t(t) \\
&= S_r +J \varphi - E t \, ,
\end{align}
where $J$ is angular momentum and $E$ is energy. The radial momentum can be extracted from the radial part of this action using
$
p_r = {\partial S_r}/{\partial r} \,
$
and the angular coordinate $\varphi$ can be found using the equation for the generalized coordinate:
$
\varphi = - {\partial S_r}/{\partial J}\, +
$ constant.
This gives the angle $\varphi$ (up to a constant) as
\begin{equation} \label{phi}
\varphi = - \int \frac{\partial p_r}{\partial J} dr \, ,
\end{equation}
where $p_r$ can be determined using the Hamiltonian (\ref{H}).

Energy in this system is conserved, so $H=E=$ constant. Solving Eq. (\ref{H}) for $p_r$ therefore gives
\begin{equation} \label{pr}
p_r^2 = -A +\frac{B}{r} -\frac{J^2}{r^2} + \frac{6 M^2 \mu^2}{r^2 }\left(1 \pm \frac{2 \hat{\alpha}}{3 r^2} \right) \, ,
\end{equation}
where we have performed a gauge transformation so that $r \rightarrow r + \mu/2$, and where
\begin{align*}
A &= - 2 (E-M) \mu + O(\eta^4) \\
B &= 2 M \mu^2 + O(\eta^4) \, .
\end{align*}
The higher-order parts of these equations are not made explicit, as they are not required in the final result.

Using Eq. (\ref{pr}) in Eq. (\ref{phi}) gives, to Newtonian order, the usual result:
\begin{equation}\label{phi2}
\varphi_{\rm N} = \cos^{-1} \left( \frac{2 J^2/r-B}{\sqrt{B^2-4 A J^2}} \right) \, ,
\end{equation}
where the constant of integration has been chosen so that the moment of periapsis occurs at $\varphi=0$, and where we have neglected the last term in Eq. (\ref{pr}), which is small compared to the first three.

The relativistic correction to $\varphi$ can then be calculated using
\begin{equation} \label{phi2r}
\varphi_{\rm R} = - \int \frac{\partial p_r^{(4)}}{\partial J} dr \Bigg\vert_{A=A^{(2)}, \, B=B^{(2)}} \, .
\end{equation}
Here there is no need to include the $\eta^4$ parts of $A$ and $B$, as the higher-order parts in the first two terms contribute in exactly the same way as the lower-order parts do to $\varphi_{\rm N}$, and the higher-order parts included in the last term would be of order $\eta^6$. 

We find that Eq. (\ref{phi2r}) evaluates to
\begin{align} \label{phi3}
\varphi_{\rm R} =& \frac{3 M^2 \mu^2}{4J^6} \left[ 4 J^4 \pm \hat{\alpha} (5 B^2 - 4 AJ^2) \right] \varphi_{\rm N} \\[5pt] \nonumber &\quad+ \,\, {\rm periodic \, \,terms} \, .
\end{align}
The constant of integration has been chosen here so that $\varphi_{\rm R}=0$ at periapsis (i.e. when $\varphi_{\rm N}=0$). The periodic terms in this expression are omitted as they do not produce a secular change that can accumulate.

To find the precession that occurs every orbit, we can put $\varphi_{\rm N}=2 \pi$ into Eq. (\ref{phi3}), and neglect the periodic terms. This gives
\begin{equation} \label{pp}
{
\delta \varphi = \frac{6 \pi M}{ a (1-e^2)} \left[ 1 \pm \frac{\hat{\alpha} (4+e^2)}{a^2(1-e^2)^2} \right] }\, ,
\end{equation}
where $a$ is the semi-major axis of the orbit and $e$ is the eccentricity. In deriving this expression we have used the Newtonian results $J^2 = M \mu^2 a (1-e^2)$  and $E =M - {M\mu}/{2a}$. This equation gives the contribution from the Gauss-Bonnet term to the advance of periapsis of a bound orbit, and can be seen to reduce to the usual expression from GR when $\hat{\alpha}=0$ (see e.g. \cite{landau1975classical}).

\subsection{Observational Constraints}


Let us now consider observational constraints that can be imposed on the coupling parameter $\hat{\alpha}$, using observations of closed orbits of time-like objects. We will first consider the classical test of Mercury's perihelion precession, before moving on to the orbits of satellites around the Earth, and the orbits of binary pulsars.\\

{\it Perihelion Precession of Mercury}: The detection of the anomalous perihelion precession of the orbit of Mercury pre-dates the discovery of relativistic gravity \cite{le1845theorie}, and was one of the original tests used to validate GR \cite{einstein1916foundation}. It therefore has important pedagogical importance.

Contributions to the perihelion precession of Mercury come from the precession of the equinoxes of the coordinates system ($\sim 5025''$ per century), from the gravitational influence of the other planets ($\sim 531''$ per century), and from the non-zero quadrupole moment of the Sun ($\sim 0.025''$ per century). A precise determination of these contributions, analysed together with the ephemeris of Mercury, gives an anomalous deficit of $\sim 43''$ per century. This can be compared with the prediction from Eq. (\ref{pp}) to constrain our theory.

For definiteness we use the anomalous perihelion precession determined by Pitjeva and Pitjev \cite{pitjeva2013relativistic}:
\begin{equation} \label{ppobs}
\delta \varphi -\delta \varphi_{\rm GR} = (- 0.0020\pm 0.0030) \,'' {\rm \,\, per \,\, century} \, ,
\end{equation}
where $\delta \varphi_{\rm GR}=42.98$ arcseconds per century is the famous prediction from GR. Other published values for this quantity exist in the literature, and can be found e.g. Ref. \cite{fienga2011inpop10a}. Taking the mass of the Sun to be $1.9884 \times 10^{30} \, {\rm kg}$, the mass of Mercury to be $3.3011 \times 10^{23} \, {\rm kg}$, and their orbit to have semi-major axis $a = 57,909,050 \, {\rm km}$ and eccentricity $e =  0.205\,630$, we find that Eq. (\ref{ppobs}) implies that the coupling parameter of 4DEGB is constrained to be
\begin{equation}
\vert \hat{\alpha} \vert = \vert (-3.54 \pm 5.31) \vert \times 10^{16} \, {\rm m^2} \, .
\end{equation}
Alternate observations will of course give alternative bounds on $\hat{\alpha}$, but for errors on $\delta \varphi$ of the order of $\sim 0.01$ arcseconds per century we can see that the constraints are going to be at the level of $\vert \hat{\alpha} \vert \lesssim 10^{17} \, {\rm m^2}$.\\

{\it LAGEOS Satellites:} The LAGEOS satellites are two man-made satellites, which are spherical in shape with a diameter of 60 cm, and which orbit the Earth at an altitude of approximately $6\, 000$ km. Lasers are reflected off the satellites from ground-based stations, which allow for precise tracking of their orbits. One of the many benefits of this is that the gravitational field of the Earth can be measured to very high accuracy.

Using 13 years of tracking data of the LAGEOS satellites, the precession of the periapsis of the LAGEOS II satellite was measured by Lucchesi and Peron to be \cite{Lucchesi:2010zzb}
\begin{equation*}
\delta \varphi = \left[ 1+(0.28 \pm 2.14) \times 10^{-3} \right] \, \delta \varphi_{\rm GR} \, ,
\end{equation*}
where $\delta \varphi_{\rm GR}$ is the prediction from GR (i.e. Eq. (\ref{pp}) with $\hat{\alpha}=0$). Taking the mass of this satellite to be $405.38 \, {\rm kg}$, and taking its orbit to have a semi-major axis $a = 5.697 \times 10^6 {\rm m}$ and eccentricity $e = 0.0135$, this corresponds to a bound on the 4DEGB coupling parameter of
\begin{equation}
\vert \hat{\alpha} \vert = \vert (0.23 \pm 1.74) \vert \times 10^{10} \, {\rm m^2}  \, ,
\end{equation}
where we have taken the mass of the Earth to be $5.9722 \times 10^{24} \, {\rm kg}$. This is a much tighter bound than that obtained from the perihelion precession of Mercury, which we take to be due to the much smaller orbital radius of the LAGEOS satellites ($\sim 10^6$m, compared to $\sim 10^{10}$m for Mercury). This being the case, the $1/r^4$ form of the gravitational potential in Eq. (\ref{gnbody}) then suppresses the contribution of the new effects in 4DEGB by a much smaller amount.\\

{\it Precession of S2 around Sgr A$^*$}: The motions of stars orbiting the central black hole of the Milky Way galaxy have now been observed for 27 years, which has enabled very accurate determinations of their orbital parameters. The GRAVITY collaboration has detected the precession of the star S2 to be~\cite{Abuter:2020dou}:
\begin{equation}
\delta \varphi = \left[1.10 \pm 0.19 \right] \, \delta \varphi_{\rm GR} \, ,
\end{equation}
where $\delta \varphi_{\rm GR}=12.1'$ per orbit is the prediction from GR. We find that this implies
\begin{equation}
\vert \hat{\alpha} \vert = \vert (2.17 \pm 4.42) \vert \times 10^{25} \, {\rm m^2} \, ,
\end{equation}
where we have taken the mass of the central black hole to be $4.261\times10^6\,M_{\odot}$, the mass of S2 to be negligible, and its orbit to have eccentricity $e =  0.884649$ and semi-major axis $a = 1.54\times 10^{14} \, {\rm m}$ (from an angular size of $125.058$ mas and a distance of $8246.7$ pc)~\cite{Abuter:2020dou}. 

This bound is weaker than that obtained from Mercury and LAGEOS, but is obtained in a very different environment. Further observations of S2, and other stars orbiting Sgr A$^*$, may provide slightly better constraints in future, but should not be expected to improve to the level of those given by LAGEOS as the $a^2$ suppression in Eq. (\ref{pp}) is orders of magnitude larger in the present case. We note while canonical scalar-tensor theories (such as Brans-Dicke) have black hole solutions with a weak field that is identical to GR \cite{hawking1972black}, this is not expected to be the case in 4DEGB. This is due to the form of the scalar field propagation equation (\ref{scalareq}), which is sourced by the Gauss-Bonnet curvature-invariant of the space-time, and not its energy-momentum content. The scalar field must therefore be non-constant, and satisfy Eq. (\ref{weakscalar}) in the weak field limit, independent of whether the gravity is due to a black hole or matter.
\\

{\it Binary Pulsars:} Binary pulsars are two-body systems that contain at least one pulsar (i.e. a rotating neutron star that emits regular pulses of radiation). These systems are excellent testing grounds for relativistic gravitational physics, as they allow precise data about orbits to be extracted from systems in which the bodies move at very high velocities.

The first, and most famous, binary pulsar system to be found was PSR B1913+16 \cite{hulse1975discovery}, also known as the {\it Hulse-Taylor binary} (after its discoverers). This system provided the first indirect evidence for the existence of gravitational waves, as the period of its orbit changed over time due to their emission. In general, binary pulsars provide the possibility to constrain relativistic gravity through five different {\it post-Keplerian} effects: the rate of advance of periapsis, the rate of change of orbital period, the gravitational redshift and two types of Shapiro time delay effect.

The most promising binary pulsar system for constraining 4DEGB is PSR J0737-3039A/B \cite{lyne2004double}, also known as the {\it double pulsar}. In this system both bodies were (for a time) emitting pulses of radiation that were visible from Earth. In addition, the system was oriented edge-on, which meant that all five post-Keplerian parameters were visible, as well as the mass ratio of the pulsars being determinable. 

As we will discuss later, the Shapiro time delay effects are unaltered in 4DEGB from their values in GR. The advance of periapsis, on the other hand, can be seen from (\ref{pp}) to be dependent on the coupling parameter $\hat{\alpha}$. We can therefore use the mass ratio, together with observations of these two relativistic effects, to determine the masses of both pulsars together with the value of $\hat{\alpha}$ (as there are three observables and three unknowns). Using the mass ratio and the periapsis advance to determine the masses, Kramer {\it et al.} find the time delay parameter $r$ to be given by \cite{Kramer:2006nb}
\begin{equation}
r = (1.009 \pm 0.055) \, r_{\rm GR}\,.
\end{equation}
where $r_{\rm GR}$ is the value predicted in GR. Assuming this combination of observables leads to a similar constraint on the advance of periapsis of the system gives the following constraint on the coupling parameter:
\begin{equation}
\vert \hat{\alpha} \vert = \vert (0.4 \pm 2.4) \vert \times 10^{15} \, \frac{\rm m^2}{\sin i} \, ,
\end{equation}
where $i$ is the inclination angle of the system, where the mass ratio has been taken to be $1.0714$ and the semi-major axis and ellipticity have been taken to be given by $a = 1.415032 \, {\rm c} \, {\rm s^{-1}}/{\sin i}$ and $e = 0.087777$. This gives a best estimate for the constraints available from these observations to be $\vert \alpha \vert \lesssim 10^{15} \, {\rm m^2}$, which is better than that available from Mercury, but worse than the constraints available from LAGEOS.

We note that while the structure of neutron stars in 4DEGB has not yet been studied, we still expect the analysis performed above to be applicable to the weak field of such bodies. This is despite non-perturbative effects, such as ``spontaneous scalarization'', being known to exist in some scalar-tensor theories \cite{damour1993nonperturbative}. This is because the only degree-of-freedom in our weak-field analysis that could be altered by such non-linear physics is the mass of the body, which is already treated as a nuisance parameter when extracting constraints from data. In particular, the coupling constant $\hat{\alpha}$ cannot be dependent on environment, and so the cannot be affected by non-linearities in the same way as the coupling function $\omega (\phi)$ of canonical scalar-tensor theories.

\section{Propagation of Radiation}
\label{sec:rad}

As well as the motion of massive bodies, there are also a number of observational tests of gravity that rely on the propagation of radiation. In the case of isolated, weakly gravitating systems, two of the most widely used tests of this type are gravitational lensing and Shapiro time delay. More recently, there is also the direct detection of gravitational waves by LIGO/VIRGO.

\subsection{Lensing and Time Delay}

The first test of relativistic gravity to be performed after the publication of GR was the observation of the gravitational lensing of light by the Sun, which is predicted by Einstein's equations to be $1.75''$ for a null trajectory that grazes its edge. This effect was observed by Eddington in May of 1919, during his famous trip to Pr\'{i}ncipe, and was of great importance in establishing the validity of GR. Today, this same effect is measured using Very Long Baseline Interferometry, with results from around 2500 days of observations taken over a period of 20 years giving the constraint \cite{shapiro2004measurement}
\begin{equation} \label{lens}
\theta = (0.99992 \pm 0.00023) \, \theta_{\rm GR} \, ,
\end{equation}
where $\theta$ is the deflection angle, and $\theta_{\rm GR} =1.75''$ is the prediction from GR. This is one of the highest precision results available on relativistic gravity, and is used to place constraints on the post-Newtonian parameter $\gamma$ to around $1$ part in $10^4$ of its value in GR.

Even tighter constraints on $\gamma$ are available from observations of the Shapiro time delay effect, which accounts for the deflection in time of a radio signal as it passes through a gravitational field. The most constraining observation of this effect in the Solar System to date was from radio signals sent to the Cassini spacecraft on its mission to Saturn. These give \cite{bertotti2003test}
\begin{equation} \label{shap}
\Delta t = (1.00001 \pm 0.00001) \, \Delta t_{\rm GR} \, ,
\end{equation}
where $\Delta t_{\rm GR}$ is the expected size of the effect from GR. This gives a bound on $\gamma$ of being within $1$ part in $10^5$ of its expected value, which is currently the best constraint available on this quantity (or on any post-Newtonian parameter associated with conservative theories of gravity).

We note that these observations, though extremely precise, provide no new constraints on the coupling parameter $\hat{\alpha}$. This can be seen from the results presented in Section \ref{sec:weak}, where the order $\eta^2$ parts of both the $00$ and $ij$ components of the metric are identical to the form they take in GR. The equations of motion of null particles are only sensitive (at leading-order) to gravitational potentials that appear in these components of the metric at this order \cite{Will:1993ns}, so the lensing and time delay effects in 4DEGB should be expected to be exactly the same as they are in GR. This means that neither effect can be used to constrain $\hat{\alpha}$, and all that can be said is that the results quoted in Eqs. (\ref{lens}) and (\ref{shap}) are consistent with this theory.

\subsection{Gravitational Waves}

A further constraint on $\hat{\alpha}$ comes from the propagation of gravitational waves  
from the double neutron star collision that resulted in the signal GW170817 \cite{TheLIGOScientific:2017qsa}.

The spatially flat Friedmann-Roberston-Walker (FRW) metric is a solution to the field equations~\eqref{feqs0}, but results in an altered Friedmann equation:
\begin{align}
\label{eq:friedmann}
H^2 + {\hat{\alpha}} H^4 = \frac{8 \pi}{3} \rho +\frac{C^4}{a^4}\,,
\end{align} 
where $H=\dot{a}/{a}$ is the Hubble rate, $\rho$ is the energy density, $a$ is the scale factor, and  $C$ is a constant of integration. 

For Horndeski theories we know that the propagation speed of gravitational waves is \cite{HorndeskiReview}
\begin{equation}
c_{T}^{2}=\frac{G_{4}-X\left(\ddot{\phi} G_{5, X}+G_{5, \phi}\right)}{G_{4}-2 X G_{4, X}-X\left(H \dot \phi G_{5, X}-G_{5, \phi}\right)}
\label{eq:GWs}
\end{equation}
where $X=-\frac{1}{2}\partial_\mu \phi \partial^\mu \phi$. The reader may note that only $G_4$ and $G_5$ are required to calculate the gravitational wave speed.

Recalling that our theory is a subset of Horndeski with $G_2=8 \hat{\alpha} X^2$, $G_3=8\hat{\alpha} X$, $G_4=1+4\hat{\alpha} X$, $G_5=4\hat{\alpha} \log X$, we can now calculate $c_T$ in 4DEGB. We find this to be given by
\begin{equation}
c_{T}^2
= 1 + \frac{4\hat{\alpha} \left(\dot H + \frac{C^2}{a^2} \right)}{1+2\hat{\alpha} \left(H^2 - \frac{C^2}{a^2}\right)}  = 1 + \frac{\dot{\Gamma}}{H \Gamma} \, , \label{ct4degb}
\end{equation}
where $\Gamma = 1 + 2 \hat{\alpha} (H^2- {C^2}/{a^2})$ and where we have used $X= \frac{1}{2}\dot \phi^2$ and $\dot \phi = -H +{C}/{a}$ in a Friedmann background. We note that the speed of propagation of gravitational waves in an FRW cosmology in 4DEGB is not equal to the speed of light, but that it reduces to the speed of light in Minkowski space (i.e. when $H=C=0$).

Now, the electromagnetic counterpart to GW170817 indicates that the deviation in the speed of gravitational waves from that of light must be less than one part in $10^{15}$ \cite{Goldstein:2017mmi, Savchenko:2017ffs, Baker:2017hug, Creminelli:2017sry, Ezquiaga:2017ekz}. From Eq.~\eqref{ct4degb} this leads to the rather weak constraint
\begin{align}
\left | \frac{\dot{\Gamma}}{H \Gamma}\right | &<10^{-15} \, ,
\end{align}
which, taking $C=0$ for simplicity, implies
\begin{align} \label{gwcon}
\vert \hat{\alpha} \vert &\lesssim  10^{36} m^2\,,
\end{align}
where 
we have taken $\dot{H}\approx H^2 \approx 5.8 \times 10^{-36} \, {\rm s}^{-2}$. Taking $C\neq0$ will change this constraint, and will correspond to cosmologies that contain a period in which the free kinetic energy of the scalar field dominates over matter. This is an intruiging possibility, which often occurs in the cosmologies of scalar-tensor theories of gravity, but which we will not consider further here.

The result (\ref{gwcon}) agrees with similar estimates in Ref.~\cite{Aoki:2020iwm}, and can be seen to be considerably weaker than the constraints imposed from the trajectories of massive bodies studied in Section \ref{sec:2bd}. It therefore appears to provide an exception to the rule that GW170817 tightly constrains Horndeski theories with non-trivial $G_4$ and $G_5$ \cite{Baker:2017hug, Creminelli:2017sry, Ezquiaga:2017ekz}.

\section{Other tests}
\label{sec:other}

We have so far considered constraints that are available on the 4DEGB theory from bound orbits of massive bodies, and from the propagation of radiation. In this section we will discuss some other tests of gravity that are available, and what they may imply for 4DEGB.

\subsection{Black Hole Shadows}

The shadow of the super-massive black hole of M87 has recently been observed ~\cite{Akiyama:2019cqa,Akiyama:2019eap}, and can be used to constrain deviations from GR~\cite{Vagnozzi:2019apd,Cunha:2019dwb,Held:2019xde,Zhu:2019ura}. Here we will recap what this implies for 4DEGB, following Refs.~\cite{EGB1,EGB2,Wei:2020ght,Jin:2020emq}.

Firstly, the simplest static, spherically symmetric solution of 4DEGB has line-element \cite{original,EGBST1}
\begin{equation} \label{dsbh}
ds^2=-f(r) dt^2+f(r)^{-1} dr^2+r^2d\Omega_2\,,
\end{equation}
where
\begin{equation} 
f(r)=1 + \frac{r^2}{2\hat\alpha} \left(1 - \sqrt{1 + \frac{8 M \hat\alpha}{r^3}}\right)\,,
\label{eq:metricbh}
\end{equation}
and where $M$ is a constant mass parameter. The scalar field profile that together with \eqref{dsbh} and \eqref{eq:metricbh} solves the field equations \eqref{feqs0} and \eqref{scalareq} is given by $\phi(r)=\int ({r \sqrt{f(r)}})^{-1} dr-\log(r/L)$ \cite{EGBST1}, where $L$ is a length scale. This is a dimensionally-reduced version of the general static and spherically symmetric vacuum solution of the Einstein-Gauss-Bonnet theory \cite{boulware1985string}, and numerical investigations suggest it is the only physically interesting static and spherically symmetric solution of 4DEGB. It has a non-trivial horizon structure, containing both black and white hole horizons, and exhibits a repulsive force at small radial distances, which may have consequences for the singularity problem in relativistic gravity. Furthermore, it presents logarithmic corrections to the Hawking-Bekenstein entropy area formula \cite{EGB3,EGBST1}, as predicted by some quantum theories of gravity (such as Loop Quantum Gravity and String Theory \cite{Rovelli:1996dv,Kaul:2000kf}).

The radius of the photon sphere, $r_{\rm ph}$, for an object described by (\ref{dsbh}), is given by the appropriate solution of $r f'(r)=2f(r)$, and the corresponding black hole shadow radius is well approximated by $R_{\rm Sh}=r_{\rm ph}/\sqrt{f(r_{\rm ph})}$. This all gives
\begin{equation} \label{rsh} \hspace{-3pt}
\frac{R_{\rm Sh}}{M}=\frac{3+\delta^{\frac23}}{\delta^{\frac13}}\left(1+\frac{(3+\delta^{\frac23})^2}{2\beta\delta^{\frac23}}\left[1-\sqrt{1+\frac{8\beta\delta}{(3+\delta^{\frac23})^3}}\right]\right)^{-\frac12}
\end{equation}
with $\delta=-4\beta+\sqrt{16\beta^2-27}$ and $\beta=\hat\alpha/M^2$. It can be shown that the radius of the shadow monotonically decreases with $\hat\alpha$.
Now, the black hole M87$^*$ 
has been measured by the Event Horizon Telescope to have a
shadow of size $R_{\rm Sh}=(4.96\pm0.75)\times 10^{13}$~m. Using Eq. (\ref{rsh}), this value, together with a value for the mass of the black hole, can be used to infer bounds on $\hat{\alpha}$.

Observations of stellar dynamics around M87$^*$ imply a mass of $M=6.14_{-0.62}^{+1.07}\times 10^9 M_{\odot}$~\cite{2011ApJ...729..119G}, which in turn allows one to derive the constraint
\begin{equation}
\hat{\alpha} =  (-0.67 \pm 1.44) \times 10^{26} \, {\rm m^2}  \, .
\end{equation}
Alternatively, measurements of gas dynamics imply $M=3.45_{-0.26}^{+0.85}\times 10^9 M_{\odot}$~\cite{2013ApJ...770...86W}, which gives\footnote{Note that the masses reported in Refs.~\cite{2011ApJ...729..119G,2013ApJ...770...86W} are not those written here, as they assumed a distance to M87$^*$ of $D=17.9$~Mpc. This does not agree with measurements made by the Event Horizon Telescope, and was corrected to the values presented above in Ref.~\cite{Akiyama:2019eap}.} 
\begin{equation}
\hat{\alpha} =  (-1.26 \pm 0.80) \times 10^{27} \, {\rm m^2}  \, .
\end{equation}
Both cases prefer negative values for the Gauss-Bonnet coupling, but clearly place much looser constraints than those obtained in Section \ref{sec:2bd}.

In addition to the constraints above, further bounds can be placed on positive values of $\hat\alpha$ by the requirement for the existence of an event horizon and a photon sphere. These require $\hat\alpha<M^2$ and $\hat\alpha<3\sqrt{3}M^2/4$, respectively. Since we observe the shadow, this alone can be used to place the constraints $\hat\alpha\leq1.07\times10^{26}$ m$^2$ for the stellar dynamics case, and $\hat\alpha\leq3.37\times10^{25}$ m$^2$ for the gas dynamics mass measurement.

We note that the consequences of 4DEGB gravity for black holes depend greatly on the ratio $\hat\alpha/M^2$, which for M87$^*$ is not large enough to place competitive constraints on $\hat{\alpha}$. However, smaller black holes will result in larger effects, due to the existence of $M$ in the denominator of this ratio. This means that one could achieve constraints on the order of $\hat{\alpha} \lesssim 10^{20}$ m$^2$ from observations of SgrA*, which are expected to be released by the EHT collaboration in the near future. Going further, it may be possible to achieve $\hat{\alpha} \lesssim 10^6$ m$^2$ from observations of a solar mass black hole, which could be possible through an analysis of gravitational wave emission from binaries.

Finally, we note that an analysis of rotating solutions in 4DEGB should be performed in order to properly understand the shadows of realistic astrophysical black holes. Our intuition from GR is that the consequences of rotation should be small in this situation, but this should be verified to be true in 4DEGB too.

\subsection{Black Hole Binaries}

The gravitational waves events that have been observed by the LIGO/VIRGO collaborations, that resulted from the inspiral and merger of binary black holes, offer the possibility of imposing tight constraints on modified theories of gravity \cite{Abbott:2016blz, TheLIGOScientific:2016wfe, TheLIGOScientific:2016src, Abbott:2016bqf}. Here we will investigate the possible bounds that such observations could impose on the coupling constant $\hat \alpha$, using simple physical arguments (as in Ref. \cite{Abbott:2016bqf}). This is not intended as a replacement for more sophisticated studies of these events in 4DEGB, which will require complicated numerical implementation, but to gain some insight into the constraining power that they offer.

The black hole described by the metric function \eqref{eq:metricbh}, with mass $m_i$, has an event horizon located at
\begin{equation}
r_H (m_i) = {m_i} + \sqrt{m_i^2-\hat \alpha} \, .
\end{equation}
Let us consider the merger event that led to GW150914, using this radius as an approximate size for the black holes\footnote{Rotation, and the perturbations due to the other black hole, will change this value, but we expect it be correct to an order of magnitude.}. The frequency at which this waveform has maximum amplitude is around $f_{\rm GW} \sim 150 Hz$, and the chirp mass of the black holes is inferred to be $M_c \sim 30 M_{\odot}$ (assuming the two black holes are of equal mass, this corresponds to masses $m_{1} \sim m_2 \sim 35 \, M_{\odot}$) \cite{Abbott:2016blz, TheLIGOScientific:2016wfe, TheLIGOScientific:2016src, Abbott:2016bqf}.
This value of $f_{\rm GW}$ is taken directly from the data, and does not require a particular theory of gravity in order to be determined. The value of $M_c$ can be determined from the inspiral, and requires only weak gravity in order to discovered. We therefore expect the orders of magnitude quoted here to be correct for both GR and 4DEGB.

Now, using a weak field analysis it can be shown that at the time of peak amplitude the two bodies have an orbital separation of
\begin{equation}
R=\left(\frac{m_1+m_2}{\pi^2 f_{\rm GW}^2}\right)^{1/3} \sim 350 km \, ,
\end{equation}
where the numerical value is obtained using the numbers above. In order to quantify the closeness of the two objects, relative to their natural gravitational radius, we introduce the compactness ratio:
\begin{equation}
\mathcal{R} = \frac{R}{r_H(m_1)+r_H(m_2)}.
\end{equation}
Assuming both masses to be $m_i\sim 35 M_\odot$, and imposing $\mathcal{R} > 1$, so that the two black holes are not overlapping, we obtain a lower bound on negative values of the coupling parameter of $\hat \alpha \gtrsim -10^{10} \,{\rm m}^2$. We can gain an upper bound on $\hat{\alpha}$ by requiring that $r_H$ is a real number. For this same black hole masses, this gives $\hat \alpha \lesssim 10^9 \, {\rm m}^2$. An order of magnitude estimate of the constraints that may be available from GW150914 is therefore
\begin{equation}
-10^{10} \,{\rm m}^2 \lesssim \hat \alpha \lesssim 10^9 \, {\rm m}^2 \, .
\end{equation}
This is among the tightest constraints we have found.

Of course, there are also many other gravitational wave events that could also be used for our current purpose, that have been detected since the discovery of GW150914. One of the more promising of these is GW170608 \cite{Abbott:2017gyy}, which is the lowest mass binary black hole merger event that has been confirmed to date. Following through the same logic as above, this system could be expected to give the constraint $-10^{10} \, {\rm m}^2 \lesssim  \hat \alpha \lesssim 10^8 \, {\rm m}^2$, which can be seen to be marginally tighter. If the sources of GW190814 are both confirmed to be black holes \cite{Abbott_2020}, this would give $\hat{\alpha} \lesssim 10^7 \, {\rm m}^2$. Combining constraints from multiple systems, or future events, may also improve on the constraining power of these observations.

We reiterate that this is only a rough estimate of the bounds that may be available from these systems, and that it may well be possible to gain significantly tighter constraints using a more thorough numerical relativity treatment. This will be a significant challenge to implement, however, and will be left for subsequent studies. In particular, as well as assuming the validity of weak-field treatments, we have neglected spin and assumed circular orbits. Such assumptions could be removed in proper numerical studies.


\subsection{Table-Top Tests of Gravity}

Due to the $r^{-4}$ scaling of the extra terms that appear in the metric in 4DEGB, measurements of gravity on small scales offer the possibility of imposing tight constraints on the theory. Of particular interest in this regard are the so-called ``tabletop'' tests of gravity. These are gravitational physics experiments that seeks to directly measure gravitational effects in the laboratory, and which are the modern counterparts of the Cavendish experiment.

Tabletop tests of gravity most commonly test for deviations from the inverse square law by modelling extra gravitational forces as being due to an additional Yukawa-type potential. However, and of more relevance for our work, there have also been studies that look for additional power-law terms, of the form~\cite{Fischbach:1999bx,Adelberger:2006dh,PhysRevLett.116.131101}
\begin{align} \label{tteq}
V =  &\frac{m_1 m_2}{r} \left(1 - \frac{A_n}{ r^n} \right)\,,
\end{align}
where no sum is implied over $n$. Such a potential can be compared to Eq.~\eqref{eq:2bpotential}, which gives
\begin{align}
\label{eq:naive1}
V \approx  
& \frac{m_1 m_2}{r} \left[1 \mp \, \frac{2\, \hat{\alpha} (m_1+m_2)}{ r^3}   \right] \, ,
\end{align}
where we have kept the additional term from 4DEGB and neglected all other post-Newtonian terms. It can be seen that Eqs. (\ref{tteq}) and (\ref{eq:naive1}) give $A_3 = \pm 2\hat{\alpha} \,  { (m_1+m_2)}$.

Now, the observational constraints from these experiments currently yield bounds of order $A_3 \lesssim 2.2 \times 10^{-14}\, {\rm m}^3$~\cite{PhysRevLett.116.131101}, which from the above implies
\begin{equation} \label{ttbound}
\vert \hat{\alpha} \vert \lesssim 10^{16}\, {\rm m}^2 \, ,
\end{equation}
where we have used a value of $\sim 1$g for the masses involved in the experiment.

This bound, while being one of the more promising we have found in this paper, should be taken with a large pinch of salt. There are number of reasons for this. Firstly, the experiments themselves do not involve two point-like masses, but are instead much more complicated set-ups. In particular, the study in Ref.~\cite{PhysRevLett.116.131101} is based on a torsion balance in which one disk of metal with holes in is hung directly above two other disks with holes. The holes in these disks are the ``masses", and the torque on the top disk is measured as the bottom disks rotate. A calculation involving multiple extended objects should therefore be performed to give more precise results. 

Secondly, the extra ``potential'' in Eq. (\ref{eq:naive1}) is a post-Newtonian term, rather than the Newtonian-level term. There is a gauge dependence on the $00$ component of the metric at this level of accuracy, which introduces an extra degree of ambiguity, and which does not occur at the Newtonian level. This is not to say that this term does not affect the equations of motion of time-like objects in the same manner as a Newtonian potential (it does), but that one would need to make sure that the gauge choice that we used to calculate this metric corresponds to the coordinates used by the experimentalists in a sensible way, in order to make precise statements.

Giving thorough answers to the questions above would require a more detailed study of the experimental set-up that is used in Ref.~\cite{Adelberger:2006dh, PhysRevLett.116.131101}, which we leave for future work. Despite these issues, we nevertheless expect the bound in Eq. (\ref{ttbound}) to give a representative order of magnitude for what these experiments should be able to achieve. This is a bound that is competitive with those achievable from analyzing the orbit of Mercury, but is considerably weaker than those from the LAGEOS satellites and binary black hole systems.

\subsection{Primordial Nucleosynthesis}

If we take $C=0$, then it can be seen from the Friedmann equation~\eqref{eq:friedmann} that the size of corrections to the Hubble rate of standard cosmology is controlled by the combination $\hat{\alpha} H^2$, with the $H^4$ term in Eq.~\eqref{eq:friedmann} becoming dominant if $\hat{\alpha} H^2 \gg 1$.  However, given that $H_0 \approx 2.4 \times 10^{-18} \, {\rm s}^{-1}$ at the present time, this suggests that a very large value of $\hat{\alpha}$ would be required to make any noticable difference to the current rate of expansion of the Universe. 

For example, for the correction term to be of order unity would require $\hat \alpha \sim 10^{52}\,{\rm m}^2$. This means that any constraints from the recent expansion history of the universe, or from structure formation, are likely to be extremely weak. This is confirmed in Ref.~\cite{Haghani:2020ynl}, where structure formation in 4DEGB leads to the constraint $\hat \alpha  \lesssim  10^{50}\, {\rm m}^2$. To do better than this we must consider the evolution of the Universe at much earlier times. 

An ideal environment for testing our theory is therefore the epoch of primoridal nucleosynthesis, which occured roughly between 10 seconds and 20 minutes after the big bang \cite{Carroll:2001bv, Arbey:2011nf}. The products of this period can be estimated by observing the abundance of the light elements in the Universe today, and can hence be used to constrain the rate of the Universe's expansion at energy levels of about $1$ MeV, corresponding to  $H^2\approx 0.14 \, {\rm s}^{-2}$.

A rough estimate of the constraints that can be imposed on the coupling parameter $\hat{\alpha}$ can be achieved by requiring that the $H^4$ term in Eq.~\eqref{eq:friedmann} does not dominate at the start of nucleosynthesis.  This implies that 
\begin{equation}
\vert \hat{\alpha} \vert \lesssim  10^{18} \, {\rm m}^2\, . 
\end{equation}
To get a more precise result we can run the open source code PRIMAT\footnote{http://www2.iap.fr/users/pitrou/primat.htm} \cite{Pitrou:2018cgg}, with an appropriately modified expansion rate, to find the Helium mass fraction $Y_P$. Comparing the result of this to the observational bound $Y_P = 0.2449 \pm 0.0040$ \cite{Aver:2015iza} then gives $\hat{\alpha} \lesssim  10^{17} \, {\rm m}^2$. Again, this bound is comparable to that which can be achieved using observations of the perihelion precession of Mercury, but is not as strong as those that can be found using LAGEOS.

\subsection{Early Universe Inflation}

At earlier times than primordial nucleosynthesis, and at higher energies, the physical processes that occurred in the Universe are less well understood. Nevertheless, the very early Universe is widely believed to have undergone a period of very rapid expansion, known as ``inflation''. It is during this epoch that the seeds of large-scale structure are believed to have been sown, and which therefore provides us with an opportunity to constrain $\hat{\alpha}$ using processes that occured at very early times, when the correction term in  Eq.~\eqref{eq:friedmann} may become more significant. 

First, it is important to note that Eq.~\eqref{eq:friedmann} in combination with the energy conservation equation, which is unchanged in this theory, indicates that $\dot H$ can become infinite at a finite value of the scale factor for negative values of $\hat \alpha$. This occurs when $\Gamma=0$ and implies that  there is an upper limit on $H$ given by $H^2 = 1/(-2 \hat \alpha) $. If inflation takes place above the TeV scale, which is consistent with the lack of new physics at the LHC and also the need for Baryogenesis, we can turn this upper limit into the constraint $\hat \alpha \gtrsim  -10^{-6} \, {\rm m}^2$, which would be an extremely tight contraint. Now let us see if anything can be said concerning positive values of $\hat \alpha$. 

Inflation occurs when $\epsilon \equiv -\dot{H}/H^2 < 1$. If we consider the matter content of the Universe during this period to be well modelled by a scalar field $\phi$ in a potential $V(\phi)$, then this translates to the condition\footnote{The reader will note that we have switched to Planck units here, as this is standard choice in this area of physics.}
\begin{equation}
\frac{2 V'^2 \hat{\alpha}^2}{9 m_{\rm pl}^2 \Gamma_V(\Gamma_V-1)^2} \lesssim 1 \, ,
\end{equation}
where  $\displaystyle \Gamma_V = \sqrt{1 + 4 V\hat{\alpha}/(3 m_{\rm pl}^2})$, and where a dash indicates a derivative with respect to $\phi$. 
This result implies that inflation is harder to achieve in 4DEGB than it is in GR, and that 
the slope of the potential (i.e. $V'/V$) must be shallower. 
However, we can always tailor the form of $V(\phi)$ in order to allow inflation to still proceed.

Let us now turn to the production of perturbations during inflation. The action for tensor perturbations in 4DEGB is given by
\begin{equation}
\label{eq:tensorf}
S_T= \frac{1}{2}\int {d}^3 x \, {d} t \,a^3 \,\Gamma \left ( \dot{h}^2  - c_T^2 \frac{(\partial h )^2}{a^2} \right)\,,
\end{equation}
where $h$ is the amplitude of either of the two gravitational wave polarisations, $c_T^2 = 1 + \dot{\Gamma}/(H \Gamma) $ is the speed of propagation, and $\Gamma = 1 + 2 \hat{\alpha} H^2$ (i.e. we are again setting $C=0$, for convenience). The evolution of the uniform density curvature perturbation $\zeta$ follows from the action
\begin{equation}
\label{eq:zetaf}
S_\zeta= \frac{1}{2}\int {d}^3 x \, {d} t \,a^3 \Gamma \epsilon \left ( \dot{\zeta}^2  -  \frac{(\partial \zeta )^2}{a^2} \right)\,.
\end{equation}
These equations imply that both the curvature and tensorial perturbations are conserved on super-horizon scales, when the wavenumber obeys $k> aH$. 

Equations~\eqref{eq:tensorf} and \eqref{eq:zetaf} allow the spectra of tensor and scalar perturbations to be calculated in the usual way, in terms of quantities at the time a given $k$ crosses the apparent horizon, which results in
\begin{equation}
P_T = \left.\frac{2}{\pi^2 m_{\rm pl}^2 }\frac{H^2}{\Gamma}\right |_*  \quad ~ ~ {\rm and} \quad ~ ~ P_\zeta = \left. \frac{1}{8\pi^2 m_{\rm pl}^2} \frac{H^2}{\epsilon \Gamma}\right |_* \, ,
\end{equation}
where the asterisk indicates quantities are to be evaluated at horizon crossing. 

As already discussed in Ref.~\cite{Casalino:2020pyv}, these expressions imply that the tensor-to-scalar ratio takes its usual form $r=16\epsilon$, while differentiating the power spectra with respect to horizon crossing scale $k=aH$ gives the spectral indices at leading-order in slow roll as
\begin{align}
\frac{\partial \log P_\zeta }{\partial \log (a H) } &= n_s -1 =  - 2 \epsilon - \dot \epsilon/(\epsilon H)- \dot \Gamma/(\Gamma H)\,,\nonumber\\
\frac{\partial \log P_T }{\partial \log (a H) } &= n_T = -2 \epsilon- \dot \Gamma/(\Gamma H)\,.
\end{align}
It is clear that $n_T \neq -r/8$, meaning that the consistency equation of canonical single field inflation is violated. The form of $n_s$ is also different from its canonical form, but even when $\hat{\alpha} H^2 \gg 1$ it seems it should still be possible to choose a $V(\phi)$ that meets the tight observational constraints on this quantity \cite{Akrami:2018odb}.

We conclude that although both the background dynamics and the spectra of perturbations are different from their canonical form, there does not appear to be any compelling reason why inflation in 4DEGB should not be considered consistent with current observations for any positive value of $\hat{\alpha}$ (assuming we are otherwise free to choose the shape of the inflationary potential). This may change in the future, when the spectrum of primordial gravitational waves is observed, and the consistency condition can be tested. It may also change when higher-order correlations are calculated, but we leave calculations of such quantities for future work.

\begin{table}
\begin{center}
 \begin{tabular}{|c | c | c |} 
 \hline
 Observation & Upper bound  & Data \\
 & on $\vert \hat{\alpha} \vert / {\rm m}^2 \phantom{*}$ & source  \\ [0.5ex] 
 \hline\hline
 GW observations & $\sim 10^{8}*$ & Ref. \cite{TheLIGOScientific:2016wfe} \\
 \hline
 LAGEOS satellites & $ \phantom{\sim} 10^{10}\phantom{*}$ & Ref.~\cite{Lucchesi:2010zzb} \\ 
 \hline
  Double Pulsar & $\sim 10^{15}\phantom{*}$ &  Ref.~\cite{Kramer:2006nb} \\
 \hline
  Tabletop experiment & $ \sim 10^{16}\phantom{*}$ &  Ref.~\cite{Adelberger:2006dh} \\
 \hline
   Orbit of Mercury & $\phantom{\sim}10^{17}\phantom{*}$ & Ref.~\cite{pitjeva2013relativistic} \\ 
 \hline
   Primordial nucleosynthesis & $ \phantom{\sim}10^{18}\phantom{*}$ &  Ref.~\cite{Aver:2015iza} \\
 \hline
 Orbits around Sgr $A^*$ & $\phantom{\sim}10^{25}\phantom{*}$ & Ref.~\cite{Abuter:2020dou} \\ 
 \hline
Event Horizon Telescope & ${\sim} 10^{26} *$ &  Ref.~\cite{TheLIGOScientific:2017qsa} \\
 \hline
   Speed of GWs & $\phantom{\sim} 10^{36}\phantom{*}$ &  Ref.~\cite{TheLIGOScientific:2017qsa} \\
 \hline
  Gravitational lensing & $\phantom{\sim}-\phantom{*}$ &   \\
 \hline
 Shapiro time delay & $\phantom{\sim}-\phantom{*}$ &  \\
 \hline
   Early Universe inflation & $\phantom{\sim}-*$ &  \\
 \hline
\end{tabular}
\end{center}
\caption{\label{table}A summary of the order-of-magnitude constraints available on $\vert \hat{\alpha} \vert / {\rm m}^2$ for the various different observables considered in this paper, ordered by stringency. A dash indicates no constraint, a $\sim$ indicates that constraints are indicative (due to simplifying assumptions), and a $*$ indicates that asymmetric bounds are available on positive and negative values of $\hat \alpha$ (the weaker of the two are shown here).}
\end{table}

\section{Discussion}

We have studied the observational constraints that can be imposed on the coupling parameter $\hat{\alpha}$ of regularized 4DEGB theory. This has included studying the weak field solutions of this theory, and calculating the equations of two body dynamics within it. It has also included the bounds that can be imposed by studying the propagation of electromagnetic and gravitational radiation, as well as black hole shadows, tabletop experiments, primordial nucleosynthesis of the elements, and early universe inflation. Our results are summarized in Table \ref{table}.

The tightest definite constraint in Table \ref{table} come from observations of the periapsis advance of the LAGEOS II satellite, which gives $\vert \hat{\alpha} \vert \lesssim 10^{10} \, {\rm m}^2$. Other observations, which often give tight contraints on alternative theories of gravity are much less constraining, with observations based on gravitational lensing and the Shapiro time delay effect giving no constraints at all. In particular, the recent constraints on the propagational velocity of gravitational waves from GW170817, which are often highly constraining for scalar-tensor theories, are found to be particularly weak in this case. 

Being less conservative, we note that early universe inflation appears to rule out all but the smallest negative values of $\hat \alpha$, and that binary black hole systems offer the possibility of strong constraints on positive values, leading to the overall range of allowed values being
\begin{equation}
0 \lesssim \hat \alpha \lesssim 10^8 \, {\rm m}^2 \, .
\end{equation}
These are the strongest constraints that we are aware of, for the regularized 4DEGB theory (\ref{eq:action}).

Taking our conservative constraint of $\vert \hat{\alpha} \vert \lesssim 10^{10} \, {\rm m}^2$, it would appear that strong deviations from GR are only possible in the very early universe (at times $t\lesssim 10^{-3} s$) or in the immediate vicinity of stellar-mass black holes ($M \lesssim 100 M_{\odot}$). This is promising in one sense, in that the merger events of such objects are now being recorded by the LIGO/Virgo collaboration with high frequency. It would be particularly interesting to run numerical simulations of such events in 4DEGB, to determine what observational signatures should be expected to result. On the other hand, our results suggest that there is unlikely to be any observable consequences from studying super-massive black holes or the expansion of the late universe. The accelerated expansion of the Universe being driven by this theory, in particular, is ruled out to extremely high significance by these bounds.

We consider this work to be a first study on the observational constraints that can be imposed on 4DEGB, with much remaining work to be done to make these bounds more precise. In particular, effects such as geodetic precession and the Nordvedt effect have not been included here at all, as they will require detailed analyses of spinning and extended bodies in order to be applied. Likewise, strong field calculations have only been estimated, with more work remaining to be done to fully understand rotating and multi-black hole systems. In the end, we expect observations of binary black hole mergers and early universe physics to produce the tightest constraints on the 4DEGB theory, as it is these regimes that the new non-linear gravitational effects of this theory will become most pronounced.

\section*{Acknowledgements}

TC and PC acknowledge financial support from the STFC under grant ST/P000592/1. PF is supported by the Royal Society grant RGF/EA/180022 and acknowledges support from the project CERN/FISPAR/0027/2019, and DJM is supported by a Royal Society University Research Fellowship.

\appendix

\section{Calculating $\Phi_{\mathcal{G}}$}
\label{appa}

Here we wish to calculate the form of $\Phi_{\mathcal{G}}$ for a system of $N$ bodies with energy density given by Eq. (\ref{ed}). We start by defining a general potential $\mathcal{V}$ sourced by a field $X$:
\begin{equation}
\mathcal{V} (X) = \int \frac{X'}{\vert {\bf x} - {\bf x}' \vert} d^3x' \, .
\end{equation}
If we use the identity $$U_{,ij} U_{,ij} =\frac{1}{2} \nabla^2 \vert \nabla U \vert^2 +4 \pi \nabla \rho \cdot \nabla U\,,$$ then we can use $\Phi_{\mathcal{G}} = \mathcal{V}(\mathcal{G})$ to write
\begin{align*}
\Phi_{\mathcal{G}}
=& 8 \mathcal{V} (U_{,ij} U_{,ij}) - 8 \mathcal{V} ((\nabla^2 U)^2) \\
=& 4 \mathcal{V} ( \nabla^2 \vert \nabla U \vert^2) + 32 \pi \mathcal{V} (\nabla \rho \cdot \nabla U) - 8 (4 \pi)^2 \mathcal{V} (\rho^2) \, .
\end{align*}
Now $\mathcal{V} ( \nabla^2 \vert \nabla U \vert^2) = -4 \pi  \vert \nabla U \vert^2$, and we can write
\begin{equation*}
\mathcal{V} (\nabla \rho \cdot \nabla U) 
= 4\pi \mathcal{V} (\rho^2) +\psi_1 \, ,
\end{equation*}
where we have discarded a surface term and where
\begin{equation}
\psi_1 = \int \frac{\rho' \rho'' ({\bf x} - {\bf x}' ) \cdot ({\bf x}' - {\bf x}'' )}{\vert {\bf x} - {\bf x}' \vert^3 \vert {\bf x}' - {\bf x}'' \vert^3} d^3 x' d^3 x'' \, .
\end{equation}
This all gives
\begin{equation}
\Phi_{\mathcal{G}} = -32 \pi \left( \frac{1}{2} \vert \nabla U \vert^2 - \psi_1 \right) \, ,
\end{equation}
which on substituting for $\tilde{\mu}$ from (\ref{ed}) into the relevant expressions for $U$ and $\psi_1$ gives Eqs. (\ref{gn1}) and (\ref{gn2}), once divergent terms are neglected.

\bibliography{biblio}

\begin{thebibliography}{151}%
\makeatletter
\providecommand \@ifxundefined [1]{%
 \@ifx{#1\undefined}
}%
\providecommand \@ifnum [1]{%
 \ifnum #1\expandafter \@firstoftwo
 \else \expandafter \@secondoftwo
 \fi
}%
\providecommand \@ifx [1]{%
 \ifx #1\expandafter \@firstoftwo
 \else \expandafter \@secondoftwo
 \fi
}%
\providecommand \natexlab [1]{#1}%
\providecommand \enquote  [1]{``#1''}%
\providecommand \bibnamefont  [1]{#1}%
\providecommand \bibfnamefont [1]{#1}%
\providecommand \citenamefont [1]{#1}%
\providecommand \href@noop [0]{\@secondoftwo}%
\providecommand \href [0]{\begingroup \@sanitize@url \@href}%
\providecommand \@href[1]{\@@startlink{#1}\@@href}%
\providecommand \@@href[1]{\endgroup#1\@@endlink}%
\providecommand \@sanitize@url [0]{\catcode `\\12\catcode `\$12\catcode
  `\&12\catcode `\#12\catcode `\^12\catcode `\_12\catcode `\%12\relax}%
\providecommand \@@startlink[1]{}%
\providecommand \@@endlink[0]{}%
\providecommand \url  [0]{\begingroup\@sanitize@url \@url }%
\providecommand \@url [1]{\endgroup\@href {#1}{\urlprefix }}%
\providecommand \urlprefix  [0]{URL }%
\providecommand \Eprint [0]{\href }%
\providecommand \doibase [0]{http://dx.doi.org/}%
\providecommand \selectlanguage [0]{\@gobble}%
\providecommand \bibinfo  [0]{\@secondoftwo}%
\providecommand \bibfield  [0]{\@secondoftwo}%
\providecommand \translation [1]{[#1]}%
\providecommand \BibitemOpen [0]{}%
\providecommand \bibitemStop [0]{}%
\providecommand \bibitemNoStop [0]{.\EOS\space}%
\providecommand \EOS [0]{\spacefactor3000\relax}%
\providecommand \BibitemShut  [1]{\csname bibitem#1\endcsname}%
\let\auto@bib@innerbib\@empty
\bibitem [{\citenamefont {Zwiebach}(1985)}]{zwiebach1985curvature}%
  \BibitemOpen
  \bibfield  {author} {\bibinfo {author} {\bibfnamefont {B.}~\bibnamefont
  {Zwiebach}},\ }\href@noop {} {\bibfield  {journal} {\bibinfo  {journal}
  {Physics Letters B}\ }\textbf {\bibinfo {volume} {156}},\ \bibinfo {pages}
  {315} (\bibinfo {year} {1985})}\BibitemShut {NoStop}%
\bibitem [{\citenamefont {Nepomechie}(1985)}]{nepomechie1985low}%
  \BibitemOpen
  \bibfield  {author} {\bibinfo {author} {\bibfnamefont {R.~I.}\ \bibnamefont
  {Nepomechie}},\ }\href@noop {} {\bibfield  {journal} {\bibinfo  {journal}
  {Physical Review D}\ }\textbf {\bibinfo {volume} {32}},\ \bibinfo {pages}
  {3201} (\bibinfo {year} {1985})}\BibitemShut {NoStop}%
\bibitem [{\citenamefont {Maldacena}(1999)}]{maldacena1999large}%
  \BibitemOpen
  \bibfield  {author} {\bibinfo {author} {\bibfnamefont {J.}~\bibnamefont
  {Maldacena}},\ }\href@noop {} {\bibfield  {journal} {\bibinfo  {journal}
  {International journal of theoretical physics}\ }\textbf {\bibinfo {volume}
  {38}},\ \bibinfo {pages} {1113} (\bibinfo {year} {1999})}\BibitemShut
  {NoStop}%
\bibitem [{\citenamefont {Gubser}\ \emph {et~al.}(1998)\citenamefont {Gubser},
  \citenamefont {Klebanov},\ and\ \citenamefont {Polyakov}}]{gubser1998gauge}%
  \BibitemOpen
  \bibfield  {author} {\bibinfo {author} {\bibfnamefont {S.~S.}\ \bibnamefont
  {Gubser}}, \bibinfo {author} {\bibfnamefont {I.~R.}\ \bibnamefont
  {Klebanov}}, \ and\ \bibinfo {author} {\bibfnamefont {A.~M.}\ \bibnamefont
  {Polyakov}},\ }\href@noop {} {\bibfield  {journal} {\bibinfo  {journal}
  {Physics Letters B}\ }\textbf {\bibinfo {volume} {428}},\ \bibinfo {pages}
  {105} (\bibinfo {year} {1998})}\BibitemShut {NoStop}%
\bibitem [{\citenamefont {Witten}(1998)}]{witten1998anti}%
  \BibitemOpen
  \bibfield  {author} {\bibinfo {author} {\bibfnamefont {E.}~\bibnamefont
  {Witten}},\ }\href@noop {} {\bibfield  {journal} {\bibinfo  {journal} {arXiv
  preprint hep-th/9802150}\ } (\bibinfo {year} {1998})}\BibitemShut {NoStop}%
\bibitem [{\citenamefont {Birrell}\ \emph {et~al.}(1984)\citenamefont
  {Birrell}, \citenamefont {Birrell},\ and\ \citenamefont
  {Davies}}]{birrell1984quantum}%
  \BibitemOpen
  \bibfield  {author} {\bibinfo {author} {\bibfnamefont {N.~D.}\ \bibnamefont
  {Birrell}}, \bibinfo {author} {\bibfnamefont {N.~D.}\ \bibnamefont
  {Birrell}}, \ and\ \bibinfo {author} {\bibfnamefont {P.}~\bibnamefont
  {Davies}},\ }\href@noop {} {\emph {\bibinfo {title} {Quantum fields in curved
  space}}},\ \bibinfo {number} {7}\ (\bibinfo  {publisher} {Cambridge
  university press},\ \bibinfo {year} {1984})\BibitemShut {NoStop}%
\bibitem [{\citenamefont {Lovelock}(1971)}]{Lovelock_original}%
  \BibitemOpen
  \bibfield  {author} {\bibinfo {author} {\bibfnamefont {D.}~\bibnamefont
  {Lovelock}},\ }\href {\doibase 10.1063/1.1665613} {\bibfield  {journal}
  {\bibinfo  {journal} {J. Math. Phys.}\ }\textbf {\bibinfo {volume} {12}},\
  \bibinfo {pages} {498} (\bibinfo {year} {1971})}\BibitemShut {NoStop}%
\bibitem [{\citenamefont {Sotiriou}\ and\ \citenamefont
  {Zhou}(2014{\natexlab{a}})}]{Sotiriou:2013qea}%
  \BibitemOpen
  \bibfield  {author} {\bibinfo {author} {\bibfnamefont {T.~P.}\ \bibnamefont
  {Sotiriou}}\ and\ \bibinfo {author} {\bibfnamefont {S.-Y.}\ \bibnamefont
  {Zhou}},\ }\href {\doibase 10.1103/PhysRevLett.112.251102} {\bibfield
  {journal} {\bibinfo  {journal} {Phys. Rev. Lett.}\ }\textbf {\bibinfo
  {volume} {112}},\ \bibinfo {pages} {251102} (\bibinfo {year}
  {2014}{\natexlab{a}})},\ \Eprint {http://arxiv.org/abs/1312.3622}
  {arXiv:1312.3622 [gr-qc]} \BibitemShut {NoStop}%
\bibitem [{\citenamefont {Sotiriou}\ and\ \citenamefont
  {Zhou}(2014{\natexlab{b}})}]{Sotiriou:2014pfa}%
  \BibitemOpen
  \bibfield  {author} {\bibinfo {author} {\bibfnamefont {T.~P.}\ \bibnamefont
  {Sotiriou}}\ and\ \bibinfo {author} {\bibfnamefont {S.-Y.}\ \bibnamefont
  {Zhou}},\ }\href {\doibase 10.1103/PhysRevD.90.124063} {\bibfield  {journal}
  {\bibinfo  {journal} {Phys. Rev. D}\ }\textbf {\bibinfo {volume} {90}},\
  \bibinfo {pages} {124063} (\bibinfo {year} {2014}{\natexlab{b}})},\ \Eprint
  {http://arxiv.org/abs/1408.1698} {arXiv:1408.1698 [gr-qc]} \BibitemShut
  {NoStop}%
\bibitem [{\citenamefont {Saravani}\ and\ \citenamefont
  {Sotiriou}(2019)}]{Saravani:2019xwx}%
  \BibitemOpen
  \bibfield  {author} {\bibinfo {author} {\bibfnamefont {M.}~\bibnamefont
  {Saravani}}\ and\ \bibinfo {author} {\bibfnamefont {T.~P.}\ \bibnamefont
  {Sotiriou}},\ }\href {\doibase 10.1103/PhysRevD.99.124004} {\bibfield
  {journal} {\bibinfo  {journal} {Phys. Rev. D}\ }\textbf {\bibinfo {volume}
  {99}},\ \bibinfo {pages} {124004} (\bibinfo {year} {2019})},\ \Eprint
  {http://arxiv.org/abs/1903.02055} {arXiv:1903.02055 [gr-qc]} \BibitemShut
  {NoStop}%
\bibitem [{\citenamefont {Delgado}\ \emph {et~al.}(2020)\citenamefont
  {Delgado}, \citenamefont {Herdeiro},\ and\ \citenamefont
  {Radu}}]{Delgado:2020rev}%
  \BibitemOpen
  \bibfield  {author} {\bibinfo {author} {\bibfnamefont {J.~F.}\ \bibnamefont
  {Delgado}}, \bibinfo {author} {\bibfnamefont {C.~A.}\ \bibnamefont
  {Herdeiro}}, \ and\ \bibinfo {author} {\bibfnamefont {E.}~\bibnamefont
  {Radu}},\ }\href {\doibase 10.1007/JHEP04(2020)180} {\bibfield  {journal}
  {\bibinfo  {journal} {JHEP}\ }\textbf {\bibinfo {volume} {04}},\ \bibinfo
  {pages} {180} (\bibinfo {year} {2020})},\ \Eprint
  {http://arxiv.org/abs/2002.05012} {arXiv:2002.05012 [gr-qc]} \BibitemShut
  {NoStop}%
\bibitem [{\citenamefont {Doneva}\ and\ \citenamefont
  {Yazadjiev}(2018)}]{Doneva:2017bvd}%
  \BibitemOpen
  \bibfield  {author} {\bibinfo {author} {\bibfnamefont {D.~D.}\ \bibnamefont
  {Doneva}}\ and\ \bibinfo {author} {\bibfnamefont {S.~S.}\ \bibnamefont
  {Yazadjiev}},\ }\href {\doibase 10.1103/PhysRevLett.120.131103} {\bibfield
  {journal} {\bibinfo  {journal} {Phys. Rev. Lett.}\ }\textbf {\bibinfo
  {volume} {120}},\ \bibinfo {pages} {131103} (\bibinfo {year} {2018})},\
  \Eprint {http://arxiv.org/abs/1711.01187} {arXiv:1711.01187 [gr-qc]}
  \BibitemShut {NoStop}%
\bibitem [{\citenamefont {Silva}\ \emph {et~al.}(2018)\citenamefont {Silva},
  \citenamefont {Sakstein}, \citenamefont {Gualtieri}, \citenamefont
  {Sotiriou},\ and\ \citenamefont {Berti}}]{Silva:2017uqg}%
  \BibitemOpen
  \bibfield  {author} {\bibinfo {author} {\bibfnamefont {H.~O.}\ \bibnamefont
  {Silva}}, \bibinfo {author} {\bibfnamefont {J.}~\bibnamefont {Sakstein}},
  \bibinfo {author} {\bibfnamefont {L.}~\bibnamefont {Gualtieri}}, \bibinfo
  {author} {\bibfnamefont {T.~P.}\ \bibnamefont {Sotiriou}}, \ and\ \bibinfo
  {author} {\bibfnamefont {E.}~\bibnamefont {Berti}},\ }\href {\doibase
  10.1103/PhysRevLett.120.131104} {\bibfield  {journal} {\bibinfo  {journal}
  {Phys. Rev. Lett.}\ }\textbf {\bibinfo {volume} {120}},\ \bibinfo {pages}
  {131104} (\bibinfo {year} {2018})},\ \Eprint
  {http://arxiv.org/abs/1711.02080} {arXiv:1711.02080 [gr-qc]} \BibitemShut
  {NoStop}%
\bibitem [{\citenamefont {Antoniou}\ \emph {et~al.}(2018)\citenamefont
  {Antoniou}, \citenamefont {Bakopoulos},\ and\ \citenamefont
  {Kanti}}]{Antoniou:2017acq}%
  \BibitemOpen
  \bibfield  {author} {\bibinfo {author} {\bibfnamefont {G.}~\bibnamefont
  {Antoniou}}, \bibinfo {author} {\bibfnamefont {A.}~\bibnamefont
  {Bakopoulos}}, \ and\ \bibinfo {author} {\bibfnamefont {P.}~\bibnamefont
  {Kanti}},\ }\href {\doibase 10.1103/PhysRevLett.120.131102} {\bibfield
  {journal} {\bibinfo  {journal} {Phys. Rev. Lett.}\ }\textbf {\bibinfo
  {volume} {120}},\ \bibinfo {pages} {131102} (\bibinfo {year} {2018})},\
  \Eprint {http://arxiv.org/abs/1711.03390} {arXiv:1711.03390 [hep-th]}
  \BibitemShut {NoStop}%
\bibitem [{\citenamefont {Cunha}\ \emph {et~al.}(2019)\citenamefont {Cunha},
  \citenamefont {Herdeiro},\ and\ \citenamefont {Radu}}]{Cunha:2019dwb}%
  \BibitemOpen
  \bibfield  {author} {\bibinfo {author} {\bibfnamefont {P.~V.}\ \bibnamefont
  {Cunha}}, \bibinfo {author} {\bibfnamefont {C.~A.}\ \bibnamefont {Herdeiro}},
  \ and\ \bibinfo {author} {\bibfnamefont {E.}~\bibnamefont {Radu}},\ }\href
  {\doibase 10.1103/PhysRevLett.123.011101} {\bibfield  {journal} {\bibinfo
  {journal} {Phys. Rev. Lett.}\ }\textbf {\bibinfo {volume} {123}},\ \bibinfo
  {pages} {011101} (\bibinfo {year} {2019})},\ \Eprint
  {http://arxiv.org/abs/1904.09997} {arXiv:1904.09997 [gr-qc]} \BibitemShut
  {NoStop}%
\bibitem [{\citenamefont {Collodel}\ \emph {et~al.}(2020)\citenamefont
  {Collodel}, \citenamefont {Kleihaus}, \citenamefont {Kunz},\ and\
  \citenamefont {Berti}}]{Collodel:2019kkx}%
  \BibitemOpen
  \bibfield  {author} {\bibinfo {author} {\bibfnamefont {L.~G.}\ \bibnamefont
  {Collodel}}, \bibinfo {author} {\bibfnamefont {B.}~\bibnamefont {Kleihaus}},
  \bibinfo {author} {\bibfnamefont {J.}~\bibnamefont {Kunz}}, \ and\ \bibinfo
  {author} {\bibfnamefont {E.}~\bibnamefont {Berti}},\ }\href {\doibase
  10.1088/1361-6382/ab74f9} {\bibfield  {journal} {\bibinfo  {journal} {Class.
  Quant. Grav.}\ }\textbf {\bibinfo {volume} {37}},\ \bibinfo {pages} {075018}
  (\bibinfo {year} {2020})},\ \Eprint {http://arxiv.org/abs/1912.05382}
  {arXiv:1912.05382 [gr-qc]} \BibitemShut {NoStop}%
\bibitem [{\citenamefont {Kanti}\ \emph {et~al.}(1996)\citenamefont {Kanti},
  \citenamefont {Mavromatos}, \citenamefont {Rizos}, \citenamefont {Tamvakis},\
  and\ \citenamefont {Winstanley}}]{Kanti:1995vq}%
  \BibitemOpen
  \bibfield  {author} {\bibinfo {author} {\bibfnamefont {P.}~\bibnamefont
  {Kanti}}, \bibinfo {author} {\bibfnamefont {N.}~\bibnamefont {Mavromatos}},
  \bibinfo {author} {\bibfnamefont {J.}~\bibnamefont {Rizos}}, \bibinfo
  {author} {\bibfnamefont {K.}~\bibnamefont {Tamvakis}}, \ and\ \bibinfo
  {author} {\bibfnamefont {E.}~\bibnamefont {Winstanley}},\ }\href {\doibase
  10.1103/PhysRevD.54.5049} {\bibfield  {journal} {\bibinfo  {journal} {Phys.
  Rev. D}\ }\textbf {\bibinfo {volume} {54}},\ \bibinfo {pages} {5049}
  (\bibinfo {year} {1996})},\ \Eprint {http://arxiv.org/abs/hep-th/9511071}
  {arXiv:hep-th/9511071} \BibitemShut {NoStop}%
\bibitem [{\citenamefont {Kleihaus}\ \emph {et~al.}(2011)\citenamefont
  {Kleihaus}, \citenamefont {Kunz},\ and\ \citenamefont
  {Radu}}]{Kleihaus:2011tg}%
  \BibitemOpen
  \bibfield  {author} {\bibinfo {author} {\bibfnamefont {B.}~\bibnamefont
  {Kleihaus}}, \bibinfo {author} {\bibfnamefont {J.}~\bibnamefont {Kunz}}, \
  and\ \bibinfo {author} {\bibfnamefont {E.}~\bibnamefont {Radu}},\ }\href
  {\doibase 10.1103/PhysRevLett.106.151104} {\bibfield  {journal} {\bibinfo
  {journal} {Phys. Rev. Lett.}\ }\textbf {\bibinfo {volume} {106}},\ \bibinfo
  {pages} {151104} (\bibinfo {year} {2011})},\ \Eprint
  {http://arxiv.org/abs/1101.2868} {arXiv:1101.2868 [gr-qc]} \BibitemShut
  {NoStop}%
\bibitem [{\citenamefont {Kleihaus}\ \emph {et~al.}(2016)\citenamefont
  {Kleihaus}, \citenamefont {Kunz}, \citenamefont {Mojica},\ and\ \citenamefont
  {Radu}}]{Kleihaus:2015aje}%
  \BibitemOpen
  \bibfield  {author} {\bibinfo {author} {\bibfnamefont {B.}~\bibnamefont
  {Kleihaus}}, \bibinfo {author} {\bibfnamefont {J.}~\bibnamefont {Kunz}},
  \bibinfo {author} {\bibfnamefont {S.}~\bibnamefont {Mojica}}, \ and\ \bibinfo
  {author} {\bibfnamefont {E.}~\bibnamefont {Radu}},\ }\href {\doibase
  10.1103/PhysRevD.93.044047} {\bibfield  {journal} {\bibinfo  {journal} {Phys.
  Rev. D}\ }\textbf {\bibinfo {volume} {93}},\ \bibinfo {pages} {044047}
  (\bibinfo {year} {2016})},\ \Eprint {http://arxiv.org/abs/1511.05513}
  {arXiv:1511.05513 [gr-qc]} \BibitemShut {NoStop}%
\bibitem [{\citenamefont {Cunha}\ \emph {et~al.}(2017)\citenamefont {Cunha},
  \citenamefont {Herdeiro}, \citenamefont {Kleihaus}, \citenamefont {Kunz},\
  and\ \citenamefont {Radu}}]{Cunha:2016wzk}%
  \BibitemOpen
  \bibfield  {author} {\bibinfo {author} {\bibfnamefont {P.~V.}\ \bibnamefont
  {Cunha}}, \bibinfo {author} {\bibfnamefont {C.~A.~R.}\ \bibnamefont
  {Herdeiro}}, \bibinfo {author} {\bibfnamefont {B.}~\bibnamefont {Kleihaus}},
  \bibinfo {author} {\bibfnamefont {J.}~\bibnamefont {Kunz}}, \ and\ \bibinfo
  {author} {\bibfnamefont {E.}~\bibnamefont {Radu}},\ }\href {\doibase
  10.1016/j.physletb.2017.03.020} {\bibfield  {journal} {\bibinfo  {journal}
  {Phys. Lett. B}\ }\textbf {\bibinfo {volume} {768}},\ \bibinfo {pages} {373}
  (\bibinfo {year} {2017})},\ \Eprint {http://arxiv.org/abs/1701.00079}
  {arXiv:1701.00079 [gr-qc]} \BibitemShut {NoStop}%
\bibitem [{\citenamefont {Blázquez-Salcedo}\ \emph {et~al.}(2017)\citenamefont
  {Blázquez-Salcedo}, \citenamefont {Khoo},\ and\ \citenamefont
  {Kunz}}]{Blazquez-Salcedo:2017txk}%
  \BibitemOpen
  \bibfield  {author} {\bibinfo {author} {\bibfnamefont {J.~L.}\ \bibnamefont
  {Blázquez-Salcedo}}, \bibinfo {author} {\bibfnamefont {F.~S.}\ \bibnamefont
  {Khoo}}, \ and\ \bibinfo {author} {\bibfnamefont {J.}~\bibnamefont {Kunz}},\
  }\href {\doibase 10.1103/PhysRevD.96.064008} {\bibfield  {journal} {\bibinfo
  {journal} {Phys. Rev. D}\ }\textbf {\bibinfo {volume} {96}},\ \bibinfo
  {pages} {064008} (\bibinfo {year} {2017})},\ \Eprint
  {http://arxiv.org/abs/1706.03262} {arXiv:1706.03262 [gr-qc]} \BibitemShut
  {NoStop}%
\bibitem [{\citenamefont {Nojiri}\ \emph {et~al.}(2005)\citenamefont {Nojiri},
  \citenamefont {Odintsov},\ and\ \citenamefont {Sasaki}}]{Nojiri:2005vv}%
  \BibitemOpen
  \bibfield  {author} {\bibinfo {author} {\bibfnamefont {S.}~\bibnamefont
  {Nojiri}}, \bibinfo {author} {\bibfnamefont {S.~D.}\ \bibnamefont
  {Odintsov}}, \ and\ \bibinfo {author} {\bibfnamefont {M.}~\bibnamefont
  {Sasaki}},\ }\href {\doibase 10.1103/PhysRevD.71.123509} {\bibfield
  {journal} {\bibinfo  {journal} {Phys. Rev. D}\ }\textbf {\bibinfo {volume}
  {71}},\ \bibinfo {pages} {123509} (\bibinfo {year} {2005})},\ \Eprint
  {http://arxiv.org/abs/hep-th/0504052} {arXiv:hep-th/0504052} \BibitemShut
  {NoStop}%
\bibitem [{\citenamefont {Jiang}\ \emph {et~al.}(2013)\citenamefont {Jiang},
  \citenamefont {Hu},\ and\ \citenamefont {Guo}}]{Jiang:2013gza}%
  \BibitemOpen
  \bibfield  {author} {\bibinfo {author} {\bibfnamefont {P.-X.}\ \bibnamefont
  {Jiang}}, \bibinfo {author} {\bibfnamefont {J.-W.}\ \bibnamefont {Hu}}, \
  and\ \bibinfo {author} {\bibfnamefont {Z.-K.}\ \bibnamefont {Guo}},\ }\href
  {\doibase 10.1103/PhysRevD.88.123508} {\bibfield  {journal} {\bibinfo
  {journal} {Phys. Rev. D}\ }\textbf {\bibinfo {volume} {88}},\ \bibinfo
  {pages} {123508} (\bibinfo {year} {2013})},\ \Eprint
  {http://arxiv.org/abs/1310.5579} {arXiv:1310.5579 [hep-th]} \BibitemShut
  {NoStop}%
\bibitem [{\citenamefont {Kanti}\ \emph {et~al.}(2015)\citenamefont {Kanti},
  \citenamefont {Gannouji},\ and\ \citenamefont {Dadhich}}]{Kanti:2015pda}%
  \BibitemOpen
  \bibfield  {author} {\bibinfo {author} {\bibfnamefont {P.}~\bibnamefont
  {Kanti}}, \bibinfo {author} {\bibfnamefont {R.}~\bibnamefont {Gannouji}}, \
  and\ \bibinfo {author} {\bibfnamefont {N.}~\bibnamefont {Dadhich}},\ }\href
  {\doibase 10.1103/PhysRevD.92.041302} {\bibfield  {journal} {\bibinfo
  {journal} {Phys. Rev. D}\ }\textbf {\bibinfo {volume} {92}},\ \bibinfo
  {pages} {041302} (\bibinfo {year} {2015})},\ \Eprint
  {http://arxiv.org/abs/1503.01579} {arXiv:1503.01579 [hep-th]} \BibitemShut
  {NoStop}%
\bibitem [{\citenamefont {Chakraborty}\ \emph {et~al.}(2018)\citenamefont
  {Chakraborty}, \citenamefont {Paul},\ and\ \citenamefont
  {SenGupta}}]{Chakraborty:2018scm}%
  \BibitemOpen
  \bibfield  {author} {\bibinfo {author} {\bibfnamefont {S.}~\bibnamefont
  {Chakraborty}}, \bibinfo {author} {\bibfnamefont {T.}~\bibnamefont {Paul}}, \
  and\ \bibinfo {author} {\bibfnamefont {S.}~\bibnamefont {SenGupta}},\ }\href
  {\doibase 10.1103/PhysRevD.98.083539} {\bibfield  {journal} {\bibinfo
  {journal} {Phys. Rev. D}\ }\textbf {\bibinfo {volume} {98}},\ \bibinfo
  {pages} {083539} (\bibinfo {year} {2018})},\ \Eprint
  {http://arxiv.org/abs/1804.03004} {arXiv:1804.03004 [gr-qc]} \BibitemShut
  {NoStop}%
\bibitem [{\citenamefont {Odintsov}\ and\ \citenamefont
  {Oikonomou}(2018)}]{Odintsov:2018zhw}%
  \BibitemOpen
  \bibfield  {author} {\bibinfo {author} {\bibfnamefont {S.}~\bibnamefont
  {Odintsov}}\ and\ \bibinfo {author} {\bibfnamefont {V.}~\bibnamefont
  {Oikonomou}},\ }\href {\doibase 10.1103/PhysRevD.98.044039} {\bibfield
  {journal} {\bibinfo  {journal} {Phys. Rev. D}\ }\textbf {\bibinfo {volume}
  {98}},\ \bibinfo {pages} {044039} (\bibinfo {year} {2018})},\ \Eprint
  {http://arxiv.org/abs/1808.05045} {arXiv:1808.05045 [gr-qc]} \BibitemShut
  {NoStop}%
\bibitem [{\citenamefont {Odintsov}\ and\ \citenamefont
  {Oikonomou}(2019)}]{Odintsov:2019clh}%
  \BibitemOpen
  \bibfield  {author} {\bibinfo {author} {\bibfnamefont {S.}~\bibnamefont
  {Odintsov}}\ and\ \bibinfo {author} {\bibfnamefont {V.}~\bibnamefont
  {Oikonomou}},\ }\href {\doibase 10.1016/j.physletb.2019.134874} {\bibfield
  {journal} {\bibinfo  {journal} {Phys. Lett. B}\ }\textbf {\bibinfo {volume}
  {797}},\ \bibinfo {pages} {134874} (\bibinfo {year} {2019})},\ \Eprint
  {http://arxiv.org/abs/1908.07555} {arXiv:1908.07555 [gr-qc]} \BibitemShut
  {NoStop}%
\bibitem [{\citenamefont {Odintsov}\ and\ \citenamefont
  {Oikonomou}(2020)}]{Odintsov:2020zkl}%
  \BibitemOpen
  \bibfield  {author} {\bibinfo {author} {\bibfnamefont {S.}~\bibnamefont
  {Odintsov}}\ and\ \bibinfo {author} {\bibfnamefont {V.}~\bibnamefont
  {Oikonomou}},\ }\href {\doibase 10.1016/j.physletb.2020.135437} {\bibfield
  {journal} {\bibinfo  {journal} {Phys. Lett. B}\ }\textbf {\bibinfo {volume}
  {805}},\ \bibinfo {pages} {135437} (\bibinfo {year} {2020})},\ \Eprint
  {http://arxiv.org/abs/2004.00479} {arXiv:2004.00479 [gr-qc]} \BibitemShut
  {NoStop}%
\bibitem [{\citenamefont {Fernandes}\ \emph {et~al.}(2020)\citenamefont
  {Fernandes}, \citenamefont {Carrilho}, \citenamefont {Clifton},\ and\
  \citenamefont {Mulryne}}]{previous}%
  \BibitemOpen
  \bibfield  {author} {\bibinfo {author} {\bibfnamefont {P.~G.}\ \bibnamefont
  {Fernandes}}, \bibinfo {author} {\bibfnamefont {P.}~\bibnamefont {Carrilho}},
  \bibinfo {author} {\bibfnamefont {T.}~\bibnamefont {Clifton}}, \ and\
  \bibinfo {author} {\bibfnamefont {D.~J.}\ \bibnamefont {Mulryne}},\
  }\href@noop {} {\  (\bibinfo {year} {2020})},\ \Eprint
  {http://arxiv.org/abs/2004.08362} {arXiv:2004.08362 [gr-qc]} \BibitemShut
  {NoStop}%
\bibitem [{\citenamefont {Hennigar}\ \emph
  {et~al.}(2020{\natexlab{a}})\citenamefont {Hennigar}, \citenamefont
  {Kubiznak}, \citenamefont {Mann},\ and\ \citenamefont {Pollack}}]{otherone}%
  \BibitemOpen
  \bibfield  {author} {\bibinfo {author} {\bibfnamefont {R.~A.}\ \bibnamefont
  {Hennigar}}, \bibinfo {author} {\bibfnamefont {D.}~\bibnamefont {Kubiznak}},
  \bibinfo {author} {\bibfnamefont {R.~B.}\ \bibnamefont {Mann}}, \ and\
  \bibinfo {author} {\bibfnamefont {C.}~\bibnamefont {Pollack}},\ }\href@noop
  {} {\  (\bibinfo {year} {2020}{\natexlab{a}})},\ \Eprint
  {http://arxiv.org/abs/2004.09472} {arXiv:2004.09472 [gr-qc]} \BibitemShut
  {NoStop}%
\bibitem [{\citenamefont {Mann}\ and\ \citenamefont {Ross}(1993)}]{2Dpaper}%
  \BibitemOpen
  \bibfield  {author} {\bibinfo {author} {\bibfnamefont {R.~B.}\ \bibnamefont
  {Mann}}\ and\ \bibinfo {author} {\bibfnamefont {S.~F.}\ \bibnamefont
  {Ross}},\ }\href {\doibase 10.1088/0264-9381/10/7/015} {\bibfield  {journal}
  {\bibinfo  {journal} {Classical and Quantum Gravity}\ }\textbf {\bibinfo
  {volume} {10}},\ \bibinfo {pages} {1405–1408} (\bibinfo {year}
  {1993})}\BibitemShut {NoStop}%
\bibitem [{\citenamefont {Glavan}\ and\ \citenamefont {Lin}(2020)}]{original}%
  \BibitemOpen
  \bibfield  {author} {\bibinfo {author} {\bibfnamefont {D.~z.}\ \bibnamefont
  {Glavan}}\ and\ \bibinfo {author} {\bibfnamefont {C.}~\bibnamefont {Lin}},\
  }\href {\doibase 10.1103/PhysRevLett.124.081301} {\bibfield  {journal}
  {\bibinfo  {journal} {Phys.\ Rev.\ Lett.}\ }\textbf {\bibinfo {volume}
  {124}},\ \bibinfo {pages} {081301} (\bibinfo {year} {2020})},\ \Eprint
  {http://arxiv.org/abs/1905.03601} {arXiv:1905.03601 [gr-qc]} \BibitemShut
  {NoStop}%
\bibitem [{\citenamefont {Nojiri}\ and\ \citenamefont
  {Odintsov}(2020)}]{EGB2DIM}%
  \BibitemOpen
  \bibfield  {author} {\bibinfo {author} {\bibfnamefont {S.}~\bibnamefont
  {Nojiri}}\ and\ \bibinfo {author} {\bibfnamefont {S.~D.}\ \bibnamefont
  {Odintsov}},\ }\href@noop {} {\enquote {\bibinfo {title} {{Novel cosmological
  and black hole solutions in Einstein and higher-derivative gravity in two
  dimensions}},}\ } (\bibinfo {year} {2020}),\ \Eprint
  {http://arxiv.org/abs/2004.01404} {arXiv:2004.01404 [hep-th]} \BibitemShut
  {NoStop}%
\bibitem [{\citenamefont {Konoplya}\ and\ \citenamefont
  {Zinhailo}(2020{\natexlab{a}})}]{EGB1}%
  \BibitemOpen
  \bibfield  {author} {\bibinfo {author} {\bibfnamefont {R.}~\bibnamefont
  {Konoplya}}\ and\ \bibinfo {author} {\bibfnamefont {A.}~\bibnamefont
  {Zinhailo}},\ }\href@noop {} {\enquote {\bibinfo {title} {{Quasinormal modes,
  stability and shadows of a black hole in the novel 4D Einstein-Gauss-Bonnet
  gravity}},}\ } (\bibinfo {year} {2020}{\natexlab{a}}),\ \Eprint
  {http://arxiv.org/abs/2003.01188} {arXiv:2003.01188 [gr-qc]} \BibitemShut
  {NoStop}%
\bibitem [{\citenamefont {Guo}\ and\ \citenamefont
  {Li}(2020{\natexlab{a}})}]{EGB2}%
  \BibitemOpen
  \bibfield  {author} {\bibinfo {author} {\bibfnamefont {M.}~\bibnamefont
  {Guo}}\ and\ \bibinfo {author} {\bibfnamefont {P.-C.}\ \bibnamefont {Li}},\
  }\href@noop {} {\enquote {\bibinfo {title} {{The innermost stable circular
  orbit and shadow in the novel $4D$ Einstein-Gauss-Bonnet gravity}},}\ }
  (\bibinfo {year} {2020}{\natexlab{a}}),\ \Eprint
  {http://arxiv.org/abs/2003.02523} {arXiv:2003.02523 [gr-qc]} \BibitemShut
  {NoStop}%
\bibitem [{\citenamefont {Wei}\ and\ \citenamefont
  {Liu}(2020{\natexlab{a}})}]{Wei:2020ght}%
  \BibitemOpen
  \bibfield  {author} {\bibinfo {author} {\bibfnamefont {S.-W.}\ \bibnamefont
  {Wei}}\ and\ \bibinfo {author} {\bibfnamefont {Y.-X.}\ \bibnamefont {Liu}},\
  }\href@noop {} {\  (\bibinfo {year} {2020}{\natexlab{a}})},\ \Eprint
  {http://arxiv.org/abs/2003.07769} {arXiv:2003.07769 [gr-qc]} \BibitemShut
  {NoStop}%
\bibitem [{\citenamefont {Jin}\ \emph {et~al.}(2020{\natexlab{a}})\citenamefont
  {Jin}, \citenamefont {Gao},\ and\ \citenamefont {Liu}}]{Jin:2020emq}%
  \BibitemOpen
  \bibfield  {author} {\bibinfo {author} {\bibfnamefont {X.-H.}\ \bibnamefont
  {Jin}}, \bibinfo {author} {\bibfnamefont {Y.-X.}\ \bibnamefont {Gao}}, \ and\
  \bibinfo {author} {\bibfnamefont {D.-J.}\ \bibnamefont {Liu}},\ }\href@noop
  {} {\  (\bibinfo {year} {2020}{\natexlab{a}})},\ \Eprint
  {http://arxiv.org/abs/2004.02261} {arXiv:2004.02261 [gr-qc]} \BibitemShut
  {NoStop}%
\bibitem [{\citenamefont {Fernandes}(2020)}]{EGB3}%
  \BibitemOpen
  \bibfield  {author} {\bibinfo {author} {\bibfnamefont {P.~G.}\ \bibnamefont
  {Fernandes}},\ }\href@noop {} {\enquote {\bibinfo {title} {{Charged Black
  Holes in AdS Spaces in $4D$ Einstein Gauss-Bonnet Gravity}},}\ } (\bibinfo
  {year} {2020}),\ \Eprint {http://arxiv.org/abs/2003.05491} {arXiv:2003.05491
  [gr-qc]} \BibitemShut {NoStop}%
\bibitem [{\citenamefont {Casalino}\ \emph {et~al.}(2020)\citenamefont
  {Casalino}, \citenamefont {Colleaux}, \citenamefont {Rinaldi},\ and\
  \citenamefont {Vicentini}}]{EGB4}%
  \BibitemOpen
  \bibfield  {author} {\bibinfo {author} {\bibfnamefont {A.}~\bibnamefont
  {Casalino}}, \bibinfo {author} {\bibfnamefont {A.}~\bibnamefont {Colleaux}},
  \bibinfo {author} {\bibfnamefont {M.}~\bibnamefont {Rinaldi}}, \ and\
  \bibinfo {author} {\bibfnamefont {S.}~\bibnamefont {Vicentini}},\ }\href@noop
  {} {\enquote {\bibinfo {title} {{Regularized Lovelock gravity}},}\ }
  (\bibinfo {year} {2020}),\ \Eprint {http://arxiv.org/abs/2003.07068}
  {arXiv:2003.07068 [gr-qc]} \BibitemShut {NoStop}%
\bibitem [{\citenamefont {Konoplya}\ and\ \citenamefont
  {Zhidenko}(2020{\natexlab{a}})}]{EGB5}%
  \BibitemOpen
  \bibfield  {author} {\bibinfo {author} {\bibfnamefont {R.}~\bibnamefont
  {Konoplya}}\ and\ \bibinfo {author} {\bibfnamefont {A.}~\bibnamefont
  {Zhidenko}},\ }\href@noop {} {\enquote {\bibinfo {title} {{Black holes in the
  four-dimensional Einstein-Lovelock gravity}},}\ } (\bibinfo {year}
  {2020}{\natexlab{a}}),\ \Eprint {http://arxiv.org/abs/2003.07788}
  {arXiv:2003.07788 [gr-qc]} \BibitemShut {NoStop}%
\bibitem [{\citenamefont {Hegde}\ \emph {et~al.}(2020)\citenamefont {Hegde},
  \citenamefont {Naveena~Kumara}, \citenamefont {Rizwan}, \citenamefont {M.},\
  and\ \citenamefont {Ali}}]{EGB6}%
  \BibitemOpen
  \bibfield  {author} {\bibinfo {author} {\bibfnamefont {K.}~\bibnamefont
  {Hegde}}, \bibinfo {author} {\bibfnamefont {A.}~\bibnamefont
  {Naveena~Kumara}}, \bibinfo {author} {\bibfnamefont {C.~A.}\ \bibnamefont
  {Rizwan}}, \bibinfo {author} {\bibfnamefont {A.~K.}\ \bibnamefont {M.}}, \
  and\ \bibinfo {author} {\bibfnamefont {M.~S.}\ \bibnamefont {Ali}},\
  }\href@noop {} {\enquote {\bibinfo {title} {{Thermodynamics, Phase Transition
  and Joule Thomson Expansion of novel 4-D Gauss Bonnet AdS Black Hole}},}\ }
  (\bibinfo {year} {2020}),\ \Eprint {http://arxiv.org/abs/2003.08778}
  {arXiv:2003.08778 [gr-qc]} \BibitemShut {NoStop}%
\bibitem [{\citenamefont {Ghosh}\ and\ \citenamefont
  {Maharaj}(2020{\natexlab{a}})}]{EGB7}%
  \BibitemOpen
  \bibfield  {author} {\bibinfo {author} {\bibfnamefont {S.~G.}\ \bibnamefont
  {Ghosh}}\ and\ \bibinfo {author} {\bibfnamefont {S.~D.}\ \bibnamefont
  {Maharaj}},\ }\href@noop {} {\enquote {\bibinfo {title} {{Radiating black
  holes in the novel 4D Einstein-Gauss-Bonnet gravity}},}\ } (\bibinfo {year}
  {2020}{\natexlab{a}}),\ \Eprint {http://arxiv.org/abs/2003.09841}
  {arXiv:2003.09841 [gr-qc]} \BibitemShut {NoStop}%
\bibitem [{\citenamefont {Doneva}\ and\ \citenamefont
  {Yazadjiev}(2020)}]{EGB8}%
  \BibitemOpen
  \bibfield  {author} {\bibinfo {author} {\bibfnamefont {D.~D.}\ \bibnamefont
  {Doneva}}\ and\ \bibinfo {author} {\bibfnamefont {S.~S.}\ \bibnamefont
  {Yazadjiev}},\ }\href@noop {} {\enquote {\bibinfo {title} {{Relativistic
  stars in 4D Einstein-Gauss-Bonnet gravity}},}\ } (\bibinfo {year} {2020}),\
  \Eprint {http://arxiv.org/abs/2003.10284} {arXiv:2003.10284 [gr-qc]}
  \BibitemShut {NoStop}%
\bibitem [{\citenamefont {Zhang}\ \emph
  {et~al.}(2020{\natexlab{a}})\citenamefont {Zhang}, \citenamefont {Wei},\ and\
  \citenamefont {Liu}}]{EGB9}%
  \BibitemOpen
  \bibfield  {author} {\bibinfo {author} {\bibfnamefont {Y.-P.}\ \bibnamefont
  {Zhang}}, \bibinfo {author} {\bibfnamefont {S.-W.}\ \bibnamefont {Wei}}, \
  and\ \bibinfo {author} {\bibfnamefont {Y.-X.}\ \bibnamefont {Liu}},\
  }\href@noop {} {\enquote {\bibinfo {title} {{Spinning test pmisc in
  four-dimensional Einstein-Gauss-Bonnet Black Hole}},}\ } (\bibinfo {year}
  {2020}{\natexlab{a}}),\ \Eprint {http://arxiv.org/abs/2003.10960}
  {arXiv:2003.10960 [gr-qc]} \BibitemShut {NoStop}%
\bibitem [{\citenamefont {Konoplya}\ and\ \citenamefont
  {Zhidenko}(2020{\natexlab{b}})}]{EGB10}%
  \BibitemOpen
  \bibfield  {author} {\bibinfo {author} {\bibfnamefont {R.}~\bibnamefont
  {Konoplya}}\ and\ \bibinfo {author} {\bibfnamefont {A.}~\bibnamefont
  {Zhidenko}},\ }\href@noop {} {\enquote {\bibinfo {title} {{BTZ black holes
  with higher curvature corrections in the 3D Einstein-Lovelock theory}},}\ }
  (\bibinfo {year} {2020}{\natexlab{b}}),\ \Eprint
  {http://arxiv.org/abs/2003.12171} {arXiv:2003.12171 [gr-qc]} \BibitemShut
  {NoStop}%
\bibitem [{\citenamefont {Singh}\ and\ \citenamefont {Siwach}(2020)}]{EGB11}%
  \BibitemOpen
  \bibfield  {author} {\bibinfo {author} {\bibfnamefont {D.~V.}\ \bibnamefont
  {Singh}}\ and\ \bibinfo {author} {\bibfnamefont {S.}~\bibnamefont {Siwach}},\
  }\href@noop {} {\enquote {\bibinfo {title} {{Thermodynamics and P-v
  criticality of Bardeen-AdS Black Hole in 4-D Einstein-Gauss-Bonnet
  Gravity}},}\ } (\bibinfo {year} {2020}),\ \Eprint
  {http://arxiv.org/abs/2003.11754} {arXiv:2003.11754 [gr-qc]} \BibitemShut
  {NoStop}%
\bibitem [{\citenamefont {Ghosh}\ and\ \citenamefont {Kumar}(2020)}]{EGB12}%
  \BibitemOpen
  \bibfield  {author} {\bibinfo {author} {\bibfnamefont {S.~G.}\ \bibnamefont
  {Ghosh}}\ and\ \bibinfo {author} {\bibfnamefont {R.}~\bibnamefont {Kumar}},\
  }\href@noop {} {\enquote {\bibinfo {title} {{Generating black holes in the
  novel $4D$ Einstein-Gauss-Bonnet gravity}},}\ } (\bibinfo {year} {2020}),\
  \Eprint {http://arxiv.org/abs/2003.12291} {arXiv:2003.12291 [gr-qc]}
  \BibitemShut {NoStop}%
\bibitem [{\citenamefont {Konoplya}\ and\ \citenamefont
  {Zhidenko}(2020{\natexlab{c}})}]{EGB13}%
  \BibitemOpen
  \bibfield  {author} {\bibinfo {author} {\bibfnamefont {R.}~\bibnamefont
  {Konoplya}}\ and\ \bibinfo {author} {\bibfnamefont {A.}~\bibnamefont
  {Zhidenko}},\ }\href@noop {} {\enquote {\bibinfo {title} {{(In)stability of
  black holes in the 4D Einstein-Gauss-Bonnet and Einstein-Lovelock
  gravities}},}\ } (\bibinfo {year} {2020}{\natexlab{c}}),\ \Eprint
  {http://arxiv.org/abs/2003.12492} {arXiv:2003.12492 [gr-qc]} \BibitemShut
  {NoStop}%
\bibitem [{\citenamefont {Kumar}\ and\ \citenamefont {Kumar}(2020)}]{EGB14}%
  \BibitemOpen
  \bibfield  {author} {\bibinfo {author} {\bibfnamefont {A.}~\bibnamefont
  {Kumar}}\ and\ \bibinfo {author} {\bibfnamefont {R.}~\bibnamefont {Kumar}},\
  }\href@noop {} {\enquote {\bibinfo {title} {{Bardeen black holes in the novel
  $4D$ Einstein-Gauss-Bonnet gravity}},}\ } (\bibinfo {year} {2020}),\ \Eprint
  {http://arxiv.org/abs/2003.13104} {arXiv:2003.13104 [gr-qc]} \BibitemShut
  {NoStop}%
\bibitem [{\citenamefont {Zhang}\ \emph
  {et~al.}(2020{\natexlab{b}})\citenamefont {Zhang}, \citenamefont {Li},\ and\
  \citenamefont {Guo}}]{EGB15}%
  \BibitemOpen
  \bibfield  {author} {\bibinfo {author} {\bibfnamefont {C.-Y.}\ \bibnamefont
  {Zhang}}, \bibinfo {author} {\bibfnamefont {P.-C.}\ \bibnamefont {Li}}, \
  and\ \bibinfo {author} {\bibfnamefont {M.}~\bibnamefont {Guo}},\ }\href@noop
  {} {\enquote {\bibinfo {title} {{Greybody factor and power spectra of the
  Hawking radiation in the novel $4D$ Einstein-Gauss-Bonnet de-Sitter
  gravity}},}\ } (\bibinfo {year} {2020}{\natexlab{b}}),\ \Eprint
  {http://arxiv.org/abs/2003.13068} {arXiv:2003.13068 [hep-th]} \BibitemShut
  {NoStop}%
\bibitem [{\citenamefont {Hosseini~Mansoori}(2020{\natexlab{a}})}]{EGB16}%
  \BibitemOpen
  \bibfield  {author} {\bibinfo {author} {\bibfnamefont {S.~A.}\ \bibnamefont
  {Hosseini~Mansoori}},\ }\href@noop {} {\enquote {\bibinfo {title}
  {{Thermodynamic geometry of novel 4-D Gauss Bonnet AdS Black Hole}},}\ }
  (\bibinfo {year} {2020}{\natexlab{a}}),\ \Eprint
  {http://arxiv.org/abs/2003.13382} {arXiv:2003.13382 [gr-qc]} \BibitemShut
  {NoStop}%
\bibitem [{\citenamefont {Wei}\ and\ \citenamefont
  {Liu}(2020{\natexlab{b}})}]{EGB17}%
  \BibitemOpen
  \bibfield  {author} {\bibinfo {author} {\bibfnamefont {S.-W.}\ \bibnamefont
  {Wei}}\ and\ \bibinfo {author} {\bibfnamefont {Y.-X.}\ \bibnamefont {Liu}},\
  }\href@noop {} {\enquote {\bibinfo {title} {{Extended thermodynamics and
  microstructures of four-dimensional charged Gauss-Bonnet black hole in AdS
  space}},}\ } (\bibinfo {year} {2020}{\natexlab{b}}),\ \Eprint
  {http://arxiv.org/abs/2003.14275} {arXiv:2003.14275 [gr-qc]} \BibitemShut
  {NoStop}%
\bibitem [{\citenamefont {Singh}\ \emph
  {et~al.}(2020{\natexlab{a}})\citenamefont {Singh}, \citenamefont {Ghosh},\
  and\ \citenamefont {Maharaj}}]{EGB18}%
  \BibitemOpen
  \bibfield  {author} {\bibinfo {author} {\bibfnamefont {D.~V.}\ \bibnamefont
  {Singh}}, \bibinfo {author} {\bibfnamefont {S.~G.}\ \bibnamefont {Ghosh}}, \
  and\ \bibinfo {author} {\bibfnamefont {S.~D.}\ \bibnamefont {Maharaj}},\
  }\href@noop {} {\enquote {\bibinfo {title} {{Clouds of string in the novel
  $4D$ Einstein-Gauss-Bonnet black holes}},}\ } (\bibinfo {year}
  {2020}{\natexlab{a}}),\ \Eprint {http://arxiv.org/abs/2003.14136}
  {arXiv:2003.14136 [gr-qc]} \BibitemShut {NoStop}%
\bibitem [{\citenamefont {Churilova}(2020{\natexlab{a}})}]{EGB19}%
  \BibitemOpen
  \bibfield  {author} {\bibinfo {author} {\bibfnamefont {M.}~\bibnamefont
  {Churilova}},\ }\href@noop {} {\enquote {\bibinfo {title} {{Quasinormal modes
  of the Dirac field in the novel 4D Einstein-Gauss-Bonnet gravity}},}\ }
  (\bibinfo {year} {2020}{\natexlab{a}}),\ \Eprint
  {http://arxiv.org/abs/2004.00513} {arXiv:2004.00513 [gr-qc]} \BibitemShut
  {NoStop}%
\bibitem [{\citenamefont {Islam}\ \emph {et~al.}(2020)\citenamefont {Islam},
  \citenamefont {Kumar},\ and\ \citenamefont {Ghosh}}]{EGB20}%
  \BibitemOpen
  \bibfield  {author} {\bibinfo {author} {\bibfnamefont {S.~U.}\ \bibnamefont
  {Islam}}, \bibinfo {author} {\bibfnamefont {R.}~\bibnamefont {Kumar}}, \ and\
  \bibinfo {author} {\bibfnamefont {S.~G.}\ \bibnamefont {Ghosh}},\ }\href@noop
  {} {\enquote {\bibinfo {title} {{Gravitational lensing by black holes in $4D$
  Einstein-Gauss-Bonnet gravity}},}\ } (\bibinfo {year} {2020}),\ \Eprint
  {http://arxiv.org/abs/2004.01038} {arXiv:2004.01038 [gr-qc]} \BibitemShut
  {NoStop}%
\bibitem [{\citenamefont {Mishra}(2020)}]{EGB21}%
  \BibitemOpen
  \bibfield  {author} {\bibinfo {author} {\bibfnamefont {A.~K.}\ \bibnamefont
  {Mishra}},\ }\href@noop {} {\enquote {\bibinfo {title} {{Quasinormal modes
  and Strong Cosmic Censorship in the novel 4D Einstein-Gauss-Bonnet
  gravity}},}\ } (\bibinfo {year} {2020}),\ \Eprint
  {http://arxiv.org/abs/2004.01243} {arXiv:2004.01243 [gr-qc]} \BibitemShut
  {NoStop}%
\bibitem [{\citenamefont {Kumar}\ and\ \citenamefont {Ghosh}(2020)}]{EGB22}%
  \BibitemOpen
  \bibfield  {author} {\bibinfo {author} {\bibfnamefont {A.}~\bibnamefont
  {Kumar}}\ and\ \bibinfo {author} {\bibfnamefont {S.~G.}\ \bibnamefont
  {Ghosh}},\ }\href@noop {} {\enquote {\bibinfo {title} {{Hayward black holes
  in the novel $4D$ Einstein-Gauss-Bonnet gravity}},}\ } (\bibinfo {year}
  {2020}),\ \Eprint {http://arxiv.org/abs/2004.01131} {arXiv:2004.01131
  [gr-qc]} \BibitemShut {NoStop}%
\bibitem [{\citenamefont {Liu}\ \emph {et~al.}(2020{\natexlab{a}})\citenamefont
  {Liu}, \citenamefont {Zhu},\ and\ \citenamefont {Wu}}]{EGB23}%
  \BibitemOpen
  \bibfield  {author} {\bibinfo {author} {\bibfnamefont {C.}~\bibnamefont
  {Liu}}, \bibinfo {author} {\bibfnamefont {T.}~\bibnamefont {Zhu}}, \ and\
  \bibinfo {author} {\bibfnamefont {Q.}~\bibnamefont {Wu}},\ }\href@noop {}
  {\enquote {\bibinfo {title} {{Thin Accretion Disk around a four-dimensional
  Einstein-Gauss-Bonnet Black Hole}},}\ } (\bibinfo {year}
  {2020}{\natexlab{a}}),\ \Eprint {http://arxiv.org/abs/2004.01662}
  {arXiv:2004.01662 [gr-qc]} \BibitemShut {NoStop}%
\bibitem [{\citenamefont {Li}\ \emph {et~al.}(2020)\citenamefont {Li},
  \citenamefont {Wu},\ and\ \citenamefont {Yu}}]{EGB24}%
  \BibitemOpen
  \bibfield  {author} {\bibinfo {author} {\bibfnamefont {S.-L.}\ \bibnamefont
  {Li}}, \bibinfo {author} {\bibfnamefont {P.}~\bibnamefont {Wu}}, \ and\
  \bibinfo {author} {\bibfnamefont {H.}~\bibnamefont {Yu}},\ }\href@noop {}
  {\enquote {\bibinfo {title} {{Stability of the Einstein Static Universe in $4
  D$ Gauss-Bonnet Gravity}},}\ } (\bibinfo {year} {2020}),\ \Eprint
  {http://arxiv.org/abs/2004.02080} {arXiv:2004.02080 [gr-qc]} \BibitemShut
  {NoStop}%
\bibitem [{\citenamefont {Konoplya}\ and\ \citenamefont
  {Zinhailo}(2020{\natexlab{b}})}]{EGB25}%
  \BibitemOpen
  \bibfield  {author} {\bibinfo {author} {\bibfnamefont {R.~A.}\ \bibnamefont
  {Konoplya}}\ and\ \bibinfo {author} {\bibfnamefont {A.~F.}\ \bibnamefont
  {Zinhailo}},\ }\href@noop {} {\enquote {\bibinfo {title} {{Grey-body factors
  and Hawking radiation of black holes in $4D$ Einstein-Gauss-Bonnet
  gravity}},}\ } (\bibinfo {year} {2020}{\natexlab{b}}),\ \Eprint
  {http://arxiv.org/abs/2004.02248} {arXiv:2004.02248 [gr-qc]} \BibitemShut
  {NoStop}%
\bibitem [{\citenamefont {Heydari-Fard}\ \emph {et~al.}(2020)\citenamefont
  {Heydari-Fard}, \citenamefont {Heydari-Fard},\ and\ \citenamefont
  {Sepangi}}]{EGB26}%
  \BibitemOpen
  \bibfield  {author} {\bibinfo {author} {\bibfnamefont {M.}~\bibnamefont
  {Heydari-Fard}}, \bibinfo {author} {\bibfnamefont {M.}~\bibnamefont
  {Heydari-Fard}}, \ and\ \bibinfo {author} {\bibfnamefont {H.}~\bibnamefont
  {Sepangi}},\ }\href@noop {} {\enquote {\bibinfo {title} {{Bending of light in
  novel 4$D$ Gauss-Bonnet-de Sitter black holes by Rindler-Ishak method}},}\ }
  (\bibinfo {year} {2020}),\ \Eprint {http://arxiv.org/abs/2004.02140}
  {arXiv:2004.02140 [gr-qc]} \BibitemShut {NoStop}%
\bibitem [{\citenamefont {Jin}\ \emph {et~al.}(2020{\natexlab{b}})\citenamefont
  {Jin}, \citenamefont {Gao},\ and\ \citenamefont {Liu}}]{EGB27}%
  \BibitemOpen
  \bibfield  {author} {\bibinfo {author} {\bibfnamefont {X.-h.}\ \bibnamefont
  {Jin}}, \bibinfo {author} {\bibfnamefont {Y.-x.}\ \bibnamefont {Gao}}, \ and\
  \bibinfo {author} {\bibfnamefont {D.-j.}\ \bibnamefont {Liu}},\ }\href@noop
  {} {\enquote {\bibinfo {title} {{Strong gravitational lensing of a 4D
  Einstein-Gauss-Bonnet black hole in homogeneous plasma}},}\ } (\bibinfo
  {year} {2020}{\natexlab{b}}),\ \Eprint {http://arxiv.org/abs/2004.02261}
  {arXiv:2004.02261 [gr-qc]} \BibitemShut {NoStop}%
\bibitem [{\citenamefont {Ai}(2020)}]{EGB28}%
  \BibitemOpen
  \bibfield  {author} {\bibinfo {author} {\bibfnamefont {W.-Y.}\ \bibnamefont
  {Ai}},\ }\href@noop {} {\enquote {\bibinfo {title} {{A note on the novel 4D
  Einstein-Gauss-Bonnet gravity}},}\ } (\bibinfo {year} {2020}),\ \Eprint
  {http://arxiv.org/abs/2004.02858} {arXiv:2004.02858 [gr-qc]} \BibitemShut
  {NoStop}%
\bibitem [{\citenamefont {Zhang}\ \emph
  {et~al.}(2020{\natexlab{c}})\citenamefont {Zhang}, \citenamefont {Zhang},
  \citenamefont {Li},\ and\ \citenamefont {Guo}}]{EGB29}%
  \BibitemOpen
  \bibfield  {author} {\bibinfo {author} {\bibfnamefont {C.-Y.}\ \bibnamefont
  {Zhang}}, \bibinfo {author} {\bibfnamefont {S.-J.}\ \bibnamefont {Zhang}},
  \bibinfo {author} {\bibfnamefont {P.-C.}\ \bibnamefont {Li}}, \ and\ \bibinfo
  {author} {\bibfnamefont {M.}~\bibnamefont {Guo}},\ }\href@noop {} {\enquote
  {\bibinfo {title} {{Superradiance and stability of the novel 4D charged
  Einstein-Gauss-Bonnet black hole}},}\ } (\bibinfo {year}
  {2020}{\natexlab{c}}),\ \Eprint {http://arxiv.org/abs/2004.03141}
  {arXiv:2004.03141 [gr-qc]} \BibitemShut {NoStop}%
\bibitem [{\citenamefont {Eslam~Panah}\ and\ \citenamefont
  {Jafarzade}(2020)}]{EGB30}%
  \BibitemOpen
  \bibfield  {author} {\bibinfo {author} {\bibfnamefont {B.}~\bibnamefont
  {Eslam~Panah}}\ and\ \bibinfo {author} {\bibfnamefont {K.}~\bibnamefont
  {Jafarzade}},\ }\href@noop {} {\enquote {\bibinfo {title} {{4D
  Einstein-Gauss-Bonnet AdS Black Holes as Heat Engine}},}\ } (\bibinfo {year}
  {2020}),\ \Eprint {http://arxiv.org/abs/2004.04058} {arXiv:2004.04058
  [hep-th]} \BibitemShut {NoStop}%
\bibitem [{\citenamefont {Gurses}\ \emph {et~al.}(2020)\citenamefont {Gurses},
  \citenamefont {Sisman},\ and\ \citenamefont {Tekin}}]{EGB31}%
  \BibitemOpen
  \bibfield  {author} {\bibinfo {author} {\bibfnamefont {M.}~\bibnamefont
  {Gurses}}, \bibinfo {author} {\bibfnamefont {T.~C.}\ \bibnamefont {Sisman}},
  \ and\ \bibinfo {author} {\bibfnamefont {B.}~\bibnamefont {Tekin}},\
  }\href@noop {} {\  (\bibinfo {year} {2020})},\ \Eprint
  {http://arxiv.org/abs/2004.03390} {arXiv:2004.03390 [gr-qc]} \BibitemShut
  {NoStop}%
\bibitem [{\citenamefont {Aragón}\ \emph {et~al.}(2020)\citenamefont
  {Aragón}, \citenamefont {Bécar}, \citenamefont {González},\ and\
  \citenamefont {Vásquez}}]{EGB32}%
  \BibitemOpen
  \bibfield  {author} {\bibinfo {author} {\bibfnamefont {A.}~\bibnamefont
  {Aragón}}, \bibinfo {author} {\bibfnamefont {R.}~\bibnamefont {Bécar}},
  \bibinfo {author} {\bibfnamefont {P.}~\bibnamefont {González}}, \ and\
  \bibinfo {author} {\bibfnamefont {Y.}~\bibnamefont {Vásquez}},\ }\href@noop
  {} {\  (\bibinfo {year} {2020})},\ \Eprint {http://arxiv.org/abs/2004.05632}
  {arXiv:2004.05632 [gr-qc]} \BibitemShut {NoStop}%
\bibitem [{\citenamefont {Guo}\ and\ \citenamefont
  {Li}(2020{\natexlab{b}})}]{Guo:2020zmf}%
  \BibitemOpen
  \bibfield  {author} {\bibinfo {author} {\bibfnamefont {M.}~\bibnamefont
  {Guo}}\ and\ \bibinfo {author} {\bibfnamefont {P.-C.}\ \bibnamefont {Li}},\
  }\href@noop {} {\  (\bibinfo {year} {2020}{\natexlab{b}})},\ \Eprint
  {http://arxiv.org/abs/2003.02523} {arXiv:2003.02523 [gr-qc]} \BibitemShut
  {NoStop}%
\bibitem [{\citenamefont
  {Hosseini~Mansoori}(2020{\natexlab{b}})}]{HosseiniMansoori:2020yfj}%
  \BibitemOpen
  \bibfield  {author} {\bibinfo {author} {\bibfnamefont {S.~A.}\ \bibnamefont
  {Hosseini~Mansoori}},\ }\href@noop {} {\  (\bibinfo {year}
  {2020}{\natexlab{b}})},\ \Eprint {http://arxiv.org/abs/2003.13382}
  {arXiv:2003.13382 [gr-qc]} \BibitemShut {NoStop}%
\bibitem [{\citenamefont {Shu}(2020)}]{Shu:2020cjw}%
  \BibitemOpen
  \bibfield  {author} {\bibinfo {author} {\bibfnamefont {F.-W.}\ \bibnamefont
  {Shu}},\ }\href@noop {} {\  (\bibinfo {year} {2020})},\ \Eprint
  {http://arxiv.org/abs/2004.09339} {arXiv:2004.09339 [gr-qc]} \BibitemShut
  {NoStop}%
\bibitem [{\citenamefont {Mahapatra}(2020)}]{Mahapatra:2020rds}%
  \BibitemOpen
  \bibfield  {author} {\bibinfo {author} {\bibfnamefont {S.}~\bibnamefont
  {Mahapatra}},\ }\href@noop {} {\  (\bibinfo {year} {2020})},\ \Eprint
  {http://arxiv.org/abs/2004.09214} {arXiv:2004.09214 [gr-qc]} \BibitemShut
  {NoStop}%
\bibitem [{\citenamefont {Bonifacio}\ \emph {et~al.}(2020)\citenamefont
  {Bonifacio}, \citenamefont {Hinterbichler},\ and\ \citenamefont
  {Johnson}}]{Bonifacio:2020vbk}%
  \BibitemOpen
  \bibfield  {author} {\bibinfo {author} {\bibfnamefont {J.}~\bibnamefont
  {Bonifacio}}, \bibinfo {author} {\bibfnamefont {K.}~\bibnamefont
  {Hinterbichler}}, \ and\ \bibinfo {author} {\bibfnamefont {L.~A.}\
  \bibnamefont {Johnson}},\ }\href@noop {} {\  (\bibinfo {year} {2020})},\
  \Eprint {http://arxiv.org/abs/2004.10716} {arXiv:2004.10716 [hep-th]}
  \BibitemShut {NoStop}%
\bibitem [{\citenamefont {Jusufi}\ \emph {et~al.}(2020)\citenamefont {Jusufi},
  \citenamefont {Banerjee},\ and\ \citenamefont {Ghosh}}]{Jusufi:2020yus}%
  \BibitemOpen
  \bibfield  {author} {\bibinfo {author} {\bibfnamefont {K.}~\bibnamefont
  {Jusufi}}, \bibinfo {author} {\bibfnamefont {A.}~\bibnamefont {Banerjee}}, \
  and\ \bibinfo {author} {\bibfnamefont {S.~G.}\ \bibnamefont {Ghosh}},\
  }\href@noop {} {\  (\bibinfo {year} {2020})},\ \Eprint
  {http://arxiv.org/abs/2004.10750} {arXiv:2004.10750 [gr-qc]} \BibitemShut
  {NoStop}%
\bibitem [{\citenamefont {Ge}\ and\ \citenamefont {Sin}(2020)}]{Ge:2020tid}%
  \BibitemOpen
  \bibfield  {author} {\bibinfo {author} {\bibfnamefont {X.-H.}\ \bibnamefont
  {Ge}}\ and\ \bibinfo {author} {\bibfnamefont {S.-J.}\ \bibnamefont {Sin}},\
  }\href@noop {} {\  (\bibinfo {year} {2020})},\ \Eprint
  {http://arxiv.org/abs/2004.12191} {arXiv:2004.12191 [hep-th]} \BibitemShut
  {NoStop}%
\bibitem [{\citenamefont {Churilova}(2020{\natexlab{b}})}]{Churilova:2020mif}%
  \BibitemOpen
  \bibfield  {author} {\bibinfo {author} {\bibfnamefont {M.}~\bibnamefont
  {Churilova}},\ }\href@noop {} {\  (\bibinfo {year} {2020}{\natexlab{b}})},\
  \Eprint {http://arxiv.org/abs/2004.14172} {arXiv:2004.14172 [gr-qc]}
  \BibitemShut {NoStop}%
\bibitem [{\citenamefont {Arrechea}\ \emph {et~al.}(2020)\citenamefont
  {Arrechea}, \citenamefont {Delhom},\ and\ \citenamefont
  {Jiménez-Cano}}]{Arrechea:2020evj}%
  \BibitemOpen
  \bibfield  {author} {\bibinfo {author} {\bibfnamefont {J.}~\bibnamefont
  {Arrechea}}, \bibinfo {author} {\bibfnamefont {A.}~\bibnamefont {Delhom}}, \
  and\ \bibinfo {author} {\bibfnamefont {A.}~\bibnamefont {Jiménez-Cano}},\
  }\href@noop {} {\  (\bibinfo {year} {2020})},\ \Eprint
  {http://arxiv.org/abs/2004.12998} {arXiv:2004.12998 [gr-qc]} \BibitemShut
  {NoStop}%
\bibitem [{\citenamefont {Kumar}\ \emph {et~al.}(2020)\citenamefont {Kumar},
  \citenamefont {Islam},\ and\ \citenamefont {Ghosh}}]{Kumar:2020sag}%
  \BibitemOpen
  \bibfield  {author} {\bibinfo {author} {\bibfnamefont {R.}~\bibnamefont
  {Kumar}}, \bibinfo {author} {\bibfnamefont {S.~U.}\ \bibnamefont {Islam}}, \
  and\ \bibinfo {author} {\bibfnamefont {S.~G.}\ \bibnamefont {Ghosh}},\
  }\href@noop {} {\  (\bibinfo {year} {2020})},\ \Eprint
  {http://arxiv.org/abs/2004.12970} {arXiv:2004.12970 [gr-qc]} \BibitemShut
  {NoStop}%
\bibitem [{\citenamefont {Ghosh}\ and\ \citenamefont
  {Maharaj}(2020{\natexlab{b}})}]{Ghosh:2020cob}%
  \BibitemOpen
  \bibfield  {author} {\bibinfo {author} {\bibfnamefont {S.~G.}\ \bibnamefont
  {Ghosh}}\ and\ \bibinfo {author} {\bibfnamefont {S.~D.}\ \bibnamefont
  {Maharaj}},\ }\href@noop {} {\  (\bibinfo {year} {2020}{\natexlab{b}})},\
  \Eprint {http://arxiv.org/abs/2004.13519} {arXiv:2004.13519 [gr-qc]}
  \BibitemShut {NoStop}%
\bibitem [{\citenamefont {Yang}\ \emph
  {et~al.}(2020{\natexlab{a}})\citenamefont {Yang}, \citenamefont {Gu},
  \citenamefont {Wei},\ and\ \citenamefont {Liu}}]{Yang:2020jno}%
  \BibitemOpen
  \bibfield  {author} {\bibinfo {author} {\bibfnamefont {K.}~\bibnamefont
  {Yang}}, \bibinfo {author} {\bibfnamefont {B.-M.}\ \bibnamefont {Gu}},
  \bibinfo {author} {\bibfnamefont {S.-W.}\ \bibnamefont {Wei}}, \ and\
  \bibinfo {author} {\bibfnamefont {Y.-X.}\ \bibnamefont {Liu}},\ }\href@noop
  {} {\  (\bibinfo {year} {2020}{\natexlab{a}})},\ \Eprint
  {http://arxiv.org/abs/2004.14468} {arXiv:2004.14468 [gr-qc]} \BibitemShut
  {NoStop}%
\bibitem [{\citenamefont {Lu}\ and\ \citenamefont {Mao}(2020)}]{Lu:2020mjp}%
  \BibitemOpen
  \bibfield  {author} {\bibinfo {author} {\bibfnamefont {H.}~\bibnamefont
  {Lu}}\ and\ \bibinfo {author} {\bibfnamefont {P.}~\bibnamefont {Mao}},\
  }\href@noop {} {\  (\bibinfo {year} {2020})},\ \Eprint
  {http://arxiv.org/abs/2004.14400} {arXiv:2004.14400 [hep-th]} \BibitemShut
  {NoStop}%
\bibitem [{\citenamefont {Ma}\ and\ \citenamefont {Lu}(2020)}]{Ma:2020ufk}%
  \BibitemOpen
  \bibfield  {author} {\bibinfo {author} {\bibfnamefont {L.}~\bibnamefont
  {Ma}}\ and\ \bibinfo {author} {\bibfnamefont {H.}~\bibnamefont {Lu}},\
  }\href@noop {} {\  (\bibinfo {year} {2020})},\ \Eprint
  {http://arxiv.org/abs/2004.14738} {arXiv:2004.14738 [gr-qc]} \BibitemShut
  {NoStop}%
\bibitem [{\citenamefont {Jusufi}(2020)}]{Jusufi:2020qyw}%
  \BibitemOpen
  \bibfield  {author} {\bibinfo {author} {\bibfnamefont {K.}~\bibnamefont
  {Jusufi}},\ }\href@noop {} {\  (\bibinfo {year} {2020})},\ \Eprint
  {http://arxiv.org/abs/2005.00360} {arXiv:2005.00360 [gr-qc]} \BibitemShut
  {NoStop}%
\bibitem [{\citenamefont {Konoplya}\ and\ \citenamefont
  {Zhidenko}(2020{\natexlab{d}})}]{Konoplya:2020der}%
  \BibitemOpen
  \bibfield  {author} {\bibinfo {author} {\bibfnamefont {R.}~\bibnamefont
  {Konoplya}}\ and\ \bibinfo {author} {\bibfnamefont {A.}~\bibnamefont
  {Zhidenko}},\ }\href@noop {} {\  (\bibinfo {year} {2020}{\natexlab{d}})},\
  \Eprint {http://arxiv.org/abs/2005.02225} {arXiv:2005.02225 [gr-qc]}
  \BibitemShut {NoStop}%
\bibitem [{\citenamefont {Qiao}\ \emph {et~al.}(2020)\citenamefont {Qiao},
  \citenamefont {OuYang}, \citenamefont {Wang}, \citenamefont {Pan},\ and\
  \citenamefont {Jing}}]{Qiao:2020hkx}%
  \BibitemOpen
  \bibfield  {author} {\bibinfo {author} {\bibfnamefont {X.}~\bibnamefont
  {Qiao}}, \bibinfo {author} {\bibfnamefont {L.}~\bibnamefont {OuYang}},
  \bibinfo {author} {\bibfnamefont {D.}~\bibnamefont {Wang}}, \bibinfo {author}
  {\bibfnamefont {Q.}~\bibnamefont {Pan}}, \ and\ \bibinfo {author}
  {\bibfnamefont {J.}~\bibnamefont {Jing}},\ }\href@noop {} {\  (\bibinfo
  {year} {2020})},\ \Eprint {http://arxiv.org/abs/2005.01007} {arXiv:2005.01007
  [hep-th]} \BibitemShut {NoStop}%
\bibitem [{\citenamefont {Haghani}(2020)}]{Haghani:2020ynl}%
  \BibitemOpen
  \bibfield  {author} {\bibinfo {author} {\bibfnamefont {Z.}~\bibnamefont
  {Haghani}},\ }\href@noop {} {\  (\bibinfo {year} {2020})},\ \Eprint
  {http://arxiv.org/abs/2005.01636} {arXiv:2005.01636 [gr-qc]} \BibitemShut
  {NoStop}%
\bibitem [{\citenamefont {Liu}\ \emph {et~al.}(2020{\natexlab{b}})\citenamefont
  {Liu}, \citenamefont {Niu},\ and\ \citenamefont {Zhang}}]{Liu:2020lwc}%
  \BibitemOpen
  \bibfield  {author} {\bibinfo {author} {\bibfnamefont {P.}~\bibnamefont
  {Liu}}, \bibinfo {author} {\bibfnamefont {C.}~\bibnamefont {Niu}}, \ and\
  \bibinfo {author} {\bibfnamefont {C.-Y.}\ \bibnamefont {Zhang}},\ }\href@noop
  {} {\  (\bibinfo {year} {2020}{\natexlab{b}})},\ \Eprint
  {http://arxiv.org/abs/2005.01507} {arXiv:2005.01507 [gr-qc]} \BibitemShut
  {NoStop}%
\bibitem [{\citenamefont {Samart}\ and\ \citenamefont
  {Channuie}(2020)}]{Samart:2020sxj}%
  \BibitemOpen
  \bibfield  {author} {\bibinfo {author} {\bibfnamefont {D.}~\bibnamefont
  {Samart}}\ and\ \bibinfo {author} {\bibfnamefont {P.}~\bibnamefont
  {Channuie}},\ }\href@noop {} {\  (\bibinfo {year} {2020})},\ \Eprint
  {http://arxiv.org/abs/2005.02826} {arXiv:2005.02826 [gr-qc]} \BibitemShut
  {NoStop}%
\bibitem [{\citenamefont {Aoki}\ \emph
  {et~al.}(2020{\natexlab{a}})\citenamefont {Aoki}, \citenamefont {Gorji},\
  and\ \citenamefont {Mukohyama}}]{Aoki:2020lig}%
  \BibitemOpen
  \bibfield  {author} {\bibinfo {author} {\bibfnamefont {K.}~\bibnamefont
  {Aoki}}, \bibinfo {author} {\bibfnamefont {M.~A.}\ \bibnamefont {Gorji}}, \
  and\ \bibinfo {author} {\bibfnamefont {S.}~\bibnamefont {Mukohyama}},\
  }\href@noop {} {\  (\bibinfo {year} {2020}{\natexlab{a}})},\ \Eprint
  {http://arxiv.org/abs/2005.03859} {arXiv:2005.03859 [gr-qc]} \BibitemShut
  {NoStop}%
\bibitem [{\citenamefont {Dadhich}(2020)}]{Dadhich:2020ukj}%
  \BibitemOpen
  \bibfield  {author} {\bibinfo {author} {\bibfnamefont {N.}~\bibnamefont
  {Dadhich}},\ }\href@noop {} {\  (\bibinfo {year} {2020})},\ \Eprint
  {http://arxiv.org/abs/2005.05757} {arXiv:2005.05757 [gr-qc]} \BibitemShut
  {NoStop}%
\bibitem [{\citenamefont {Aoki}\ \emph
  {et~al.}(2020{\natexlab{b}})\citenamefont {Aoki}, \citenamefont {Gorji},\
  and\ \citenamefont {Mukohyama}}]{Aoki:2020iwm}%
  \BibitemOpen
  \bibfield  {author} {\bibinfo {author} {\bibfnamefont {K.}~\bibnamefont
  {Aoki}}, \bibinfo {author} {\bibfnamefont {M.~A.}\ \bibnamefont {Gorji}}, \
  and\ \bibinfo {author} {\bibfnamefont {S.}~\bibnamefont {Mukohyama}},\
  }\href@noop {} {\  (\bibinfo {year} {2020}{\natexlab{b}})},\ \Eprint
  {http://arxiv.org/abs/2005.08428} {arXiv:2005.08428 [gr-qc]} \BibitemShut
  {NoStop}%
\bibitem [{\citenamefont {Easson}\ \emph {et~al.}(2020)\citenamefont {Easson},
  \citenamefont {Manton},\ and\ \citenamefont {Svesko}}]{Easson:2020mpq}%
  \BibitemOpen
  \bibfield  {author} {\bibinfo {author} {\bibfnamefont {D.~A.}\ \bibnamefont
  {Easson}}, \bibinfo {author} {\bibfnamefont {T.}~\bibnamefont {Manton}}, \
  and\ \bibinfo {author} {\bibfnamefont {A.}~\bibnamefont {Svesko}},\
  }\href@noop {} {\  (\bibinfo {year} {2020})},\ \Eprint
  {http://arxiv.org/abs/2005.12292} {arXiv:2005.12292 [hep-th]} \BibitemShut
  {NoStop}%
\bibitem [{\citenamefont {Hennigar}\ \emph
  {et~al.}(2020{\natexlab{b}})\citenamefont {Hennigar}, \citenamefont
  {Kubiznak},\ and\ \citenamefont {Mann}}]{Hennigar:2020zif}%
  \BibitemOpen
  \bibfield  {author} {\bibinfo {author} {\bibfnamefont {R.~A.}\ \bibnamefont
  {Hennigar}}, \bibinfo {author} {\bibfnamefont {D.}~\bibnamefont {Kubiznak}},
  \ and\ \bibinfo {author} {\bibfnamefont {R.~B.}\ \bibnamefont {Mann}},\
  }\href@noop {} {\  (\bibinfo {year} {2020}{\natexlab{b}})},\ \Eprint
  {http://arxiv.org/abs/2005.13732} {arXiv:2005.13732 [gr-qc]} \BibitemShut
  {NoStop}%
\bibitem [{\citenamefont {Singh}\ \emph
  {et~al.}(2020{\natexlab{b}})\citenamefont {Singh}, \citenamefont {Kumar},
  \citenamefont {Ghosh},\ and\ \citenamefont {Maharaj}}]{Singh:2020mty}%
  \BibitemOpen
  \bibfield  {author} {\bibinfo {author} {\bibfnamefont {D.~V.}\ \bibnamefont
  {Singh}}, \bibinfo {author} {\bibfnamefont {R.}~\bibnamefont {Kumar}},
  \bibinfo {author} {\bibfnamefont {S.~G.}\ \bibnamefont {Ghosh}}, \ and\
  \bibinfo {author} {\bibfnamefont {S.~D.}\ \bibnamefont {Maharaj}},\
  }\href@noop {} {\  (\bibinfo {year} {2020}{\natexlab{b}})},\ \Eprint
  {http://arxiv.org/abs/2006.00594} {arXiv:2006.00594 [gr-qc]} \BibitemShut
  {NoStop}%
\bibitem [{\citenamefont {Lin}\ \emph {et~al.}(2020)\citenamefont {Lin},
  \citenamefont {Yang}, \citenamefont {Wei}, \citenamefont {Wang},\ and\
  \citenamefont {Liu}}]{Lin:2020kqe}%
  \BibitemOpen
  \bibfield  {author} {\bibinfo {author} {\bibfnamefont {Z.-C.}\ \bibnamefont
  {Lin}}, \bibinfo {author} {\bibfnamefont {K.}~\bibnamefont {Yang}}, \bibinfo
  {author} {\bibfnamefont {S.-W.}\ \bibnamefont {Wei}}, \bibinfo {author}
  {\bibfnamefont {Y.-Q.}\ \bibnamefont {Wang}}, \ and\ \bibinfo {author}
  {\bibfnamefont {Y.-X.}\ \bibnamefont {Liu}},\ }\href@noop {} {\  (\bibinfo
  {year} {2020})},\ \Eprint {http://arxiv.org/abs/2006.07913} {arXiv:2006.07913
  [gr-qc]} \BibitemShut {NoStop}%
\bibitem [{\citenamefont {Yang}\ \emph
  {et~al.}(2020{\natexlab{b}})\citenamefont {Yang}, \citenamefont {Wan},
  \citenamefont {Chen}, \citenamefont {Yang},\ and\ \citenamefont
  {Wang}}]{yang2020weak}%
  \BibitemOpen
  \bibfield  {author} {\bibinfo {author} {\bibfnamefont {S.-J.}\ \bibnamefont
  {Yang}}, \bibinfo {author} {\bibfnamefont {J.-J.}\ \bibnamefont {Wan}},
  \bibinfo {author} {\bibfnamefont {J.}~\bibnamefont {Chen}}, \bibinfo {author}
  {\bibfnamefont {J.}~\bibnamefont {Yang}}, \ and\ \bibinfo {author}
  {\bibfnamefont {Y.-Q.}\ \bibnamefont {Wang}},\ }\href@noop {} {\bibfield
  {journal} {\bibinfo  {journal} {arXiv preprint arXiv:2004.07934}\ } (\bibinfo
  {year} {2020}{\natexlab{b}})}\BibitemShut {NoStop}%
\bibitem [{\citenamefont {Feng}\ \emph {et~al.}(2020)\citenamefont {Feng},
  \citenamefont {Gu},\ and\ \citenamefont {Shu}}]{feng2020theoretical}%
  \BibitemOpen
  \bibfield  {author} {\bibinfo {author} {\bibfnamefont {J.-X.}\ \bibnamefont
  {Feng}}, \bibinfo {author} {\bibfnamefont {B.-M.}\ \bibnamefont {Gu}}, \ and\
  \bibinfo {author} {\bibfnamefont {F.-W.}\ \bibnamefont {Shu}},\ }\href@noop
  {} {\bibfield  {journal} {\bibinfo  {journal} {arXiv preprint
  arXiv:2006.16751}\ } (\bibinfo {year} {2020})}\BibitemShut {NoStop}%
\bibitem [{\citenamefont {Hennigar}\ \emph
  {et~al.}(2020{\natexlab{c}})\citenamefont {Hennigar}, \citenamefont
  {Kubiznak}, \citenamefont {Mann},\ and\ \citenamefont
  {Pollack}}]{hennigar2020lower}%
  \BibitemOpen
  \bibfield  {author} {\bibinfo {author} {\bibfnamefont {R.~A.}\ \bibnamefont
  {Hennigar}}, \bibinfo {author} {\bibfnamefont {D.}~\bibnamefont {Kubiznak}},
  \bibinfo {author} {\bibfnamefont {R.~B.}\ \bibnamefont {Mann}}, \ and\
  \bibinfo {author} {\bibfnamefont {C.}~\bibnamefont {Pollack}},\ }\href@noop
  {} {\bibfield  {journal} {\bibinfo  {journal} {arXiv preprint
  arXiv:2004.12995}\ } (\bibinfo {year} {2020}{\natexlab{c}})}\BibitemShut
  {NoStop}%
\bibitem [{\citenamefont {Horndeski}(1974)}]{Horndeski:1974wa}%
  \BibitemOpen
  \bibfield  {author} {\bibinfo {author} {\bibfnamefont {G.~W.}\ \bibnamefont
  {Horndeski}},\ }\href {\doibase 10.1007/BF01807638} {\bibfield  {journal}
  {\bibinfo  {journal} {Int. J. Theor. Phys.}\ }\textbf {\bibinfo {volume}
  {10}},\ \bibinfo {pages} {363} (\bibinfo {year} {1974})}\BibitemShut
  {NoStop}%
\bibitem [{\citenamefont {Kobayashi}(2019)}]{HorndeskiReview}%
  \BibitemOpen
  \bibfield  {author} {\bibinfo {author} {\bibfnamefont {T.}~\bibnamefont
  {Kobayashi}},\ }\href {\doibase 10.1088/1361-6633/ab2429} {\bibfield
  {journal} {\bibinfo  {journal} {Rept. Prog. Phys.}\ }\textbf {\bibinfo
  {volume} {82}},\ \bibinfo {pages} {086901} (\bibinfo {year} {2019})},\
  \Eprint {http://arxiv.org/abs/1901.07183} {arXiv:1901.07183 [gr-qc]}
  \BibitemShut {NoStop}%
\bibitem [{\citenamefont {Lu}\ and\ \citenamefont {Pang}(2020)}]{EGBST1}%
  \BibitemOpen
  \bibfield  {author} {\bibinfo {author} {\bibfnamefont {H.}~\bibnamefont
  {Lu}}\ and\ \bibinfo {author} {\bibfnamefont {Y.}~\bibnamefont {Pang}},\
  }\href@noop {} {\enquote {\bibinfo {title} {{Horndeski Gravity as
  $D\rightarrow4$ Limit of Gauss-Bonnet}},}\ } (\bibinfo {year} {2020}),\
  \Eprint {http://arxiv.org/abs/2003.11552} {arXiv:2003.11552 [gr-qc]}
  \BibitemShut {NoStop}%
\bibitem [{\citenamefont {Kobayashi}(2020)}]{EGBST2}%
  \BibitemOpen
  \bibfield  {author} {\bibinfo {author} {\bibfnamefont {T.}~\bibnamefont
  {Kobayashi}},\ }\href@noop {} {\enquote {\bibinfo {title} {{Effective
  scalar-tensor description of regularized Lovelock gravity in four
  dimensions}},}\ } (\bibinfo {year} {2020}),\ \Eprint
  {http://arxiv.org/abs/2003.12771} {arXiv:2003.12771 [gr-qc]} \BibitemShut
  {NoStop}%
\bibitem [{\citenamefont {Riegert}(1984)}]{Riegert:1984kt}%
  \BibitemOpen
  \bibfield  {author} {\bibinfo {author} {\bibfnamefont {R.}~\bibnamefont
  {Riegert}},\ }\href {\doibase 10.1016/0370-2693(84)90983-3} {\bibfield
  {journal} {\bibinfo  {journal} {Phys. Lett. B}\ }\textbf {\bibinfo {volume}
  {134}},\ \bibinfo {pages} {56} (\bibinfo {year} {1984})}\BibitemShut
  {NoStop}%
\bibitem [{\citenamefont {Komargodski}\ and\ \citenamefont
  {Schwimmer}(2011)}]{Komargodski:2011vj}%
  \BibitemOpen
  \bibfield  {author} {\bibinfo {author} {\bibfnamefont {Z.}~\bibnamefont
  {Komargodski}}\ and\ \bibinfo {author} {\bibfnamefont {A.}~\bibnamefont
  {Schwimmer}},\ }\href {\doibase 10.1007/JHEP12(2011)099} {\bibfield
  {journal} {\bibinfo  {journal} {JHEP}\ }\textbf {\bibinfo {volume} {12}},\
  \bibinfo {pages} {099} (\bibinfo {year} {2011})},\ \Eprint
  {http://arxiv.org/abs/1107.3987} {arXiv:1107.3987 [hep-th]} \BibitemShut
  {NoStop}%
\bibitem [{\citenamefont {Shapiro}\ and\ \citenamefont
  {Jacksenaev}(1994)}]{shapiro1994gauge}%
  \BibitemOpen
  \bibfield  {author} {\bibinfo {author} {\bibfnamefont {I.}~\bibnamefont
  {Shapiro}}\ and\ \bibinfo {author} {\bibfnamefont {A.}~\bibnamefont
  {Jacksenaev}},\ }\href@noop {} {\bibfield  {journal} {\bibinfo  {journal}
  {Physics Letters B}\ }\textbf {\bibinfo {volume} {324}},\ \bibinfo {pages}
  {286} (\bibinfo {year} {1994})}\BibitemShut {NoStop}%
\bibitem [{\citenamefont {Rovelli}(1996)}]{Rovelli:1996dv}%
  \BibitemOpen
  \bibfield  {author} {\bibinfo {author} {\bibfnamefont {C.}~\bibnamefont
  {Rovelli}},\ }\href {\doibase 10.1103/PhysRevLett.77.3288} {\bibfield
  {journal} {\bibinfo  {journal} {Phys. Rev. Lett.}\ }\textbf {\bibinfo
  {volume} {77}},\ \bibinfo {pages} {3288} (\bibinfo {year} {1996})},\ \Eprint
  {http://arxiv.org/abs/gr-qc/9603063} {arXiv:gr-qc/9603063} \BibitemShut
  {NoStop}%
\bibitem [{\citenamefont {Kaul}\ and\ \citenamefont
  {Majumdar}(2000)}]{Kaul:2000kf}%
  \BibitemOpen
  \bibfield  {author} {\bibinfo {author} {\bibfnamefont {R.~K.}\ \bibnamefont
  {Kaul}}\ and\ \bibinfo {author} {\bibfnamefont {P.}~\bibnamefont
  {Majumdar}},\ }\href {\doibase 10.1103/PhysRevLett.84.5255} {\bibfield
  {journal} {\bibinfo  {journal} {Phys. Rev. Lett.}\ }\textbf {\bibinfo
  {volume} {84}},\ \bibinfo {pages} {5255} (\bibinfo {year} {2000})},\ \Eprint
  {http://arxiv.org/abs/gr-qc/0002040} {arXiv:gr-qc/0002040} \BibitemShut
  {NoStop}%
\bibitem [{\citenamefont {Poisson}\ and\ \citenamefont
  {Will}(2014)}]{poisson2014gravity}%
  \BibitemOpen
  \bibfield  {author} {\bibinfo {author} {\bibfnamefont {E.}~\bibnamefont
  {Poisson}}\ and\ \bibinfo {author} {\bibfnamefont {C.~M.}\ \bibnamefont
  {Will}},\ }\href@noop {} {\emph {\bibinfo {title} {Gravity: Newtonian,
  post-newtonian, relativistic}}}\ (\bibinfo  {publisher} {Cambridge University
  Press},\ \bibinfo {year} {2014})\BibitemShut {NoStop}%
\bibitem [{\citenamefont {Will}(1993)}]{Will:1993ns}%
  \BibitemOpen
  \bibfield  {author} {\bibinfo {author} {\bibfnamefont {C.}~\bibnamefont
  {Will}},\ }\href@noop {} {\emph {\bibinfo {title} {{Theory and experiment in
  gravitational physics}}}}\ (\bibinfo {year} {1993})\BibitemShut {NoStop}%
\bibitem [{\citenamefont {Landau}\ and\ \citenamefont
  {Lifschitz}(1975)}]{landau1975classical}%
  \BibitemOpen
  \bibfield  {author} {\bibinfo {author} {\bibfnamefont {L.}~\bibnamefont
  {Landau}}\ and\ \bibinfo {author} {\bibfnamefont {E.}~\bibnamefont
  {Lifschitz}},\ }\href {https://books.google.co.uk/books?id=X18PF4oKyrUC}
  {\emph {\bibinfo {title} {The Classical Theory of Fields: Volume 2}}},\
  Course of theoretical physics\ (\bibinfo  {publisher} {Elsevier Science},\
  \bibinfo {year} {1975})\BibitemShut {NoStop}%
\bibitem [{\citenamefont {Le~Verrier}(1845)}]{le1845theorie}%
  \BibitemOpen
  \bibfield  {author} {\bibinfo {author} {\bibfnamefont {U.~J.}\ \bibnamefont
  {Le~Verrier}},\ }\href@noop {} {\emph {\bibinfo {title} {Theorie du mouvement
  de Mercure}}}\ (\bibinfo  {publisher} {Bachelier},\ \bibinfo {year}
  {1845})\BibitemShut {NoStop}%
\bibitem [{\citenamefont {Einstein}\ \emph {et~al.}(1916)\citenamefont
  {Einstein} \emph {et~al.}}]{einstein1916foundation}%
  \BibitemOpen
  \bibfield  {author} {\bibinfo {author} {\bibfnamefont {A.}~\bibnamefont
  {Einstein}} \emph {et~al.},\ }\href@noop {} {\bibfield  {journal} {\bibinfo
  {journal} {Annalen der Physik}\ }\textbf {\bibinfo {volume} {49}},\ \bibinfo
  {pages} {769} (\bibinfo {year} {1916})}\BibitemShut {NoStop}%
\bibitem [{\citenamefont {Pitjeva}\ and\ \citenamefont
  {Pitjev}(2013)}]{pitjeva2013relativistic}%
  \BibitemOpen
  \bibfield  {author} {\bibinfo {author} {\bibfnamefont {E.}~\bibnamefont
  {Pitjeva}}\ and\ \bibinfo {author} {\bibfnamefont {N.}~\bibnamefont
  {Pitjev}},\ }\href@noop {} {\bibfield  {journal} {\bibinfo  {journal}
  {Monthly Notices of the Royal Astronomical Society}\ }\textbf {\bibinfo
  {volume} {432}},\ \bibinfo {pages} {3431} (\bibinfo {year}
  {2013})}\BibitemShut {NoStop}%
\bibitem [{\citenamefont {Fienga}\ \emph {et~al.}(2011)\citenamefont {Fienga},
  \citenamefont {Laskar}, \citenamefont {Kuchynka}, \citenamefont {Manche},
  \citenamefont {Desvignes}, \citenamefont {Gastineau}, \citenamefont
  {Cognard},\ and\ \citenamefont {Theureau}}]{fienga2011inpop10a}%
  \BibitemOpen
  \bibfield  {author} {\bibinfo {author} {\bibfnamefont {A.}~\bibnamefont
  {Fienga}}, \bibinfo {author} {\bibfnamefont {J.}~\bibnamefont {Laskar}},
  \bibinfo {author} {\bibfnamefont {P.}~\bibnamefont {Kuchynka}}, \bibinfo
  {author} {\bibfnamefont {H.}~\bibnamefont {Manche}}, \bibinfo {author}
  {\bibfnamefont {G.}~\bibnamefont {Desvignes}}, \bibinfo {author}
  {\bibfnamefont {M.}~\bibnamefont {Gastineau}}, \bibinfo {author}
  {\bibfnamefont {I.}~\bibnamefont {Cognard}}, \ and\ \bibinfo {author}
  {\bibfnamefont {G.}~\bibnamefont {Theureau}},\ }\href@noop {} {\bibfield
  {journal} {\bibinfo  {journal} {Celestial Mechanics and Dynamical Astronomy}\
  }\textbf {\bibinfo {volume} {111}},\ \bibinfo {pages} {363} (\bibinfo {year}
  {2011})}\BibitemShut {NoStop}%
\bibitem [{\citenamefont {Lucchesi}\ and\ \citenamefont
  {Peron}(2010)}]{Lucchesi:2010zzb}%
  \BibitemOpen
  \bibfield  {author} {\bibinfo {author} {\bibfnamefont {D.~M.}\ \bibnamefont
  {Lucchesi}}\ and\ \bibinfo {author} {\bibfnamefont {R.}~\bibnamefont
  {Peron}},\ }\href {\doibase 10.1103/PhysRevLett.105.231103} {\bibfield
  {journal} {\bibinfo  {journal} {Phys. Rev. Lett.}\ }\textbf {\bibinfo
  {volume} {105}},\ \bibinfo {pages} {231103} (\bibinfo {year} {2010})},\
  \Eprint {http://arxiv.org/abs/1106.2905} {arXiv:1106.2905 [gr-qc]}
  \BibitemShut {NoStop}%
\bibitem [{\citenamefont {Abuter}\ \emph {et~al.}(2020)\citenamefont {Abuter}
  \emph {et~al.}}]{Abuter:2020dou}%
  \BibitemOpen
  \bibfield  {author} {\bibinfo {author} {\bibfnamefont {R.}~\bibnamefont
  {Abuter}} \emph {et~al.} (\bibinfo {collaboration} {GRAVITY}),\ }\href
  {\doibase 10.1051/0004-6361/202037813} {\bibfield  {journal} {\bibinfo
  {journal} {Astron. Astrophys.}\ }\textbf {\bibinfo {volume} {636}},\ \bibinfo
  {pages} {L5} (\bibinfo {year} {2020})},\ \Eprint
  {http://arxiv.org/abs/2004.07187} {arXiv:2004.07187 [astro-ph.GA]}
  \BibitemShut {NoStop}%
\bibitem [{\citenamefont {Hawking}(1972)}]{hawking1972black}%
  \BibitemOpen
  \bibfield  {author} {\bibinfo {author} {\bibfnamefont {S.}~\bibnamefont
  {Hawking}},\ }\href@noop {} {\bibfield  {journal} {\bibinfo  {journal}
  {Communications in Mathematical Physics}\ }\textbf {\bibinfo {volume} {25}},\
  \bibinfo {pages} {167} (\bibinfo {year} {1972})}\BibitemShut {NoStop}%
\bibitem [{\citenamefont {Hulse}\ and\ \citenamefont
  {Taylor}(1975)}]{hulse1975discovery}%
  \BibitemOpen
  \bibfield  {author} {\bibinfo {author} {\bibfnamefont {R.~A.}\ \bibnamefont
  {Hulse}}\ and\ \bibinfo {author} {\bibfnamefont {J.~H.}\ \bibnamefont
  {Taylor}},\ }\href@noop {} {\bibfield  {journal} {\bibinfo  {journal} {The
  Astrophysical Journal}\ }\textbf {\bibinfo {volume} {195}},\ \bibinfo {pages}
  {L51} (\bibinfo {year} {1975})}\BibitemShut {NoStop}%
\bibitem [{\citenamefont {Lyne}\ \emph {et~al.}(2004)\citenamefont {Lyne},
  \citenamefont {Burgay}, \citenamefont {Kramer}, \citenamefont {Possenti},
  \citenamefont {Manchester}, \citenamefont {Camilo}, \citenamefont
  {McLaughlin}, \citenamefont {Lorimer}, \citenamefont {D'Amico}, \citenamefont
  {Joshi} \emph {et~al.}}]{lyne2004double}%
  \BibitemOpen
  \bibfield  {author} {\bibinfo {author} {\bibfnamefont {A.~G.}\ \bibnamefont
  {Lyne}}, \bibinfo {author} {\bibfnamefont {M.}~\bibnamefont {Burgay}},
  \bibinfo {author} {\bibfnamefont {M.}~\bibnamefont {Kramer}}, \bibinfo
  {author} {\bibfnamefont {A.}~\bibnamefont {Possenti}}, \bibinfo {author}
  {\bibfnamefont {R.}~\bibnamefont {Manchester}}, \bibinfo {author}
  {\bibfnamefont {F.}~\bibnamefont {Camilo}}, \bibinfo {author} {\bibfnamefont
  {M.}~\bibnamefont {McLaughlin}}, \bibinfo {author} {\bibfnamefont
  {D.}~\bibnamefont {Lorimer}}, \bibinfo {author} {\bibfnamefont
  {N.}~\bibnamefont {D'Amico}}, \bibinfo {author} {\bibfnamefont
  {B.}~\bibnamefont {Joshi}},  \emph {et~al.},\ }\href@noop {} {\bibfield
  {journal} {\bibinfo  {journal} {Science}\ }\textbf {\bibinfo {volume}
  {303}},\ \bibinfo {pages} {1153} (\bibinfo {year} {2004})}\BibitemShut
  {NoStop}%
\bibitem [{\citenamefont {Kramer}\ \emph {et~al.}(2006)\citenamefont {Kramer}
  \emph {et~al.}}]{Kramer:2006nb}%
  \BibitemOpen
  \bibfield  {author} {\bibinfo {author} {\bibfnamefont {M.}~\bibnamefont
  {Kramer}} \emph {et~al.},\ }\href {\doibase 10.1126/science.1132305}
  {\bibfield  {journal} {\bibinfo  {journal} {Science}\ }\textbf {\bibinfo
  {volume} {314}},\ \bibinfo {pages} {97} (\bibinfo {year} {2006})},\ \Eprint
  {http://arxiv.org/abs/astro-ph/0609417} {arXiv:astro-ph/0609417} \BibitemShut
  {NoStop}%
\bibitem [{\citenamefont {Damour}\ and\ \citenamefont
  {Esposito-Farese}(1993)}]{damour1993nonperturbative}%
  \BibitemOpen
  \bibfield  {author} {\bibinfo {author} {\bibfnamefont {T.}~\bibnamefont
  {Damour}}\ and\ \bibinfo {author} {\bibfnamefont {G.}~\bibnamefont
  {Esposito-Farese}},\ }\href@noop {} {\bibfield  {journal} {\bibinfo
  {journal} {Physical Review Letters}\ }\textbf {\bibinfo {volume} {70}},\
  \bibinfo {pages} {2220} (\bibinfo {year} {1993})}\BibitemShut {NoStop}%
\bibitem [{\citenamefont {Shapiro}\ \emph {et~al.}(2004)\citenamefont
  {Shapiro}, \citenamefont {Davis}, \citenamefont {Lebach},\ and\ \citenamefont
  {Gregory}}]{shapiro2004measurement}%
  \BibitemOpen
  \bibfield  {author} {\bibinfo {author} {\bibfnamefont {S.~S.}\ \bibnamefont
  {Shapiro}}, \bibinfo {author} {\bibfnamefont {J.~L.}\ \bibnamefont {Davis}},
  \bibinfo {author} {\bibfnamefont {D.~E.}\ \bibnamefont {Lebach}}, \ and\
  \bibinfo {author} {\bibfnamefont {J.}~\bibnamefont {Gregory}},\ }\href@noop
  {} {\bibfield  {journal} {\bibinfo  {journal} {Physical Review Letters}\
  }\textbf {\bibinfo {volume} {92}},\ \bibinfo {pages} {121101} (\bibinfo
  {year} {2004})}\BibitemShut {NoStop}%
\bibitem [{\citenamefont {Bertotti}\ \emph {et~al.}(2003)\citenamefont
  {Bertotti}, \citenamefont {Iess},\ and\ \citenamefont
  {Tortora}}]{bertotti2003test}%
  \BibitemOpen
  \bibfield  {author} {\bibinfo {author} {\bibfnamefont {B.}~\bibnamefont
  {Bertotti}}, \bibinfo {author} {\bibfnamefont {L.}~\bibnamefont {Iess}}, \
  and\ \bibinfo {author} {\bibfnamefont {P.}~\bibnamefont {Tortora}},\
  }\href@noop {} {\bibfield  {journal} {\bibinfo  {journal} {Nature}\ }\textbf
  {\bibinfo {volume} {425}},\ \bibinfo {pages} {374} (\bibinfo {year}
  {2003})}\BibitemShut {NoStop}%
\bibitem [{\citenamefont {Abbott}\ \emph
  {et~al.}(2017{\natexlab{a}})\citenamefont {Abbott} \emph
  {et~al.}}]{TheLIGOScientific:2017qsa}%
  \BibitemOpen
  \bibfield  {author} {\bibinfo {author} {\bibfnamefont {B.}~\bibnamefont
  {Abbott}} \emph {et~al.} (\bibinfo {collaboration} {LIGO Scientific,
  Virgo}),\ }\href {\doibase 10.1103/PhysRevLett.119.161101} {\bibfield
  {journal} {\bibinfo  {journal} {Phys. Rev. Lett.}\ }\textbf {\bibinfo
  {volume} {119}},\ \bibinfo {pages} {161101} (\bibinfo {year}
  {2017}{\natexlab{a}})},\ \Eprint {http://arxiv.org/abs/1710.05832}
  {arXiv:1710.05832 [gr-qc]} \BibitemShut {NoStop}%
\bibitem [{\citenamefont {Goldstein}\ \emph {et~al.}(2017)\citenamefont
  {Goldstein} \emph {et~al.}}]{Goldstein:2017mmi}%
  \BibitemOpen
  \bibfield  {author} {\bibinfo {author} {\bibfnamefont {A.}~\bibnamefont
  {Goldstein}} \emph {et~al.},\ }\href {\doibase 10.3847/2041-8213/aa8f41}
  {\bibfield  {journal} {\bibinfo  {journal} {Astrophys. J. Lett.}\ }\textbf
  {\bibinfo {volume} {848}},\ \bibinfo {pages} {L14} (\bibinfo {year}
  {2017})},\ \Eprint {http://arxiv.org/abs/1710.05446} {arXiv:1710.05446
  [astro-ph.HE]} \BibitemShut {NoStop}%
\bibitem [{\citenamefont {Savchenko}\ \emph {et~al.}(2017)\citenamefont
  {Savchenko} \emph {et~al.}}]{Savchenko:2017ffs}%
  \BibitemOpen
  \bibfield  {author} {\bibinfo {author} {\bibfnamefont {V.}~\bibnamefont
  {Savchenko}} \emph {et~al.},\ }\href {\doibase 10.3847/2041-8213/aa8f94}
  {\bibfield  {journal} {\bibinfo  {journal} {Astrophys. J. Lett.}\ }\textbf
  {\bibinfo {volume} {848}},\ \bibinfo {pages} {L15} (\bibinfo {year}
  {2017})},\ \Eprint {http://arxiv.org/abs/1710.05449} {arXiv:1710.05449
  [astro-ph.HE]} \BibitemShut {NoStop}%
\bibitem [{\citenamefont {Baker}\ \emph {et~al.}(2017)\citenamefont {Baker},
  \citenamefont {Bellini}, \citenamefont {Ferreira}, \citenamefont {Lagos},
  \citenamefont {Noller},\ and\ \citenamefont {Sawicki}}]{Baker:2017hug}%
  \BibitemOpen
  \bibfield  {author} {\bibinfo {author} {\bibfnamefont {T.}~\bibnamefont
  {Baker}}, \bibinfo {author} {\bibfnamefont {E.}~\bibnamefont {Bellini}},
  \bibinfo {author} {\bibfnamefont {P.}~\bibnamefont {Ferreira}}, \bibinfo
  {author} {\bibfnamefont {M.}~\bibnamefont {Lagos}}, \bibinfo {author}
  {\bibfnamefont {J.}~\bibnamefont {Noller}}, \ and\ \bibinfo {author}
  {\bibfnamefont {I.}~\bibnamefont {Sawicki}},\ }\href {\doibase
  10.1103/PhysRevLett.119.251301} {\bibfield  {journal} {\bibinfo  {journal}
  {Phys. Rev. Lett.}\ }\textbf {\bibinfo {volume} {119}},\ \bibinfo {pages}
  {251301} (\bibinfo {year} {2017})},\ \Eprint
  {http://arxiv.org/abs/1710.06394} {arXiv:1710.06394 [astro-ph.CO]}
  \BibitemShut {NoStop}%
\bibitem [{\citenamefont {Creminelli}\ and\ \citenamefont
  {Vernizzi}(2017)}]{Creminelli:2017sry}%
  \BibitemOpen
  \bibfield  {author} {\bibinfo {author} {\bibfnamefont {P.}~\bibnamefont
  {Creminelli}}\ and\ \bibinfo {author} {\bibfnamefont {F.}~\bibnamefont
  {Vernizzi}},\ }\href {\doibase 10.1103/PhysRevLett.119.251302} {\bibfield
  {journal} {\bibinfo  {journal} {Phys. Rev. Lett.}\ }\textbf {\bibinfo
  {volume} {119}},\ \bibinfo {pages} {251302} (\bibinfo {year} {2017})},\
  \Eprint {http://arxiv.org/abs/1710.05877} {arXiv:1710.05877 [astro-ph.CO]}
  \BibitemShut {NoStop}%
\bibitem [{\citenamefont {Ezquiaga}\ and\ \citenamefont
  {Zumalacárregui}(2017)}]{Ezquiaga:2017ekz}%
  \BibitemOpen
  \bibfield  {author} {\bibinfo {author} {\bibfnamefont {J.~M.}\ \bibnamefont
  {Ezquiaga}}\ and\ \bibinfo {author} {\bibfnamefont {M.}~\bibnamefont
  {Zumalacárregui}},\ }\href {\doibase 10.1103/PhysRevLett.119.251304}
  {\bibfield  {journal} {\bibinfo  {journal} {Phys. Rev. Lett.}\ }\textbf
  {\bibinfo {volume} {119}},\ \bibinfo {pages} {251304} (\bibinfo {year}
  {2017})},\ \Eprint {http://arxiv.org/abs/1710.05901} {arXiv:1710.05901
  [astro-ph.CO]} \BibitemShut {NoStop}%
\bibitem [{\citenamefont {Akiyama}\ \emph
  {et~al.}(2019{\natexlab{a}})\citenamefont {Akiyama} \emph
  {et~al.}}]{Akiyama:2019cqa}%
  \BibitemOpen
  \bibfield  {author} {\bibinfo {author} {\bibfnamefont {K.}~\bibnamefont
  {Akiyama}} \emph {et~al.} (\bibinfo {collaboration} {Event Horizon
  Telescope}),\ }\href {\doibase 10.3847/2041-8213/ab0ec7} {\bibfield
  {journal} {\bibinfo  {journal} {Astrophys. J.}\ }\textbf {\bibinfo {volume}
  {875}},\ \bibinfo {pages} {L1} (\bibinfo {year} {2019}{\natexlab{a}})},\
  \Eprint {http://arxiv.org/abs/1906.11238} {arXiv:1906.11238 [astro-ph.GA]}
  \BibitemShut {NoStop}%
\bibitem [{\citenamefont {Akiyama}\ \emph
  {et~al.}(2019{\natexlab{b}})\citenamefont {Akiyama} \emph
  {et~al.}}]{Akiyama:2019eap}%
  \BibitemOpen
  \bibfield  {author} {\bibinfo {author} {\bibfnamefont {K.}~\bibnamefont
  {Akiyama}} \emph {et~al.} (\bibinfo {collaboration} {Event Horizon
  Telescope}),\ }\href {\doibase 10.3847/2041-8213/ab1141} {\bibfield
  {journal} {\bibinfo  {journal} {Astrophys. J. Lett.}\ }\textbf {\bibinfo
  {volume} {875}},\ \bibinfo {pages} {L6} (\bibinfo {year}
  {2019}{\natexlab{b}})},\ \Eprint {http://arxiv.org/abs/1906.11243}
  {arXiv:1906.11243 [astro-ph.GA]} \BibitemShut {NoStop}%
\bibitem [{\citenamefont {Vagnozzi}\ and\ \citenamefont
  {Visinelli}(2019)}]{Vagnozzi:2019apd}%
  \BibitemOpen
  \bibfield  {author} {\bibinfo {author} {\bibfnamefont {S.}~\bibnamefont
  {Vagnozzi}}\ and\ \bibinfo {author} {\bibfnamefont {L.}~\bibnamefont
  {Visinelli}},\ }\href {\doibase 10.1103/PhysRevD.100.024020} {\bibfield
  {journal} {\bibinfo  {journal} {Phys. Rev. D}\ }\textbf {\bibinfo {volume}
  {100}},\ \bibinfo {pages} {024020} (\bibinfo {year} {2019})},\ \Eprint
  {http://arxiv.org/abs/1905.12421} {arXiv:1905.12421 [gr-qc]} \BibitemShut
  {NoStop}%
\bibitem [{\citenamefont {Held}\ \emph {et~al.}(2019)\citenamefont {Held},
  \citenamefont {Gold},\ and\ \citenamefont {Eichhorn}}]{Held:2019xde}%
  \BibitemOpen
  \bibfield  {author} {\bibinfo {author} {\bibfnamefont {A.}~\bibnamefont
  {Held}}, \bibinfo {author} {\bibfnamefont {R.}~\bibnamefont {Gold}}, \ and\
  \bibinfo {author} {\bibfnamefont {A.}~\bibnamefont {Eichhorn}},\ }\href
  {\doibase 10.1088/1475-7516/2019/06/029} {\bibfield  {journal} {\bibinfo
  {journal} {JCAP}\ }\textbf {\bibinfo {volume} {06}},\ \bibinfo {pages} {029}
  (\bibinfo {year} {2019})},\ \Eprint {http://arxiv.org/abs/1904.07133}
  {arXiv:1904.07133 [gr-qc]} \BibitemShut {NoStop}%
\bibitem [{\citenamefont {Zhu}\ \emph {et~al.}(2019)\citenamefont {Zhu},
  \citenamefont {Wu}, \citenamefont {Jamil},\ and\ \citenamefont
  {Jusufi}}]{Zhu:2019ura}%
  \BibitemOpen
  \bibfield  {author} {\bibinfo {author} {\bibfnamefont {T.}~\bibnamefont
  {Zhu}}, \bibinfo {author} {\bibfnamefont {Q.}~\bibnamefont {Wu}}, \bibinfo
  {author} {\bibfnamefont {M.}~\bibnamefont {Jamil}}, \ and\ \bibinfo {author}
  {\bibfnamefont {K.}~\bibnamefont {Jusufi}},\ }\href {\doibase
  10.1103/PhysRevD.100.044055} {\bibfield  {journal} {\bibinfo  {journal}
  {Phys. Rev. D}\ }\textbf {\bibinfo {volume} {100}},\ \bibinfo {pages}
  {044055} (\bibinfo {year} {2019})},\ \Eprint
  {http://arxiv.org/abs/1906.05673} {arXiv:1906.05673 [gr-qc]} \BibitemShut
  {NoStop}%
\bibitem [{\citenamefont {Boulware}\ and\ \citenamefont
  {Deser}(1985)}]{boulware1985string}%
  \BibitemOpen
  \bibfield  {author} {\bibinfo {author} {\bibfnamefont {D.~G.}\ \bibnamefont
  {Boulware}}\ and\ \bibinfo {author} {\bibfnamefont {S.}~\bibnamefont
  {Deser}},\ }\href@noop {} {\bibfield  {journal} {\bibinfo  {journal}
  {Physical Review Letters}\ }\textbf {\bibinfo {volume} {55}},\ \bibinfo
  {pages} {2656} (\bibinfo {year} {1985})}\BibitemShut {NoStop}%
\bibitem [{\citenamefont {{Gebhardt}}\ \emph {et~al.}(2011)\citenamefont
  {{Gebhardt}}, \citenamefont {{Adams}}, \citenamefont {{Richstone}},
  \citenamefont {{Lauer}}, \citenamefont {{Faber}}, \citenamefont
  {{G{\"u}ltekin}}, \citenamefont {{Murphy}},\ and\ \citenamefont
  {{Tremaine}}}]{2011ApJ...729..119G}%
  \BibitemOpen
  \bibfield  {author} {\bibinfo {author} {\bibfnamefont {K.}~\bibnamefont
  {{Gebhardt}}}, \bibinfo {author} {\bibfnamefont {J.}~\bibnamefont {{Adams}}},
  \bibinfo {author} {\bibfnamefont {D.}~\bibnamefont {{Richstone}}}, \bibinfo
  {author} {\bibfnamefont {T.~R.}\ \bibnamefont {{Lauer}}}, \bibinfo {author}
  {\bibfnamefont {S.~M.}\ \bibnamefont {{Faber}}}, \bibinfo {author}
  {\bibfnamefont {K.}~\bibnamefont {{G{\"u}ltekin}}}, \bibinfo {author}
  {\bibfnamefont {J.}~\bibnamefont {{Murphy}}}, \ and\ \bibinfo {author}
  {\bibfnamefont {S.}~\bibnamefont {{Tremaine}}},\ }\href {\doibase
  10.1088/0004-637X/729/2/119} {\bibfield  {journal} {\bibinfo  {journal}
  {\apj}\ }\textbf {\bibinfo {volume} {729}},\ \bibinfo {eid} {119} (\bibinfo
  {year} {2011})},\ \Eprint {http://arxiv.org/abs/1101.1954} {arXiv:1101.1954
  [astro-ph.CO]} \BibitemShut {NoStop}%
\bibitem [{\citenamefont {{Walsh}}\ \emph {et~al.}(2013)\citenamefont
  {{Walsh}}, \citenamefont {{Barth}}, \citenamefont {{Ho}},\ and\ \citenamefont
  {{Sarzi}}}]{2013ApJ...770...86W}%
  \BibitemOpen
  \bibfield  {author} {\bibinfo {author} {\bibfnamefont {J.~L.}\ \bibnamefont
  {{Walsh}}}, \bibinfo {author} {\bibfnamefont {A.~J.}\ \bibnamefont
  {{Barth}}}, \bibinfo {author} {\bibfnamefont {L.~C.}\ \bibnamefont {{Ho}}}, \
  and\ \bibinfo {author} {\bibfnamefont {M.}~\bibnamefont {{Sarzi}}},\ }\href
  {\doibase 10.1088/0004-637X/770/2/86} {\bibfield  {journal} {\bibinfo
  {journal} {\apj}\ }\textbf {\bibinfo {volume} {770}},\ \bibinfo {eid} {86}
  (\bibinfo {year} {2013})},\ \Eprint {http://arxiv.org/abs/1304.7273}
  {arXiv:1304.7273 [astro-ph.CO]} \BibitemShut {NoStop}%
\bibitem [{\citenamefont {Abbott}\ \emph
  {et~al.}(2016{\natexlab{a}})\citenamefont {Abbott} \emph
  {et~al.}}]{Abbott:2016blz}%
  \BibitemOpen
  \bibfield  {author} {\bibinfo {author} {\bibfnamefont {B.}~\bibnamefont
  {Abbott}} \emph {et~al.} (\bibinfo {collaboration} {LIGO Scientific,
  Virgo}),\ }\href {\doibase 10.1103/PhysRevLett.116.061102} {\bibfield
  {journal} {\bibinfo  {journal} {Phys. Rev. Lett.}\ }\textbf {\bibinfo
  {volume} {116}},\ \bibinfo {pages} {061102} (\bibinfo {year}
  {2016}{\natexlab{a}})},\ \Eprint {http://arxiv.org/abs/1602.03837}
  {arXiv:1602.03837 [gr-qc]} \BibitemShut {NoStop}%
\bibitem [{\citenamefont {Abbott}\ \emph
  {et~al.}(2016{\natexlab{b}})\citenamefont {Abbott} \emph
  {et~al.}}]{TheLIGOScientific:2016wfe}%
  \BibitemOpen
  \bibfield  {author} {\bibinfo {author} {\bibfnamefont {B.}~\bibnamefont
  {Abbott}} \emph {et~al.} (\bibinfo {collaboration} {LIGO Scientific,
  Virgo}),\ }\href {\doibase 10.1103/PhysRevLett.116.241102} {\bibfield
  {journal} {\bibinfo  {journal} {Phys. Rev. Lett.}\ }\textbf {\bibinfo
  {volume} {116}},\ \bibinfo {pages} {241102} (\bibinfo {year}
  {2016}{\natexlab{b}})},\ \Eprint {http://arxiv.org/abs/1602.03840}
  {arXiv:1602.03840 [gr-qc]} \BibitemShut {NoStop}%
\bibitem [{\citenamefont {Abbott}\ \emph
  {et~al.}(2016{\natexlab{c}})\citenamefont {Abbott} \emph
  {et~al.}}]{TheLIGOScientific:2016src}%
  \BibitemOpen
  \bibfield  {author} {\bibinfo {author} {\bibfnamefont {B.}~\bibnamefont
  {Abbott}} \emph {et~al.} (\bibinfo {collaboration} {LIGO Scientific,
  Virgo}),\ }\href {\doibase 10.1103/PhysRevLett.116.221101} {\bibfield
  {journal} {\bibinfo  {journal} {Phys. Rev. Lett.}\ }\textbf {\bibinfo
  {volume} {116}},\ \bibinfo {pages} {221101} (\bibinfo {year}
  {2016}{\natexlab{c}})},\ \bibinfo {note} {[Erratum: Phys.Rev.Lett. 121,
  129902 (2018)]},\ \Eprint {http://arxiv.org/abs/1602.03841} {arXiv:1602.03841
  [gr-qc]} \BibitemShut {NoStop}%
\bibitem [{\citenamefont {Abbott}\ \emph
  {et~al.}(2017{\natexlab{b}})\citenamefont {Abbott} \emph
  {et~al.}}]{Abbott:2016bqf}%
  \BibitemOpen
  \bibfield  {author} {\bibinfo {author} {\bibfnamefont {B.~P.}\ \bibnamefont
  {Abbott}} \emph {et~al.} (\bibinfo {collaboration} {LIGO Scientific,
  Virgo}),\ }\href {\doibase 10.1002/andp.201600209} {\bibfield  {journal}
  {\bibinfo  {journal} {Annalen Phys.}\ }\textbf {\bibinfo {volume} {529}},\
  \bibinfo {pages} {1600209} (\bibinfo {year} {2017}{\natexlab{b}})},\ \Eprint
  {http://arxiv.org/abs/1608.01940} {arXiv:1608.01940 [gr-qc]} \BibitemShut
  {NoStop}%
\bibitem [{\citenamefont {Abbott}\ \emph
  {et~al.}(2017{\natexlab{c}})\citenamefont {Abbott} \emph
  {et~al.}}]{Abbott:2017gyy}%
  \BibitemOpen
  \bibfield  {author} {\bibinfo {author} {\bibfnamefont {B.~P.}\ \bibnamefont
  {Abbott}} \emph {et~al.} (\bibinfo {collaboration} {LIGO Scientific,
  Virgo}),\ }\href {\doibase 10.3847/2041-8213/aa9f0c} {\bibfield  {journal}
  {\bibinfo  {journal} {Astrophys. J.}\ }\textbf {\bibinfo {volume} {851}},\
  \bibinfo {pages} {L35} (\bibinfo {year} {2017}{\natexlab{c}})},\ \Eprint
  {http://arxiv.org/abs/1711.05578} {arXiv:1711.05578 [astro-ph.HE]}
  \BibitemShut {NoStop}%
\bibitem [{\citenamefont {Abbott}\ \emph {et~al.}(2020)\citenamefont {Abbott}
  \emph {et~al.}}]{Abbott_2020}%
  \BibitemOpen
  \bibfield  {author} {\bibinfo {author} {\bibfnamefont {B.}~\bibnamefont
  {Abbott}} \emph {et~al.},\ }\href {\doibase 10.3847/2041-8213/ab960f}
  {\bibfield  {journal} {\bibinfo  {journal} {The Astrophysical Journal}\
  }\textbf {\bibinfo {volume} {896}},\ \bibinfo {pages} {L44} (\bibinfo {year}
  {2020})}\BibitemShut {NoStop}%
\bibitem [{\citenamefont {Fischbach}\ and\ \citenamefont
  {Krause}(1999)}]{Fischbach:1999bx}%
  \BibitemOpen
  \bibfield  {author} {\bibinfo {author} {\bibfnamefont {E.}~\bibnamefont
  {Fischbach}}\ and\ \bibinfo {author} {\bibfnamefont {D.~E.}\ \bibnamefont
  {Krause}},\ }\href {\doibase 10.1103/PhysRevLett.83.3593} {\bibfield
  {journal} {\bibinfo  {journal} {Phys. Rev. Lett.}\ }\textbf {\bibinfo
  {volume} {83}},\ \bibinfo {pages} {3593} (\bibinfo {year} {1999})},\ \Eprint
  {http://arxiv.org/abs/hep-ph/9906240} {arXiv:hep-ph/9906240} \BibitemShut
  {NoStop}%
\bibitem [{\citenamefont {Adelberger}\ \emph {et~al.}(2007)\citenamefont
  {Adelberger}, \citenamefont {Heckel}, \citenamefont {Hoedl}, \citenamefont
  {Hoyle}, \citenamefont {Kapner},\ and\ \citenamefont
  {Upadhye}}]{Adelberger:2006dh}%
  \BibitemOpen
  \bibfield  {author} {\bibinfo {author} {\bibfnamefont {E.}~\bibnamefont
  {Adelberger}}, \bibinfo {author} {\bibfnamefont {B.~R.}\ \bibnamefont
  {Heckel}}, \bibinfo {author} {\bibfnamefont {S.~A.}\ \bibnamefont {Hoedl}},
  \bibinfo {author} {\bibfnamefont {C.}~\bibnamefont {Hoyle}}, \bibinfo
  {author} {\bibfnamefont {D.}~\bibnamefont {Kapner}}, \ and\ \bibinfo {author}
  {\bibfnamefont {A.}~\bibnamefont {Upadhye}},\ }\href {\doibase
  10.1103/PhysRevLett.98.131104} {\bibfield  {journal} {\bibinfo  {journal}
  {Phys. Rev. Lett.}\ }\textbf {\bibinfo {volume} {98}},\ \bibinfo {pages}
  {131104} (\bibinfo {year} {2007})},\ \Eprint
  {http://arxiv.org/abs/hep-ph/0611223} {arXiv:hep-ph/0611223} \BibitemShut
  {NoStop}%
\bibitem [{\citenamefont {Tan}\ \emph {et~al.}(2016)\citenamefont {Tan},
  \citenamefont {Yang}, \citenamefont {Shao}, \citenamefont {Li}, \citenamefont
  {Du}, \citenamefont {Zhan}, \citenamefont {Wang}, \citenamefont {Luo},
  \citenamefont {Tu},\ and\ \citenamefont {Luo}}]{PhysRevLett.116.131101}%
  \BibitemOpen
  \bibfield  {author} {\bibinfo {author} {\bibfnamefont {W.-H.}\ \bibnamefont
  {Tan}}, \bibinfo {author} {\bibfnamefont {S.-Q.}\ \bibnamefont {Yang}},
  \bibinfo {author} {\bibfnamefont {C.-G.}\ \bibnamefont {Shao}}, \bibinfo
  {author} {\bibfnamefont {J.}~\bibnamefont {Li}}, \bibinfo {author}
  {\bibfnamefont {A.-B.}\ \bibnamefont {Du}}, \bibinfo {author} {\bibfnamefont
  {B.-F.}\ \bibnamefont {Zhan}}, \bibinfo {author} {\bibfnamefont {Q.-L.}\
  \bibnamefont {Wang}}, \bibinfo {author} {\bibfnamefont {P.-S.}\ \bibnamefont
  {Luo}}, \bibinfo {author} {\bibfnamefont {L.-C.}\ \bibnamefont {Tu}}, \ and\
  \bibinfo {author} {\bibfnamefont {J.}~\bibnamefont {Luo}},\ }\href {\doibase
  10.1103/PhysRevLett.116.131101} {\bibfield  {journal} {\bibinfo  {journal}
  {Phys. Rev. Lett.}\ }\textbf {\bibinfo {volume} {116}},\ \bibinfo {pages}
  {131101} (\bibinfo {year} {2016})}\BibitemShut {NoStop}%
\bibitem [{\citenamefont {Carroll}\ and\ \citenamefont
  {Kaplinghat}(2002)}]{Carroll:2001bv}%
  \BibitemOpen
  \bibfield  {author} {\bibinfo {author} {\bibfnamefont {S.~M.}\ \bibnamefont
  {Carroll}}\ and\ \bibinfo {author} {\bibfnamefont {M.}~\bibnamefont
  {Kaplinghat}},\ }\href {\doibase 10.1103/PhysRevD.65.063507} {\bibfield
  {journal} {\bibinfo  {journal} {Phys. Rev. D}\ }\textbf {\bibinfo {volume}
  {65}},\ \bibinfo {pages} {063507} (\bibinfo {year} {2002})},\ \Eprint
  {http://arxiv.org/abs/astro-ph/0108002} {arXiv:astro-ph/0108002} \BibitemShut
  {NoStop}%
\bibitem [{\citenamefont {Arbey}(2012)}]{Arbey:2011nf}%
  \BibitemOpen
  \bibfield  {author} {\bibinfo {author} {\bibfnamefont {A.}~\bibnamefont
  {Arbey}},\ }\href {\doibase 10.1016/j.cpc.2012.03.018} {\bibfield  {journal}
  {\bibinfo  {journal} {Comput. Phys. Commun.}\ }\textbf {\bibinfo {volume}
  {183}},\ \bibinfo {pages} {1822} (\bibinfo {year} {2012})},\ \Eprint
  {http://arxiv.org/abs/1106.1363} {arXiv:1106.1363 [astro-ph.CO]} \BibitemShut
  {NoStop}%
\bibitem [{\citenamefont {Pitrou}\ \emph {et~al.}(2018)\citenamefont {Pitrou},
  \citenamefont {Coc}, \citenamefont {Uzan},\ and\ \citenamefont
  {Vangioni}}]{Pitrou:2018cgg}%
  \BibitemOpen
  \bibfield  {author} {\bibinfo {author} {\bibfnamefont {C.}~\bibnamefont
  {Pitrou}}, \bibinfo {author} {\bibfnamefont {A.}~\bibnamefont {Coc}},
  \bibinfo {author} {\bibfnamefont {J.-P.}\ \bibnamefont {Uzan}}, \ and\
  \bibinfo {author} {\bibfnamefont {E.}~\bibnamefont {Vangioni}},\ }\href
  {\doibase 10.1016/j.physrep.2018.04.005} {\bibfield  {journal} {\bibinfo
  {journal} {Phys. Rept.}\ }\textbf {\bibinfo {volume} {754}},\ \bibinfo
  {pages} {1} (\bibinfo {year} {2018})},\ \Eprint
  {http://arxiv.org/abs/1801.08023} {arXiv:1801.08023 [astro-ph.CO]}
  \BibitemShut {NoStop}%
\bibitem [{\citenamefont {Aver}\ \emph {et~al.}(2015)\citenamefont {Aver},
  \citenamefont {Olive},\ and\ \citenamefont {Skillman}}]{Aver:2015iza}%
  \BibitemOpen
  \bibfield  {author} {\bibinfo {author} {\bibfnamefont {E.}~\bibnamefont
  {Aver}}, \bibinfo {author} {\bibfnamefont {K.~A.}\ \bibnamefont {Olive}}, \
  and\ \bibinfo {author} {\bibfnamefont {E.~D.}\ \bibnamefont {Skillman}},\
  }\href {\doibase 10.1088/1475-7516/2015/07/011} {\bibfield  {journal}
  {\bibinfo  {journal} {JCAP}\ }\textbf {\bibinfo {volume} {07}},\ \bibinfo
  {pages} {011} (\bibinfo {year} {2015})},\ \Eprint
  {http://arxiv.org/abs/1503.08146} {arXiv:1503.08146 [astro-ph.CO]}
  \BibitemShut {NoStop}%
\bibitem [{\citenamefont {Casalino}\ and\ \citenamefont
  {Sebastiani}(2020)}]{Casalino:2020pyv}%
  \BibitemOpen
  \bibfield  {author} {\bibinfo {author} {\bibfnamefont {A.}~\bibnamefont
  {Casalino}}\ and\ \bibinfo {author} {\bibfnamefont {L.}~\bibnamefont
  {Sebastiani}},\ }\href@noop {} {\  (\bibinfo {year} {2020})},\ \Eprint
  {http://arxiv.org/abs/2004.10229} {arXiv:2004.10229 [gr-qc]} \BibitemShut
  {NoStop}%
\bibitem [{\citenamefont {Akrami}\ \emph {et~al.}(2018)\citenamefont {Akrami}
  \emph {et~al.}}]{Akrami:2018odb}%
  \BibitemOpen
  \bibfield  {author} {\bibinfo {author} {\bibfnamefont {Y.}~\bibnamefont
  {Akrami}} \emph {et~al.} (\bibinfo {collaboration} {Planck}),\ }\href@noop {}
  {\  (\bibinfo {year} {2018})},\ \Eprint {http://arxiv.org/abs/1807.06211}
  {arXiv:1807.06211 [astro-ph.CO]} \BibitemShut {NoStop}%
\end{thebibliography}%

\end{document}